\documentclass{new_tlp}

\usepackage{xspace,xcolor,cmll}
\usepackage{tikz}
\usetikzlibrary{cd} %DM%
\usepackage{enumitem}

\usepackage{amssymb}
\usepackage{stmaryrd}
\usepackage{amsmath}
\usepackage{proof}
\usepackage{url,breakurl}
\usepackage{booktabs}
\usepackage{hyperref} % Allowing this line, causes arxiv to overfill lines
\hypersetup{
     colorlinks=true,
     linkcolor=darkblue,
     filecolor=darkblue,
     citecolor = darkblue,
     urlcolor=darkblue,
 }

\usepackage{listings}
\usepackage{letltxmacro}
\colorlet{lprolog}{blue!70!black}
\colorlet{abellatop}{blue!70!green}
\colorlet{abellatac}{orange!30!black}
\colorlet{abellabad}{red!80!yellow}
\definecolor{bleu}{HTML}{000DB3}

\lstdefinelanguage{lprolog}{%
  alsoletter={-},
  classoffset=0,%
  morekeywords={sig,module,type,kind,pi,sigma,end,infixl,infixr,o},%
  keywordstyle=\color{lprolog},%
  classoffset=0,%
  otherkeywords={:-,=>,<=,\&},%
  sensitive=true,%
  morestring=[bd]",%
  morecomment=[l]\%,%
  morecomment=[n]{/*}{*/},%
}

\lstdefinelanguage{abella}[]{lprolog}{%
  alsoletter={-},
  classoffset=1,%
  morekeywords={Close,CoDefine,Define,Kind,Query,Quit,Specification,
    Set,Split,Theorem,Type,Undo,by,as,prop,true,false,forall,exists},%
  keywordstyle=\color{abellatop},%
  classoffset=2,%
  morekeywords={abbrev,apply,backchain,case,coinduction,cut,
    induction,inst,intros,monotone,permute,rename,%left,right,
    witness,
    search,split,to,unabbrev,unfold,assert,with},%
  keywordstyle=\color{abellatac},%
  classoffset=3,%
  morekeywords={undo,abort,skip,clear},%
  keywordstyle=\color{abellabad}\underbar,%
  classoffset=0,%
}

\lstdefinelanguage{mlts}[Objective]{Caml}{%
  morekeywords={nab, new},%
  otherkeywords={@,\\, =>},%
}

\newcommand*{\SavedLstInline}{}
\LetLtxMacro\SavedLstInline\lstinline
\DeclareRobustCommand*{\lstinline}{%
  \ifmmode
    \let\SavedBGroup\bgroup
    \def\bgroup{%
      \let\bgroup\SavedBGroup
      \hbox\bgroup
    }%
  \fi
  \SavedLstInline
}
\DeclareRobustCommand\lsti[1][]{\lstinline[basicstyle=\ttfamily,keepspaces=true,#1]}

\definecolor{darkblue}{RGB}{0, 0, 185}

\newcommand{\spec}{SL\xspace}
\newcommand{\reas}{RL\xspace}
\newcommand{\ie}{\emph{i.e.},\xspace}

\newcommand{\lP}{$\lambda$Prolog\xspace}

\newcommand{\reverse}{\hbox{\sl reverse}}

\newcommand{\true}{tt}

\newcommand{\Iscr}{{\mathcal I}}
\newcommand{\Pscr}{{\mathcal P}}

\newcommand{\trueExpert }[1]{{\true_e}(#1)}
\newcommand{\eqExpert }[1]{{=_e}(#1)}
\newcommand{\unfoldExpert}[3]{{\hbox{\sl backchain}_e}(#1,#2,#3)}
\newcommand{\andExpert}[3]{{\wedge_e}(#1,#2,#3)}

\newcommand{\orExpert  }[3]{{\vee_e}(#1,#2,#3)}
\newcommand{\someExpert}[3]{\exists_e(#1,#2,#3)}

\newcommand{\step}{\longrightarrow_{\beta}}

\newcommand{\sep}{~\vert~}
\newcommand{\limp}{\multimap}
\newcommand{\imp}{\supset}
\newcommand{\namefp}{\Iscr}
\newcommand{\Interp}[1]{(\lsti{interp}~(#1))}
\newcommand{\triple}[3]{#1;#2\vdash #3}

\def\bnfas{\mathrel{::=}}

\def\arrow{\rightarrow}

\newcommand{\red}{\longrightarrow_{\eta}}
 \newcommand{\unit}{\mathbf{1}}
\newcommand{\ep}{\langle\rangle}
\newcommand{\betar}{\eta}
\newcommand{\epr}{\eta-\unit}
\newcommand{\appl}{\texttt{E-App-L}}
\newcommand{\appr}{\texttt{E-App-R}}
\newcommand{\rabs}{\texttt{E-$\xi$}}

\newcommand{\counter}[1]{\hbox{\textsl{counter}}~#1}

\newcommand{\bang}{\mathop{\oc}}

\newcommand{\io}[5]{#2\setminus #3 \vdash_{#4} #5}
\newcommand{\XXi}{{\color{blue}{\Xi}}}

\newenvironment{modified}{}{}

\begin{document}
%\title{Property-Based Testing via Proof Reconstruction}
% Alternative title
\title{Property-Based Testing by Elaborating Proof Outlines}
\author[D.~Miller and A.~Momigliano]{
  DALE MILLER\\
INRIA Saclay \& LIX, \'Ecole Polytechnique, France\\
\email{dale.miller@inria.fr}
\and ALBERTO MOMIGLIANO\\
DI, Universit\`a degli Studi di Milano, Italy\\
\email{momigliano@di.unimi.it}
}

\maketitle

\begin{abstract}
  Property-based testing (PBT) is a technique for validating code
  against an executable specification by automatically generating
  test-data.  We present a proof-theoretical reconstruction of this
  style of testing for relational specifications and employ the
  Foundational Proof Certificate framework to describe test
  generators.  We do this by encoding certain kinds of ``proof
  outlines'' as proof certificates that can describe various common
  generation strategies in the PBT literature, ranging from random to
  exhaustive, including their combination. We also address the
  \emph{shrinking} of counterexamples as a first step toward their
  explanation.  Once generation is accomplished, the testing phase is
  a standard logic programming search.  After illustrating our
  techniques on simple, first-order (algebraic) data structures, we
  lift it to data structures containing bindings by using the
  $\lambda$-tree syntax approach to encode bindings.  The \lP
  programming language can perform both generating and checking of
  tests using this approach to syntax.  We then further extend PBT to
  specifications in a fragment of linear logic.
  
  \smallskip\noindent Under consideration in Theory and Practice of
    Logic Programming (TPLP).
\end{abstract}


\section{Introduction}
\label{sec:intro}

\emph{Property-based testing} (PBT) is a technique for validating code
that successfully combines two well-trodden ideas: \emph{automatic}
test data generation trying to refute \emph{executable}
specifications. Pioneered by \emph{QuickCheck} for functional
programming~\cite{claessen00icfp}, PBT tools are now available for
most programming languages and are having a growing impact in
industry~\cite{quickcheckfunprofit}. Moreover, this idea has spread to
several proof assistants (see~\cite{BlanchetteBN11,QChick} to name the
main players) to complement (interactive) theorem proving with a
preliminary phase of conjecture testing. The synergy of PBT with proof
assistants is so accomplished that PBT is now a part of the
\emph{Software Foundations}'s curriculum~\cite{Lampropoulos:SF4}.

%(\url{https://softwarefoundations.cis.upenn.edu/qc-current}).

In our opinion, this tumultuous rate of growth is characterized by a
lack of common (logical) foundation. For one, PBT comes in different
flavors as far as data generation is concerned: while random
generation is the most common one, other tools employ exhaustive
generation~\cite{smallcheck,cheney_momigliano_2017} or a combination
thereof~\cite{feat}.  At the same time, one could say that PBT is
rediscovering logic and, in particular, logic programming: to begin
with, QuickCheck's DSL is based on Horn clause logic;
\emph{LazySmallCheck}~\cite{smallcheck} has adopted \emph{narrowing}
to permit less redundant testing over partial rather than ground
terms; a recent version of PLT-Redex~\cite{PLTbook} contains a
re-implementation of constraint logic programming in order to better
generate well-typed $\lambda$-terms~\cite{pltredexconstraintlogic}.
Finally, PBT in Isabelle/HOL features the notion of \emph{smart} test
generators~\cite{Bulwahn12}, and this is achieved by turning the
functional code into logic programs and inferring through mode
analysis their data-flow behavior.  We refer to the Related Work
(Section~\ref{sec:rel}) for more recent examples of this phenomenon, together
with the relevant citations.

This paper considers the general setting of applying PBT
techniques to \emph{logic specifications}.  In doing so, 
we also insist on the need to involve \emph{two levels} of logic.
\begin{enumerate}
  \item The \emph{specification-level logic} is the logic of
    entailment between a logic program and a goal.  In this paper,
    logic programs can be Horn clauses, \emph{hereditary Harrop
    formulas}, or a linear logic extension of the latter.  The
    entailment use at the specification level is classical,
    intuitionistic, or linear.

  \item The \emph{reasoning-level logic} is the logic where statements
    about the specification level entailment are made.  For example,
    in this logic, one might try to argue that a certain
    specification-level entailment \emph{does not hold}.  This level
    of logic can also be called \emph{arithmetic} since it involves
    least fixed points.  In particular, our use of arithmetic fits
    within the $I\Sigma_1$ fragment of Peano arithmetic, which is
    known to coincide with Heyting arithmetic~\cite{friedman78hol}.
    As a result, we can consider our uses of the reasoning-level logic
    to be either classical or intuitionistic.
\end{enumerate}

We shall use proof-theoretic techniques to deal with both of these
logic levels.  In particular, instead of attempting some kind of
amalgamation of these two levels, we will encode into the reasoning
logic inductively defined predicates that faithfully capture
specification-level terms, formulas, and provability.  One of the
strengths of using proof theory (in particular, the sequent calculus)
is that it allows for an elegant treatment of syntactic structures
with bindings (such as quantified formulas) at both logic levels.  As
a result, our approach to PBT lifts from the conventional suite of
examples to meta-programming examples without significant
complications.

Property-based testing requires a flexible way to specify what tests
should be generated.  This flexibility arises here from our use of
\emph{foundational proof certificates} (FPC)~\cite{chihani17jar}.
Arising from proof-theoretic considerations, FPCs can describe proofs
with varying degrees of detail: in this paper, we use FPCs
to describe proofs in the specification logic.
For example, an FPC can specify that a proof has a certain
height, size, or satisfies a specific outline~\cite{blanco15wof}.  It
can also specify 
that all instantiations for quantifiers have a specific property.
By employing standard logic programming techniques (e.g., unification
and backtracking search), the very process of \emph{checking} that a
(specification-level) sequent has a proof that satisfies an FPC is a
process that \emph{generates} such proofs, or more in general, results
in an attempt to fill in the missing details.  In this way, a proof
certificate goes through a proof reconstruction to yield a fully
\emph{elaborated} proof that a trusted proof-checking kernel can
accept.
%% \ednote{DM I'm not sure that the expression ``proof
%%   reconstruction'' is the right one (although it's in the title).
%%   Maybe ``proof elaboration'' is better.}
As we shall see, small
localized changes in the specification of relevant FPCs allow us to
account for both exhaustive and random generation. We can also use
FPCs to perform \emph{shrinking}: this is an indispensable ingredient
in the random generation approach, whereby counterexamples are
minimized to be more easily understandable by the user.

Throughout this paper, we use \lP~\cite{miller12proghol} to specify
and prototype all aspects of our PBT project. One role for \lP is as
an example of computing within the Specification Logic. However, since
the kinds of queries that PBT requires of our Reasoning Logic are
relatively weak, it is possible to use \lP to implement a prover for
the needed properties at the reasoning level. The typing system of \lP
will help clarify when we are using it at these two different levels:
in particular, logic program clauses are used as specifications within
a reasoning level prover are given the type \lsti{sl} instead of the
usual type \lsti{o} of \lP clauses.

If we are only concerned with PBT for logic specifications written
using first-order Horn clauses (as is the case for
Sections~\ref{sec:pbt-pt} and~\ref{sec:basic}), then the \lP
specifications can be replaced with Prolog specifications without much
change. However, this interchangeability between \lP and Prolog
disappears when we turn our attention to applying PBT to
meta-programming or, equivalently, \emph{meta-theory
model-checking}~\cite{cheney_momigliano_2017}. Although PBT has been
used extensively with meta-theory specifications, there are many
difficulties~\cite{Klein12}, mainly dealing with the \emph{binding
structures} one finds within the syntax of many programming language
constructs. In that setting, \lP's support of $\lambda$-tree
syntax~\cite{miller18jar}, which is
not supported by Prolog, allows us to be competitive with specialized
tools, such as $\alpha$Check~\cite{cheney_momigliano_2017}.

% Map of paper

\begin{modified}
  This paper and its contributions are organized as follows.  In % the
  Sections~\ref{sec:spec} and~\ref{sec:reas}, we describe the two
  levels of logic---the specification level and the reasoning
  level---whose importance we wish to underline when applying PBT to
  logic programs.  Section~\ref{sec:pbt-pt} gives a gentle
  introduction to foundational proof certificates (FPCs).  We show
  there that proof checking in logic programming can serve as a
  nondeterministic elaboration of proof outlines into complete proof
  structures, which themselves can be used to generate test cases.  We
  use proof outlines (formalized using FPCs) to provide a flexible
  description of the space of terms in which to search for
  counterexamples.  Section~\ref{sec:basic} shows how FPCs can be used
  to specify many flavors of PBT flexibly.  The programmable nature of
  the FPC framework allows us not only to describe the search space of
  possible counterexamples, but also to specify a wide range of
  popular approaches to PBT, including exhaustive, bounded, and
  randomized search, as well as the shrinking of counterexamples.  In
  Section~\ref{sec:mmm}, we lift our approach to meta-programming with
  applications to the problem of analyzing confluence in the
  $\lambda$-calculus.  When implemented in \lP, our proof checker can
  also treat the search for counterexamples that contain bindings, a
  feature important when data such as programs are the possible
  counterexamples under study.  For example, we illustrate how the \lP
  implementation of this framework easily discovers a counterexample
  to the (false) claim that, in the untyped $\lambda$-calculus,
  beta-conversion satisfies the diamond property.
  Section~\ref{sec:linear} extends our approach to a fragment of
  linear logic.  We provide one proof checker (displayed in
  Figure~\ref{fig:llinterp}) that can deal with the three
  specification logics that we employ here, namely (certain restricted
  subsets of) classical, intuitionistic, and linear logic.  The
  correctness of this checker depends only on its dozen-clause
  specification and the soundness of the underlying \lP
  implementation.  We conclude with a review of related work
  (Section~\ref{sec:rel} and~\ref{sec:concl}).
\end{modified}

This paper significantly extends our conference
paper~\cite{blanco19ppdp} by clarifying the relationship between
specification and reasoning logics, by tackling new examples and by
including logic specifications based on linear logic.  The code
mentioned in this paper can be found at
\begin{center}
  \url{https://github.com/proofcert/pbt/tree/journal}
\end{center}

\section{The specification logic \spec}
\label{sec:spec}

Originally, logic programming was  based on relational
specifications (\ie formulas in first-order predicate logic) given as
Horn clauses.  Such clauses can be defined as closed formulas of the
form $\forall \bar x [G\imp A]$ where $\bar x$ is a list of
variables (all universally quantified), $A$ is an atomic
formula, and $G$ (a \emph{goal} formula) is a formula built using
disjunctions, conjunctions, existential quantifiers, true (the unit
for conjunction), and atomic formulas.
An early extension of the logic programming paradigm, called the
\emph{hereditary Harrop formulas}, allowed both universal
quantification and certain restricted forms of the
\emph{intuitionistic implication} ($\imp$) in goal
formulas~\cite{miller91apal}.
A subsequent extension of that paradigm, called
Lolli~\cite{Hodas1994}, also allowed certain uses
of the \emph{linear implication} ($\limp$) from Girard's linear
logic~\cite{girard87tcs}.

Except for Section~\ref{sec:linear}, we shall consider the following
two classes of formulas.
\[
\begin{array}[t]{rl}
  D~\bnfas& G \imp A \sep \forall x:\tau. D\\
  G~\bnfas& A \sep \true \sep G_1\vee G_2\sep G_1\wedge G_2\sep\exists x:\tau. G
                    \sep\forall x:\tau. G\sep A\imp G
\end{array}
\]
The $D$-formulas are also called \emph{program clauses} and
\emph{definite clauses} while $G$-formulas are \emph{goal formulas}.
We will omit type information when not relevant.
Here, $A$ is a schematic variable that ranges over atomic formulas.
Given the $D$-formula $\forall x_1\ldots\forall x_n[G\imp A]$, we say
that $G$ is the \emph{body} and $A$ is the \emph{head} of that program
clause. 
In general, every $D$-formula is an \emph{hereditary Harrop
formula}~\cite{miller91apal}, although the latter is a much richer set
of formulas.

\begin{figure}[t]
\begin{lstlisting}[linerange={start-end}]
module clark.
/* start */
kind nat              type.
type z                nat.
type s                nat -> nat.
type is_nat           nat -> o.
type nlist            list nat -> o.
type append, rev_acc  list A -> list A -> list A -> o.
type reverse          list A -> list A -> o.

is_nat z.
is_nat (s X) :- is_nat X.

nlist nil.
nlist (N::Ns) :- is_nat N, nlist Ns.

append nil K K.
append (X::L) K (X::M) :- append L K M.

reverse L K  :- rev_acc L K nil.
rev_acc nil A A.
rev_acc (X::L) K A :- rev L K (X::A).
/* end */

% Previous version

is_nat X        :- X = z ;
                sigma X'\ X  = (s X'), is_nat X'.
nlist Ns     :- Ns = nil ; 
                sigma Ns'\ Ns = (N::Ns'), is_nat N, nlist Ns'.
append L K M :- L = nil, K = M ; 
                sigma L'\ sigma M'\ L = (X::L'), M = (X::M'),
                                    append L' K M'.
reverse L K  :- rev L K nil.
rev L K A    :- L = nil, K = A;
                sigma L'\ L = (X::L'), rev L' K (X::A).
\end{lstlisting}
\caption{The \lP specification of five predicates.}
\label{fig:clark}
\end{figure}

We use \lP to display logic programs in this paper.  For example,
Figure~\ref{fig:clark} contains the Horn clause specifications of five
predicates related to natural numbers and lists.  The main difference
between Prolog and \lP that appears from this example is the fact that
\lP is explicitly polymorphically typed.  Another difference
% between these two languages
is that \lP allows goal formulas to
contain universal quantification and implications.

Traditionally, entailment between a logic program and a goal has been
described using classical logic and a theorem proving technique called
SLD-resolution refutations~\cite{apt82jacm}.
As  now common when intuitionistic (and linear) logics are used within
logic programming, refutations are replaced with proofs using
Gentzen's sequent calculus~\cite{gentzen35}.
Let $\Pscr$ be a finite set of $D$-formulas and let $G$ be a goal
formula.  We are interested in finding proofs of the \emph{sequent}
$\Pscr\longrightarrow G$.
As it has been shown in~\cite{miller91apal}, a simple, two-phase proof
search strategy based on \emph{goal reduction} and \emph{backchaining}
forms a complete proof system when that logic is intuitionistic logic.
--- for a survey of how Gentzen's proof theory has been applied to logic
programming, see~\cite{miller22tplp}.
In the next section, we will write an interpreter for such sequents:
the  code for that interpreter is taken directly from that two-phase
proof system.

%% AM: commented out after chat of 29-5
% The completeness of the two phases strategy of goal reduction and
% backchaining will be reflected immediately into the structure of
% interpreters we shall need to write for these logic programming
% languages.

\section{The reasoning logic \reas}
\label{sec:reas}

% Weakness of SL: no negation

The specification logic \spec that we presented in the previous section is
not capable of proving the negation of any atomic formula.
This observation is an immediate consequence of the
\emph{monotoncity} of \spec: that is, if
$\Pscr\subseteq\Pscr'$ and $A$ follows from $\Pscr$ then $A$ follows
from $\Pscr'$.
If $\neg A$ is provable from $\Pscr$ then both $A$ and $\neg A$ are
provable from $\Pscr\cup\{A\}$. Thus, we must conclude that
$\Pscr\cup\{A\}$ is inconsistent, but that is not possible since the
set $\Pscr\cup\{A\}$ is satisfiable (interpret all the predicates to
be the universally true property for their corresponding arity).
For example, neither $(\reverse~(z::nil)~nil)$ nor its
negation are provable from the clauses in Figure~\ref{fig:clark}.

%% Similar arguments can show that it is
%% impossible to prove the formula \( \forall L\forall K~ [\reverse~L~K
%%   \supset \reverse~K~L]\) although this formula appears to state the
%% obvious fact that the $\reverse$ predicate is symmetric.

Clearly, any PBT setting must permit proving the negation of some
formulas, since, for example, a counterexample to the claim $\forall
x.[P(x)\supset Q(x)]$ is a term $t$ such that $P(t)$ is provable and
$\neg Q(t)$ is provable.  At least two different approaches have been
used to move to a stronger logic in which such negations are provable.
The Clark completion of Horn clause
programs can be used for this purpose~\cite{clark78}.  An advantage of that approach
is that it requires only using first-order classical logic (with an
appropriate specification of equality)~\cite{apt94jlpb}.  A
disadvantage is that it only seems to work for Horn
clauses: this approach does not seem appropriate when working with
intuitionistic and linear logics.

\begin{figure}
\begin{lstlisting}[linerange={interp-end}]
sig interp.
/* interp */
kind sl       type.            % Specification logic formulas
type tt,ff    sl.              % True, False
type and, or  sl -> sl -> sl.  % Conjunction and disjunction
type some     (A -> sl) -> sl. % Existential quantifier
type eq       A  -> A  -> sl.  % Equality

infixr and    50.
infixr or     40.
infix  eq     60.

type   <>==   sl -> sl -> o.  % Predicate for storing sl clauses
infix  <>==   30.
type interp   sl -> o.        % Interpreter of sl goals
/* end */
/* nabla */
type nabla,all     (A -> sl) -> sl.
/* end */
type progs    sl -> list sl -> o.  % list based

kind nat   type.
type z     nat.
type s     nat -> nat.
/* examples */
type isnat                                      nat -> sl.
type nlist                                 list nat -> sl.
type append, rev_acc     list A -> list A -> list A -> sl.
type revApp,reverse                list A -> list A -> sl.
/* end */
type   leq        nat -> nat -> sl.
type   gt         nat -> nat -> sl.
type   ord          list nat -> sl.
type   ord_bad      list nat -> sl.
type   ins          nat -> list nat -> list nat -> sl.
type  member        A -> list A -> o.
type nlistp         list nat -> sl. %with prog
\end{lstlisting}
\begin{lstlisting}[linerange={interp-end}]
module interp.

/* interp */
interp tt.
interp (T eq T).
interp (G1 and G2) :- interp G1, interp G2.
interp (G1 or G2)  :- interp G1 ; interp G2.
interp (some G)    :- interp (G T).
interp A           :- (A <>== G), interp G.
/* end */
/* nabla */
interp (nabla G)     :- pi x\ interp (G x).
interp (all G)     :- pi x\ interp (G x).
/* end */
% interp A                  :- progs A Gs, member G GS, interp G.

% (nat z)     <>== tt.
% (nat (s X)) <>== (nat X).
% (nlist nil)     <>== tt.
% (nlist (N::Ns)) <>== ((nat N) and (nlist Ns)).
% (append nil K K)         <>== tt.
% (append (X::L) K (X::M)) <>== (append L K M).
% (reverse L K)    <>== (rev L K nil).
% (rev nil A A)    <>== tt.
% (rev (X::L) K A) <>== (rev L K (X::A)).
/* examples */
(isnat N) <>== (N eq z) or
               (some N'\ N eq (s N') and (isnat N')) or ff.
(nlist L) <>== (L eq nil) or 
               (some N\ some Ns\ L eq (N::Ns) and (isnat N) 
                                   and (nlist Ns)) or ff.
/* end */

% There was a mistake in append (which I think was not intended (X for X').  I fixed it.
% AM: not adding or ff unless it is used within a check in test-lst
(append Xs Ys Zs) <>== ((Xs eq nil) and (Ys eq Zs)) or
                        (some X'\ some Xs'\ some Zs'\ Xs eq (X':: Xs') and (append Xs' Ys Zs') and (Zs eq (X' :: Zs'))).
(revApp L R) <>==
   ((L eq nil) and (R eq nil)) or
    some X\ some Xs\ some Ts\ (L eq (X::Xs)) and ((reverse Xs Ts) and (append Ts (X:: nil) R)).

(reverse L K) <>== (rev_acc L K nil).
(rev_acc L K A) <>== ((L eq nil) and (K eq A)) or 
                   some L'\ ( L eq (X::L')) and (rev_acc L' K (X::A)).

member A (A:: _).
member A (_ :: Rs) :- member A Rs.

(leq N M) <>==
      ( N eq z) or
       some X\ (some Y\ (N eq  (s X)) and  (M eq  (s Y)) and (leq X Y)).

(gt N M) <>==
      (M eq z) and some K\ (N eq (s K)) or
       some X\ some Y\ ((N eq (s X)) and (( M eq (s Y)) and (gt X Y))).

(ord_bad L)  <>==
      (L eq nil) or
      some X\ ((L eq (X :: nil)) and (nat X)) or
      some Y\ some Rs\ ((L eq (X :: ( Y :: Rs))) and (leq X Y) and (ord_bad Rs))  or ff.

(ins X L R) <>==
       ( (L eq nil) and ( (R eq ( X :: nil)) and (nat X))) or
       some Y\ some Ys\ ((L eq ( Y :: Ys)) and (R eq  (X :: ( Y :: Ys)) and (leq X Y))) or
       some Y\ some Ys\  some Rs\ (L eq ( Y ::  Ys)) and
       	    (R eq (X :: Rs)) and (gt X Y)  and (ins X Ys Rs).

\end{lstlisting}
\caption{The basic interpreter for Horn clause specifications}
\label{fig:interp}
\begin{lstlisting}[linerange={examples-end}]
sig interp.
/* interp */
kind sl       type.            % Specification logic formulas
type tt,ff    sl.              % True, False
type and, or  sl -> sl -> sl.  % Conjunction and disjunction
type some     (A -> sl) -> sl. % Existential quantifier
type eq       A  -> A  -> sl.  % Equality

infixr and    50.
infixr or     40.
infix  eq     60.

type   <>==   sl -> sl -> o.  % Predicate for storing sl clauses
infix  <>==   30.
type interp   sl -> o.        % Interpreter of sl goals
/* end */
/* nabla */
type nabla,all     (A -> sl) -> sl.
/* end */
type progs    sl -> list sl -> o.  % list based

kind nat   type.
type z     nat.
type s     nat -> nat.
/* examples */
type isnat                                      nat -> sl.
type nlist                                 list nat -> sl.
type append, rev_acc     list A -> list A -> list A -> sl.
type revApp,reverse                list A -> list A -> sl.
/* end */
type   leq        nat -> nat -> sl.
type   gt         nat -> nat -> sl.
type   ord          list nat -> sl.
type   ord_bad      list nat -> sl.
type   ins          nat -> list nat -> list nat -> sl.
type  member        A -> list A -> o.
type nlistp         list nat -> sl. %with prog
\end{lstlisting}
\begin{lstlisting}[linerange={examples-end}]
module interp.

/* interp */
interp tt.
interp (T eq T).
interp (G1 and G2) :- interp G1, interp G2.
interp (G1 or G2)  :- interp G1 ; interp G2.
interp (some G)    :- interp (G T).
interp A           :- (A <>== G), interp G.
/* end */
/* nabla */
interp (nabla G)     :- pi x\ interp (G x).
interp (all G)     :- pi x\ interp (G x).
/* end */
% interp A                  :- progs A Gs, member G GS, interp G.

% (nat z)     <>== tt.
% (nat (s X)) <>== (nat X).
% (nlist nil)     <>== tt.
% (nlist (N::Ns)) <>== ((nat N) and (nlist Ns)).
% (append nil K K)         <>== tt.
% (append (X::L) K (X::M)) <>== (append L K M).
% (reverse L K)    <>== (rev L K nil).
% (rev nil A A)    <>== tt.
% (rev (X::L) K A) <>== (rev L K (X::A)).
/* examples */
(isnat N) <>== (N eq z) or
               (some N'\ N eq (s N') and (isnat N')) or ff.
(nlist L) <>== (L eq nil) or 
               (some N\ some Ns\ L eq (N::Ns) and (isnat N) 
                                   and (nlist Ns)) or ff.
/* end */

% There was a mistake in append (which I think was not intended (X for X').  I fixed it.
% AM: not adding or ff unless it is used within a check in test-lst
(append Xs Ys Zs) <>== ((Xs eq nil) and (Ys eq Zs)) or
                        (some X'\ some Xs'\ some Zs'\ Xs eq (X':: Xs') and (append Xs' Ys Zs') and (Zs eq (X' :: Zs'))).
(revApp L R) <>==
   ((L eq nil) and (R eq nil)) or
    some X\ some Xs\ some Ts\ (L eq (X::Xs)) and ((reverse Xs Ts) and (append Ts (X:: nil) R)).

(reverse L K) <>== (rev_acc L K nil).
(rev_acc L K A) <>== ((L eq nil) and (K eq A)) or 
                   some L'\ ( L eq (X::L')) and (rev_acc L' K (X::A)).

member A (A:: _).
member A (_ :: Rs) :- member A Rs.

(leq N M) <>==
      ( N eq z) or
       some X\ (some Y\ (N eq  (s X)) and  (M eq  (s Y)) and (leq X Y)).

(gt N M) <>==
      (M eq z) and some K\ (N eq (s K)) or
       some X\ some Y\ ((N eq (s X)) and (( M eq (s Y)) and (gt X Y))).

(ord_bad L)  <>==
      (L eq nil) or
      some X\ ((L eq (X :: nil)) and (nat X)) or
      some Y\ some Rs\ ((L eq (X :: ( Y :: Rs))) and (leq X Y) and (ord_bad Rs))  or ff.

(ins X L R) <>==
       ( (L eq nil) and ( (R eq ( X :: nil)) and (nat X))) or
       some Y\ some Ys\ ((L eq ( Y :: Ys)) and (R eq  (X :: ( Y :: Ys)) and (leq X Y))) or
       some Y\ some Ys\  some Rs\ (L eq ( Y ::  Ys)) and
       	    (R eq (X :: Rs)) and (gt X Y)  and (ins X Ys Rs).

\end{lstlisting}
\caption{The encoding of the Horn clause definitions of two predicates
  in Figure~\ref{fig:clark} as atomic formulas in \reas.}
\label{fig:examples}
\begin{lstlisting}[basicstyle=\small\ttfamily]
interp G :- (G = tt);  (sigma T\ G = (T eq T));
 (sigma H\ sigma K\ G = (H and K), interp H, interp K);
 (sigma H\ sigma K\ G = (H or  K), interp H; interp K);
 (sigma H\ sigma T\ G = (some H),  interp (H T)) ;
 (sigma B\ (G <>== B), interp B).
\end{lstlisting}
\caption{An equivalent specification of \lsti{interp} as one clause.}
\label{fig:interp1}
\begin{align*}
\namefp = 
  \mu\lambda I\lambda g~
  [g = \mathtt{tt}
  & \vee~(\exists t.~g = (t~\mathtt{eq}~ t))\\
  & \vee~(\exists h\exists k.~g = (h~\texttt{and}~k)\land (I~h)\land (I~k))\\
  & \vee~(\exists h\exists k.~g = (h~\texttt{or}~k)\land (I~h)\lor (I~k))\\
  & \vee~(\exists h\exists t.~g = (\lsti{some}~h)\land(I~(h~t)))\\
  & \vee~(\exists b.~(g~\lsti{<>==} b)\land(I~b))]
\end{align*}
\label{fig:fpex}
\caption{The least fixed point expression for \lsti{interp}.}
\end{figure}

In this paper, we follow an approach used in both the Abella proof
assistant~\cite{baelde14jfr,gacek12jar} and the Hybrid~\cite{FeltyM12}
library for Coq and Isabelle.  In those systems, a \emph{second
logic}, called the \emph{reasoning logic} (\reas, for short), is used
to give an \emph{inductive definition} for the provability for a
specification logic (in those cases, the specification logic is a
fragment of higher-order intuitionistic logic).
In particular, consider the \lP specification in Figure~\ref{fig:interp}.
Here, terms of type \lsti{sl} denote formulas in \spec.  The predicate
\lsti{<>==} is used to encode the \spec-level program clauses: for
example, the specification in Figure~\ref{fig:examples} encodes the
Horn clause programs in Figure~\ref{fig:clark}.
Note that we are able to simplify our specification of the
\lsti{<>==} predicate; the universal quantification at the \reas level
can be used to encode the (implicit) quantifiers in the \spec level.

Figure~\ref{fig:interp1} contains the single clause specification of
\lsti{interp} that corresponds to its Clark's completion --- in \lP,
the existential quantifier $\exists X$ is written as
\lsti{sigma X}.  This single clause can be turned directly into the
 least fixed point expression displayed in
Figure~\ref{fig:fpex}.  The proof theory for \reas specifications
using fixed points and equality has been developing since the 1990s.
Early partial steps were taken by Girard~\cite{girard92mail} and
Schroeder-Heister~\cite{schroeder-heister93lics}.  Their approach was
strengthened to include least and greatest fixed points for
intuitionistic logic~\cite{mcdowell00tcs,momigliano12jal}.
Applications of this fixed point logic were made to both model
checking~\cite{heath19jar,mcdowell03tcs} and to interactive theorem
proving~\cite{baelde14jfr}.

The proof search problem for the full \reas is truly difficult to
automate since proofs involving least and greatest fixed points
require the proof search mechanism to invent invariants (pre- and
post-fixed points), a problem which is far outside the usual logic
programming search paradigm.
% The full \reas has a proof search problem that is truly difficult to
% automate since proofs involving least and greatest fixed points
% require the proof search mechanism to invent invariants (pre- and
% post-fixed points), a problem that is far outside the usual logic
% programming search paradigm.
Fortunately, for our purposes here, we
will be attempting to prove simple theorems in \reas that are strictly
related to  queries about \spec formulas.  In particular, we
consider only the following kinds of theorems in \reas.  Let $A$ and
$\Pscr$ be, respectively, an atomic formula and a finite set of
$D$-formulas in \spec.  Also, let $\hat A$ be the direct encodings of
$A$ into a term of type \lsti{sl}, and let $\hat\Pscr$ be a set of
\reas atomic formulas using the \lsti{<>==} predicate that encodes the
Horn clauses in $\Pscr$.
\begin{enumerate}
  \item $(\namefp~\hat A)$ is \reas-provable from $\hat\Pscr$ if and only
    if $A$ is \spec-provable from $\Pscr$.  In addition, these are
    also equivalent to the fact that (\lsti{interp }$\hat A$) is
    provable from $\hat\Pscr$ using the logic program for the
    interpreter in Figure~\ref{fig:interp}.  This statement is proved
    by a simple induction on \reas and \spec proofs.

  \item If \lP's negation-as-finite-failure procedure succeeds for the
    goal (\lsti{interp}~$\hat A$) with respect to the program $\hat\Pscr$
    (using the logic program for the interpreter in
    Figure~\ref{fig:interp}), then $\neg(\namefp~\hat A)$ is
    \reas-provable from $\hat\Pscr$~\cite{HallnasS91}.  In this case, there is no
    \spec-proof of $A$ from $\Pscr$.
\end{enumerate}
Note that the second statement above is not an equivalence: that is,
there may be proofs of $\neg(\namefp~\hat A)$ in \reas, which
will not be captured by negation-as-finite-failure.  For example, if
$p$ is an atomic \spec formula and $\Pscr$ is the set containing just
$p\imp p$, then the usual notion of negation-as-finite-failure will
not succeed with the goal $\hat p$ and logic program containing just
$\hat p~$\lsti{<>==}$~\hat p$, while there would be a proof
(using induction) that $\neg(\namefp~\hat p)$.

Thus, we can use \lP as follows.  If \lP proves 
(\lsti{interp}~$\hat A$) then we know that $A$ is provable from
$\Pscr$ in \spec.  Also, if \lP's negation-as-finite-failure proves that
(\lsti{interp}~$\hat A$) does not hold, then we know that $A$ is not
provable from $\Pscr$ in \spec.  
In conclusion, although \lP has limited abilities to prove formulas in
\reas, it can still be used in the context of PBT where we require
limited forms of inference.

\section{Controlling the generation of tests}
\label{sec:pbt-pt}

\subsection{Generate-and-test as a proof-search strategy}

% DM Introduce $\Pscr\vdash G$ to mean that the \reas interpreter proves
% $G$ from $\Pscr$.  Maybe not needed or precise enough?

Imagine that we wish to write a relational specification for reversing
lists.  There are, of course, many ways to write such a specification,
say $\Pscr$, but in every case it should be the case that if
$\Pscr\vdash (\reverse~L~R)$ then $\Pscr\vdash (\reverse~R~L)$: that
is, \textit{reverse} is symmetric. 

In the \reas setting that we have described in the last section, this property
can be written as the formula 
\[
  \forall L\forall R. \Interp{\reverse~L~R}\imp\Interp{\reverse~R~L}
\]
where we forgo the $\hat{(\_)}$ notation. If a formula like this is
provable, it is likely that the such a proof would involve finding
an appropriate induction invariant (and possibly additional lemmas).
In the property-based testing setting, we are willing to look,
instead, for counterexamples to a proposed property.  In other words,
we are willing to consider searching for proofs of the negation of
this formula, namely
\[
  \exists L\exists R. \Interp{\reverse~L~R}\wedge\neg\Interp{\reverse~R~L}.
\]
This formula might be easier to prove since it involves only
standard logic programming search mechanisms.  Of course, proving this
second formula would mean that the first formula is not provable.

More generally,  we might wish to prove a number of formulas of the form
\[\forall x\colon\tau\ [\Interp{P(x)} \supset \Interp{Q(x)}]\]
where both $P$ and $Q$ are \spec-level relations (predicates) of a
single argument (it is an easy matter to deal with more than one
argument).
Often, it can be important in this setting to move the type
judgment $x\colon\tau$ into the logic by turning the type into a
predicate:
\(\forall x [\Interp{\tau(x)\wedge P(x)} \supset \Interp{Q(x)}]\).
As we mentioned above, it can be valuable to first attempt to find
counterexamples to such formulas prior to pursuing a proof.
That is, we might try to prove formulas of the form
\[
 \exists x [\Interp{\tau(x)\wedge P(x)} \wedge \neg\Interp{Q(x)}]
  \tag{*}\label{eq:full}
\]
instead.
If a term $t$ of type $\tau$ can be discovered such that $P(t)$ holds
while $Q(t)$ does not, then one can return to the specifications
in $P$ and $Q$ and revise them using the concrete evidence in $t$
as a witness of  how the specifications are wrong.
The process of writing and revising relational specifications should
be smoother if such counterexamples are discovered quickly and
automatically.

The search for a proof of $(\ref{eq:full})$ can be seen as essentially
the \emph{generate-and-test} paradigm that is behind much of
property-based testing.
In particular, a proof of $(\ref{eq:full})$ contains the information
needed to prove 
\[
 \exists x [\Interp{\tau(x)\wedge P(x)}]
  \tag{**}\label{eq:short}
\]
Conversely, not every proof of $(\ref{eq:short})$ yields, in fact, a
proof of $(\ref{eq:full})$.  However, if we are willing to generate
proofs of $(\ref{eq:short})$, each such proof yields a term $t$ such
that $[\tau(t)\wedge P(t)]$ is provable.  In order to achieve a proof
of $(\ref{eq:full})$, we only need to prove $\neg Q(t)$.  In the proof
systems used for \reas, such a proof involves only
negation-as-finite-failure: that is, all possible paths for proving
$Q(t)$ must be attempted and all of these must end in failures.

Of course, there can be many possible terms $t$ that are generated in
the first steps of this process.
We next show how the notion of a \emph{proof certificate} can be used
to flexibly constraint the space of terms that can  generated for
 consideration against the testing phase.

\subsection{Proof certificate checking with expert predicates}
\label{ssec:fpc}

\begin{figure}[t]
\renewcommand{\XXi}{{\color{blue}{\Xi}}}
\[
\infer{\XXi\vdash \true}
      {\trueExpert{\XXi}}
\qquad
\infer{\XXi\vdash G_1\wedge G_2}
      {\XXi_1\vdash G_1\qquad \XXi_2\vdash G_2\qquad \andExpert{\XXi}{\XXi_1}{\XXi_2}}
\qquad
\infer{\XXi\vdash G_1\vee G_2}
      {\XXi'\vdash G_i\qquad \orExpert{\XXi}{\XXi'}{i}}
\]
\[
\infer[(1)]{\XXi\vdash \exists x:\tau. G}
      {\XXi'\vdash G[t/x]\qquad \someExpert{\XXi}{\XXi'}{t}}
\qquad
\infer{\XXi\vdash t = t}
      {\eqExpert{\XXi}}
\qquad
\infer[(2)]{\XXi \vdash A}
      {\XXi'\vdash G 
                   \quad \unfoldExpert{A}{\XXi}{\XXi'}}
      % {\XXi'\vdash \theta G \quad \forall(A~\hbox{\lsti{<>==}}~G)\in\Pscr
      %              \quad \unfoldExpert{\XXi}{\XXi'}}
\]
% \begin{flushleft}
%   The two provisos (1) and (2) are:
% %  the standard ones, namely:
% \end{flushleft}
\begin{enumerate}[label=(\arabic*)]
  \item The term $t$ is of type $\tau$.
  \item There is a program clause $\forall \bar x (G' \imp
    A')\in\Pscr$ and a substitution for the variables $\bar x$ such
    that $A$ is $A'\theta$ and $G$ is $G'\theta$.
\end{enumerate}
\caption{A proof system augmented with proof certificates and
  expert predicates.}
\label{fig:augmented}
% DM Note that we use the Xi |- G syntax here for proof certificates
% Xi.  This is different from its uses in, say, Abella, where the lhs
% is a logic program.
\begin{lstlisting}[linerange={sigs-end}]
sig kernel.
accum_sig interp.
accum_sig debug.

/*
% Object-level formulas
kind oo        type.
type tt        oo.
type and, or   oo  -> oo -> oo.
type some      (A -> oo) -> oo.
type eq        A -> A  -> oo.

% Object-level Prolog clauses are
% stored as facts (prog A Body).
type prog      oo -> oo -> o.
*/
/* sigs */
% Certificates
kind cert          type.
kind choice        type.
type left, right   choice.

% The types for the expert predicates
type ttE, eqE                     cert -> o.
type backchainE     sl -> cert -> cert -> o.
type someE           cert -> cert -> A -> o.
type andE         cert -> cert -> cert -> o.
type orE        cert -> cert -> choice -> o.
/* end */

/*
/* checkx */
% type backchainEx  sl -> cert -> cert -> o.
/* end */
*/
type check    cert -> sl -> o.
% A sample program
kind   nat      type.
type   zero     nat.
type   succ     nat -> nat.
type   is_nat   nat -> sl.
type   leq      nat -> nat -> sl.
type   gt       nat -> nat -> sl.
type   is_natlist   lst nat -> sl.
kind   lst          type -> type.
type   null         lst A.
type   cons         A -> lst A -> lst A.
\end{lstlisting}
\begin{lstlisting}[linerange={check-end}]
module kernel.

%check   A B       :- announce (check A B).

/* check */
% Certificate checker
type   check    cert -> sl -> o.

check Cert tt          :- ttE Cert.
check Cert (T eq T)    :- eqE Cert.
check Cert (G1 and G2) :- andE Cert Cert1 Cert2,
                          check Cert1 G1, check Cert2 G2.
check Cert (G1 or  G2) :- orE Cert Cert' LR, 
                          ((LR = left,  check Cert' G1);
                           (LR = right, check Cert' G2)).
check Cert (some G)     :- someE Cert Cert1 T, check Cert1 (G T).
check Cert A            :- backchainE  A Cert Cert', (A <>== G), 
                           check Cert' G.
/* end */

% /* checkx */
% check Cert A :- backchainEx A Cert Cert', (A <>== G), 
%                 check Cert' G.
% /* end */

/* nabla */
check Cert (all G) :-
      pi x\ check Cert (G x).

check Cert (nabla G) :-
      pi x\ check Cert (G x).
/* end */
/*
prog (is_nat zero) (tt).
prog (is_nat (succ N)) (is_nat N).
prog (is_natlist null) (tt).
prog (is_natlist (cons Hd Tl)) (and (is_nat Hd) (is_natlist Tl)).
*/
\end{lstlisting}
\caption{A simple proof checking kernel.}
\label{fig:kernel}
\end{figure}

% \begin{metanote}
%   AM: we already made this assumption when implementing \texttt{interp}
% \end{metanote}
Recall that  we  assume that the \spec is limited so that $D$-formulas
are Horn clauses% : that is, $G$-formulas are not allow to be either
% universally quantified nor implications
.  We shall consider the full
range of $D$ and $G$-formulas in Section~\ref{sec:mmm} when we
examine PBT for metaprogramming.

Figure~\ref{fig:augmented} contains a simple proof system for Horn
clause provability that is augmented with \emph{proof certificates}
(using the schematic variable $\Xi$) and additional premises involving
\emph{expert predicates}.  The intended meaning of these augmentations
is the following: proof certificates contain some description of a
proof.  That outline might be detailed or it might just provide
some hints or constraints on the proofs it describes.  The expert
predicates provide additional premises that are used to ``extract''
information from proof certificates (in the case of $\vee$ and
$\exists$) and to provide continuation certificates where needed.
% In the last rule, $\Pscr$ is a set of Horn clauses
% definitions. %, that is, clauses of the form $A~\hbox{\lsti{<>==}}~G$.
%% DM Is the following restriction really needed?
%% where $A$ is an atomic formula that comprises a predicate applied to
%% distinct variables than may be free in $G$ (see
%% Figure~\ref{fig:examples} for examples of such clauses).

Figure~\ref{fig:kernel} contains the \lP implementation of the
inference rules in Figure~\ref{fig:augmented}: here the infix
turnstile $\vdash$ symbol is replaced by the \lsti{check} predicate
and the predicates \mbox{\lsti{ttE}, \lsti{andE}, \lsti{orE}, \lsti{someE},
\lsti{eqE}, and \lsti{backchainE}} name the predicates
$\trueExpert{\cdot}$, $\andExpert{\cdot}{\cdot}{\cdot}$,
$\orExpert{\cdot}{\cdot}{\cdot}$, $\someExpert{\cdot}{\cdot}{\cdot}$,
$\eqExpert{\cdot}$, and $\unfoldExpert{\cdot}{\cdot}{\cdot}$ used as premises
in the inference rules in Figure~\ref{fig:augmented}.
The intended meaning of the predicate \lsti{check Cert G} is that
there exists a proof of \lsti{G} from the Horn clauses in $\Pscr$ that fits
the outline prescribed by the proof certificate \lsti{Cert}.
Note that it is easy to show that no matter how the expert
predicates are defined, if the goal \lsti{check Cert G} is provable in
\lP then the goal \lsti{interp G} is provable in \lP and, therefore,
\lsti{G} is a consequence of the program clauses stored in $\Pscr$.

A proof certificate is a term of type \lsti{cert} (see
Figure~\ref{fig:kernel}) and an FPC is a logic program that specifies
the meaning of the (six) expert predicates.  Two useful proof
certificates are based on the measurements of the \emph{height} and
the \emph{size} of a proof.  In our examples here, we choose to only
count the invocations of the \lsti{backchainE} predicate when defining
these measurements.  (One could also choose to count the invocations
of all or some other subset of expert predicates.)  The height
measurement counts the maximum number of backchains on branches in a
proof\footnote{This measurement has also been called the
\emph{decide-depth} of a proof~\cite{miller11cpp}.}.  The size
measurement counts the total number of backchain steps within a
given proof.  Proof certificates based on these two measurements are
specified in Figure~\ref{fig:resources} with the declaration of two
constructors for certificates and the listing of clauses describing
how the expert predicates treat certificates based on these two
measurements.  Specifically, the first FPC defines the experts for
treating certificates that are constructed using the \lsti{height}
constructor.  As is easy to verify, the query \mbox{\lsti{?- check
    (height 5) G}} (for the encoding \lsti{G} of a goal formula) is
provable in \lP using the clauses in Figures~\ref{fig:kernel}
and~\ref{fig:resources} if and only if the height of that proof is 5
or less.  Similarly, the second FPC uses the constructor
\lsti{sze}\footnote{This spelling is used since ``size'' is a reserved
word in the \emph{Teyjus} compiler for \lP~\cite{teyjus.website}.}
(with two integers) and can be used to bound the total number of
instances of backchaining steps in a proof.  In particular, the query
\mbox{\lsti{?- sigma H\\ check (sze 5 H) G}} is provable if and only if
the total number is 5 or less.

\begin{figure}
\begin{lstlisting}[linerange={resources-end}]
sig fpcs.
accum_sig kernel.
accum_sig random.
accum_sig debug.

/* resources */
type   height   int -> cert.
type   sze     int -> int -> cert.
/* end */

/* random */
type   random   cert.
/* end */
type   rtries    int -> cert.
type   rtriesW   string -> int -> cert.

type   iterate   int -> o.

/* max */
kind max     type.
type max     max    -> cert.
type binary  max    -> max -> max.
type choose  choice -> max -> max.
type term    A      -> max -> max.
type empty   max.
/* end */

/* pairing */
type   <c>     cert ->  cert ->  cert.
infixr <c>     5.
/* end */

type l-or-r    list choice -> 
               list choice -> cert.

kind nat     type.
type zero    nat.
type succ    nat -> nat.

/* collect */
kind item                     type.
type c_nat             nat -> item.
type c_list_nat   list nat -> item.

type subterm      item -> item -> o.
type collect      list item -> list item -> cert.
/* end */
% type c1      list item -> cert.
/* huniv */
type huniv   (item -> o) -> cert.
/* end */
type tries int -> o.

/* weights */
kind qform       type.
%type qtt, qeq    qform.
type qid         qform.
type qand        qform -> qform -> qform.
type qor         int -> int -> qform -> qform -> qform.
%type qsome       qform -> qform.
type qprog       string -> qform.

type qdistr    string -> qform -> o.
type qchoose   int -> int -> choice -> o.

type qw   qform -> cert.
/* end */

% Another approach to weighing clause

/* weight_cert */
type noweight      cert.
type cases         int -> list int -> int -> cert.
/* end */

/* weight_pred */
type weights       sl -> list int -> o.
/* end */
\end{lstlisting}
\begin{lstlisting}[linerange={resources-end}]
module fpcs.

%backchainEx A B C :- announce (backchainEx A B C).
%weights A B       :- announce (weights A B).

/* iterate */
iterate N :- N > 0.
iterate N :- N > 0, N' is N - 1, iterate N'.
/* end */

/* resources */
ttE     (height _).
eqE     (height _).
orE     (height H) (height H) _.
someE   (height H) (height H) _.
andE    (height H) (height H) (height H).
backchainE _ (height H) (height H') :- H  > 0, H' is H  - 1.

ttE     (sze In In).
eqE     (sze In In).
orE     (sze In Out) (sze In Out) _.
someE   (sze In Out) (sze In Out) _.
andE    (sze In Out) (sze In Mid) (sze Mid Out).
backchainE _  (sze In Out) (sze In' Out) :-
         In > 0, In' is In - 1.
/* end */
%backchainE (qtriesW Id N) Cont :- N > 0, N' is N - 1, backchainE (qtriesW Id N') Cont.
/* random */
ttE random.
eqE random.
orE random random Choice :- next_bit I,
  ((I = 0, Choice = left); (I = 1, Choice = right)).
someE random random _.
andE  random random random.
backchainE _ random random.
/* end */

% DM Another approach to putting weights on disjunctive cases.
% Assume that when we use case-weight, our clauses are of the form
%      head <>== b1 or (b2 or ... or (bn or tt)).
% Here, the disjunction only appears at the top-level (no bi formula
% contains disjunction) and that the disjunctions are associated to
% the right (which is the default declaration).  Note that the
% disjunction must end with tt.

% Weights for the disjunctions should always add up to 128.  There
% must be a top-level disjunction in the body (otherwise, don't use
% weights).

% Since backchainE needs to know the predicate on which it is being
% invoked, it seems that we need another backchain expert that can
% examine the atom it is openning.

/* weighty */
ttE     noweight.
eqE     noweight.
andE    noweight noweight noweight.
someE   noweight noweight _.

backchainE Atom noweight (cases Rnd Ws 0) :-
   weights Atom Ws, read_7_bits Rnd.

orE (cases Rnd (W::Ws) Acc) Cert Choice :- Acc' is Acc + W,
  ((Acc'  > Rnd, Choice = left,  Cert = noweight) ;
   (Acc' =< Rnd, Choice = right, Cert = (cases Rnd Ws Acc'))).

weights (nat _)    [32,96].
weights (nlist _)  [32,96].
/* end */

weights (ord_bad _)   [20, 40, 68].
weights (reverse _ _) [20,108].

% DM end of example with weights

% Quick trick: iterate at the starting point of the evaluation
% backchainE (rtries N) random :- iterate N.
% backchainE (rtriesW Id N) (qw Cont) :- qdistr Id Cont, iterate N.
%backchainE (qtriesW Id N) (qw Cont) :- qdistr Id Cont, N > 0.

/* max */
ttE     (max empty).
eqE     (max empty).
orE     (max (choose C M)) (max M) C.
someE   (max (term   T M)) (max M) T.
andE    (max (binary M N)) (max M) (max N).
backchainE _ (max M) (max M).
/* end */
/* pairing */
ttE     (A <c> B) :- ttE A, ttE B.
eqE     (A <c> B) :- eqE A, eqE B.
someE   (A <c> B) (C <c> D) T         :- someE A C T, someE B D T.
orE     (A <c> B) (C <c> D) E         :- orE A C E, orE B D E. 
andE    (A <c> B) (C <c> D) (E <c> F) :- andE A C E, andE B D F. 
backchainE At (A <c> B) (C <c> D)  :-
        backchainE At A C, backchainE At B D.
/* end */

ttE   (l-or-r In In).
eqE   (l-or-r In In).
orE   (l-or-r [C|In] Out)
      (l-or-r In Out) C.
someE (l-or-r In Out) (l-or-r In Out) _.
andE  (l-or-r In Out) (l-or-r In Mid)
                      (l-or-r Mid Out).
backchainE _ (l-or-r In Out) (l-or-r In Out).

/* collect */
ttE   (collect In In).
eqE   (collect In In).
orE   (collect In Out)
      (collect In Out) C.
andE  (collect In Out) (collect In Mid) (collect Mid Out).
backchainE _ (collect In Out) (collect In Out).
someE (collect [(c_nat T) | In] Out)
      (collect In Out) (T : nat).
someE (collect [(c_list_nat T)|In] Out)
      (collect In Out) (T : list nat).

subterm Item Item.
subterm Item (c_nat (succ M)) :- subterm Item (c_nat M).
subterm Item (c_list_nat (Nat::L)) :- subterm Item (c_nat Nat) ;
                                      subterm Item (c_list_nat L).
/* end */
/* huniv */
ttE     (huniv _).
eqE     (huniv _).
orE     (huniv Pred) (huniv Pred) _.
andE    (huniv Pred) (huniv Pred) (huniv Pred).
backchainE _ (huniv Pred) (huniv Pred).
someE (huniv Pred) (huniv Pred) (T:nat) :-
  Pred (c_nat T).
someE (huniv Pred) (huniv Pred) (T:list nat) :-
  Pred (c_list_nat T).
/* end */

/*
/* weights */
ttE     (qw qid).
eqE     (qw qid).
andE    (qw (qand Left Right)) (qw Left) (qw Right).
orE     (qw (qor WL WR L R)) (qw Cont) Side :-
  qchoose WL WR Side,
  ((Side = left, Cont = L);
   (Side = right, Cont = R)).
someE   (qw Cont) (qw Cont) _.
backchainE (qw (qprog Id)) (qw Cont) :- qdistr Id Cont.
/* end */
qchoose WL WR Choice :-
  Rng is WL + WR, term_to_string Rng RngStr,
  Cmd is "shuf -i 1-" ^ RngStr ^ " -n 1 -o rand.txt",
  system Cmd _, system "sed -i 's/$/\\./' rand.txt" _,
  open_in "rand.txt" In, readterm In Rnd, close_in In,
  ((Rnd <= WL, Choice = left); (Rnd > WL, Choice = right)).
%  next_bit I0, next_bit I1, next_bit I2, next_bit I3,
%  next_bit I4, next_bit I5, next_bit I6, next_bit I7,
%  R is I0 + I1 * 2 + I2 * 4 + I3 * 8 + I4 * 16 + I5 * 32 + I6 * 64 + I7 * 128,
%  T is (WL * 256) div (WL + WR),
%  ((R <= T, Choice = left); (R > T, Choice = right)).

*/
\end{lstlisting}
\caption{Two FPCs that describe proofs that are limited in either
  height or in size.}
\label{fig:resources}
\begin{lstlisting}[linerange={max-end}]
sig fpcs.
accum_sig kernel.
accum_sig random.
accum_sig debug.

/* resources */
type   height   int -> cert.
type   sze     int -> int -> cert.
/* end */

/* random */
type   random   cert.
/* end */
type   rtries    int -> cert.
type   rtriesW   string -> int -> cert.

type   iterate   int -> o.

/* max */
kind max     type.
type max     max    -> cert.
type binary  max    -> max -> max.
type choose  choice -> max -> max.
type term    A      -> max -> max.
type empty   max.
/* end */

/* pairing */
type   <c>     cert ->  cert ->  cert.
infixr <c>     5.
/* end */

type l-or-r    list choice -> 
               list choice -> cert.

kind nat     type.
type zero    nat.
type succ    nat -> nat.

/* collect */
kind item                     type.
type c_nat             nat -> item.
type c_list_nat   list nat -> item.

type subterm      item -> item -> o.
type collect      list item -> list item -> cert.
/* end */
% type c1      list item -> cert.
/* huniv */
type huniv   (item -> o) -> cert.
/* end */
type tries int -> o.

/* weights */
kind qform       type.
%type qtt, qeq    qform.
type qid         qform.
type qand        qform -> qform -> qform.
type qor         int -> int -> qform -> qform -> qform.
%type qsome       qform -> qform.
type qprog       string -> qform.

type qdistr    string -> qform -> o.
type qchoose   int -> int -> choice -> o.

type qw   qform -> cert.
/* end */

% Another approach to weighing clause

/* weight_cert */
type noweight      cert.
type cases         int -> list int -> int -> cert.
/* end */

/* weight_pred */
type weights       sl -> list int -> o.
/* end */
\end{lstlisting}
\begin{lstlisting}[linerange={max-end}]
module fpcs.

%backchainEx A B C :- announce (backchainEx A B C).
%weights A B       :- announce (weights A B).

/* iterate */
iterate N :- N > 0.
iterate N :- N > 0, N' is N - 1, iterate N'.
/* end */

/* resources */
ttE     (height _).
eqE     (height _).
orE     (height H) (height H) _.
someE   (height H) (height H) _.
andE    (height H) (height H) (height H).
backchainE _ (height H) (height H') :- H  > 0, H' is H  - 1.

ttE     (sze In In).
eqE     (sze In In).
orE     (sze In Out) (sze In Out) _.
someE   (sze In Out) (sze In Out) _.
andE    (sze In Out) (sze In Mid) (sze Mid Out).
backchainE _  (sze In Out) (sze In' Out) :-
         In > 0, In' is In - 1.
/* end */
%backchainE (qtriesW Id N) Cont :- N > 0, N' is N - 1, backchainE (qtriesW Id N') Cont.
/* random */
ttE random.
eqE random.
orE random random Choice :- next_bit I,
  ((I = 0, Choice = left); (I = 1, Choice = right)).
someE random random _.
andE  random random random.
backchainE _ random random.
/* end */

% DM Another approach to putting weights on disjunctive cases.
% Assume that when we use case-weight, our clauses are of the form
%      head <>== b1 or (b2 or ... or (bn or tt)).
% Here, the disjunction only appears at the top-level (no bi formula
% contains disjunction) and that the disjunctions are associated to
% the right (which is the default declaration).  Note that the
% disjunction must end with tt.

% Weights for the disjunctions should always add up to 128.  There
% must be a top-level disjunction in the body (otherwise, don't use
% weights).

% Since backchainE needs to know the predicate on which it is being
% invoked, it seems that we need another backchain expert that can
% examine the atom it is openning.

/* weighty */
ttE     noweight.
eqE     noweight.
andE    noweight noweight noweight.
someE   noweight noweight _.

backchainE Atom noweight (cases Rnd Ws 0) :-
   weights Atom Ws, read_7_bits Rnd.

orE (cases Rnd (W::Ws) Acc) Cert Choice :- Acc' is Acc + W,
  ((Acc'  > Rnd, Choice = left,  Cert = noweight) ;
   (Acc' =< Rnd, Choice = right, Cert = (cases Rnd Ws Acc'))).

weights (nat _)    [32,96].
weights (nlist _)  [32,96].
/* end */

weights (ord_bad _)   [20, 40, 68].
weights (reverse _ _) [20,108].

% DM end of example with weights

% Quick trick: iterate at the starting point of the evaluation
% backchainE (rtries N) random :- iterate N.
% backchainE (rtriesW Id N) (qw Cont) :- qdistr Id Cont, iterate N.
%backchainE (qtriesW Id N) (qw Cont) :- qdistr Id Cont, N > 0.

/* max */
ttE     (max empty).
eqE     (max empty).
orE     (max (choose C M)) (max M) C.
someE   (max (term   T M)) (max M) T.
andE    (max (binary M N)) (max M) (max N).
backchainE _ (max M) (max M).
/* end */
/* pairing */
ttE     (A <c> B) :- ttE A, ttE B.
eqE     (A <c> B) :- eqE A, eqE B.
someE   (A <c> B) (C <c> D) T         :- someE A C T, someE B D T.
orE     (A <c> B) (C <c> D) E         :- orE A C E, orE B D E. 
andE    (A <c> B) (C <c> D) (E <c> F) :- andE A C E, andE B D F. 
backchainE At (A <c> B) (C <c> D)  :-
        backchainE At A C, backchainE At B D.
/* end */

ttE   (l-or-r In In).
eqE   (l-or-r In In).
orE   (l-or-r [C|In] Out)
      (l-or-r In Out) C.
someE (l-or-r In Out) (l-or-r In Out) _.
andE  (l-or-r In Out) (l-or-r In Mid)
                      (l-or-r Mid Out).
backchainE _ (l-or-r In Out) (l-or-r In Out).

/* collect */
ttE   (collect In In).
eqE   (collect In In).
orE   (collect In Out)
      (collect In Out) C.
andE  (collect In Out) (collect In Mid) (collect Mid Out).
backchainE _ (collect In Out) (collect In Out).
someE (collect [(c_nat T) | In] Out)
      (collect In Out) (T : nat).
someE (collect [(c_list_nat T)|In] Out)
      (collect In Out) (T : list nat).

subterm Item Item.
subterm Item (c_nat (succ M)) :- subterm Item (c_nat M).
subterm Item (c_list_nat (Nat::L)) :- subterm Item (c_nat Nat) ;
                                      subterm Item (c_list_nat L).
/* end */
/* huniv */
ttE     (huniv _).
eqE     (huniv _).
orE     (huniv Pred) (huniv Pred) _.
andE    (huniv Pred) (huniv Pred) (huniv Pred).
backchainE _ (huniv Pred) (huniv Pred).
someE (huniv Pred) (huniv Pred) (T:nat) :-
  Pred (c_nat T).
someE (huniv Pred) (huniv Pred) (T:list nat) :-
  Pred (c_list_nat T).
/* end */

/*
/* weights */
ttE     (qw qid).
eqE     (qw qid).
andE    (qw (qand Left Right)) (qw Left) (qw Right).
orE     (qw (qor WL WR L R)) (qw Cont) Side :-
  qchoose WL WR Side,
  ((Side = left, Cont = L);
   (Side = right, Cont = R)).
someE   (qw Cont) (qw Cont) _.
backchainE (qw (qprog Id)) (qw Cont) :- qdistr Id Cont.
/* end */
qchoose WL WR Choice :-
  Rng is WL + WR, term_to_string Rng RngStr,
  Cmd is "shuf -i 1-" ^ RngStr ^ " -n 1 -o rand.txt",
  system Cmd _, system "sed -i 's/$/\\./' rand.txt" _,
  open_in "rand.txt" In, readterm In Rnd, close_in In,
  ((Rnd <= WL, Choice = left); (Rnd > WL, Choice = right)).
%  next_bit I0, next_bit I1, next_bit I2, next_bit I3,
%  next_bit I4, next_bit I5, next_bit I6, next_bit I7,
%  R is I0 + I1 * 2 + I2 * 4 + I3 * 8 + I4 * 16 + I5 * 32 + I6 * 64 + I7 * 128,
%  T is (WL * 256) div (WL + WR),
%  ((R <= T, Choice = left); (R > T, Choice = right)).

*/
\end{lstlisting}
\caption{The \lsti{max} FPC}
% DM Notice that in the max FPC, the ttE and eqE are not trivially
% true!  It's best to keep these explicit.  Also true for the size
% fpc.
% AM: should the evidence for ttE and eqE be the same? Different in type theory
\label{fig:max}
\end{figure}

Figure~\ref{fig:max} contains the FPC based on the constructor
\lsti{max} that is used to record explicitly all information within a
proof, not unlike a proof-term in type theory: in particular, all
disjunctive choices and all substitution instances for existential
quantifiers are collected into a binary tree structure of type
\lsti{max}.
In this sense, proof certificates built with this constructor are \emph{maximally} explicit.
Such proof certificates are used, for example, in~\cite{Pair}; it is
important to note that proof checking with such maximally explicit
certificates can be done with much simpler proof-checkers than those
used in logic programming since backtracking search and unification
are not needed. % to check such certificates.

A characteristic of the FPCs that we have presented here is that none
contain the substitution terms used in backchaining. At the same time,
they may choose to explicitly store substitution information for the
existential quantifiers in goals (see the \lsti{max} FPC above).
While there is no problem in reorganizing our setting so that the
substitution instances used in the backchaining inference are stored
explicitly (see, for example,~\cite{chihani17jar}), we find our
particular design convenient. Furthermore, if we wish to record all
the substitution instances used in a proof, we can write logic
programs in the Clark completion style. In that case,
all substitutions used in specifying backchaining are also captured by
explicit existential quantifiers in the body of those clauses.

If we view a particular FPC as a means of \emph{restricting} proofs, it is
possible to build an FPC that {restricts} proofs satisfying two FPCs
simultaneously.
In particular, Figure~\ref{fig:pairing} defines an FPC based on the
(infix) constructor \lsti{<c>}, which \emph{pairs} two terms of type
\lsti{cert}.
The pairing experts for the certificate \lsti{Cert1 <c> Cert2} simply
request that the corresponding experts succeed for both
\lsti{Cert1} and \lsti{Cert2} and, in the case of the \lsti{orE} and
\lsti{someE}, also return the same choice and substitution term,
respectively.
Thus, the query 
\begin{lstlisting}[basicstyle=\small\ttfamily]
?- check ((height 4) <c> (sze 10 H)) G
\end{lstlisting}
will succeed if there is a proof of \lsti{G} that has a height less
than or equal to 4 while also being of size less than or equal to 10.
A related use of the pairing of two proof certificates is to %use it to
\emph{distill} or \emph{elaborate} proof certificates. 
For example, the proof certificate \lsti{(sze 5 0)} is rather implicit
since it will match any proof that used backchain exactly 5 times.
However, the query
\begin{lstlisting}[basicstyle=\small\ttfamily]
?- check ((sze 5 0) <c> (max Max)) G.
\end{lstlisting}
will store into the \lP variable \lsti{Max} more complete details of
any proof that satisfies the \lsti{(sze 5 0)} constraint.
These maximal certificates are an appropriate starting point for
documenting both the counterexample and why it serves as such. In particular, this forms the infrastructure of an
\emph{explanation} tool for attributing ``blame'' for the origin of a
counterexample.

\begin{figure}
\begin{lstlisting}[linerange={pairing-end}]
sig fpcs.
accum_sig kernel.
accum_sig random.
accum_sig debug.

/* resources */
type   height   int -> cert.
type   sze     int -> int -> cert.
/* end */

/* random */
type   random   cert.
/* end */
type   rtries    int -> cert.
type   rtriesW   string -> int -> cert.

type   iterate   int -> o.

/* max */
kind max     type.
type max     max    -> cert.
type binary  max    -> max -> max.
type choose  choice -> max -> max.
type term    A      -> max -> max.
type empty   max.
/* end */

/* pairing */
type   <c>     cert ->  cert ->  cert.
infixr <c>     5.
/* end */

type l-or-r    list choice -> 
               list choice -> cert.

kind nat     type.
type zero    nat.
type succ    nat -> nat.

/* collect */
kind item                     type.
type c_nat             nat -> item.
type c_list_nat   list nat -> item.

type subterm      item -> item -> o.
type collect      list item -> list item -> cert.
/* end */
% type c1      list item -> cert.
/* huniv */
type huniv   (item -> o) -> cert.
/* end */
type tries int -> o.

/* weights */
kind qform       type.
%type qtt, qeq    qform.
type qid         qform.
type qand        qform -> qform -> qform.
type qor         int -> int -> qform -> qform -> qform.
%type qsome       qform -> qform.
type qprog       string -> qform.

type qdistr    string -> qform -> o.
type qchoose   int -> int -> choice -> o.

type qw   qform -> cert.
/* end */

% Another approach to weighing clause

/* weight_cert */
type noweight      cert.
type cases         int -> list int -> int -> cert.
/* end */

/* weight_pred */
type weights       sl -> list int -> o.
/* end */
\end{lstlisting}
\begin{lstlisting}[linerange={pairing-end}]
module fpcs.

%backchainEx A B C :- announce (backchainEx A B C).
%weights A B       :- announce (weights A B).

/* iterate */
iterate N :- N > 0.
iterate N :- N > 0, N' is N - 1, iterate N'.
/* end */

/* resources */
ttE     (height _).
eqE     (height _).
orE     (height H) (height H) _.
someE   (height H) (height H) _.
andE    (height H) (height H) (height H).
backchainE _ (height H) (height H') :- H  > 0, H' is H  - 1.

ttE     (sze In In).
eqE     (sze In In).
orE     (sze In Out) (sze In Out) _.
someE   (sze In Out) (sze In Out) _.
andE    (sze In Out) (sze In Mid) (sze Mid Out).
backchainE _  (sze In Out) (sze In' Out) :-
         In > 0, In' is In - 1.
/* end */
%backchainE (qtriesW Id N) Cont :- N > 0, N' is N - 1, backchainE (qtriesW Id N') Cont.
/* random */
ttE random.
eqE random.
orE random random Choice :- next_bit I,
  ((I = 0, Choice = left); (I = 1, Choice = right)).
someE random random _.
andE  random random random.
backchainE _ random random.
/* end */

% DM Another approach to putting weights on disjunctive cases.
% Assume that when we use case-weight, our clauses are of the form
%      head <>== b1 or (b2 or ... or (bn or tt)).
% Here, the disjunction only appears at the top-level (no bi formula
% contains disjunction) and that the disjunctions are associated to
% the right (which is the default declaration).  Note that the
% disjunction must end with tt.

% Weights for the disjunctions should always add up to 128.  There
% must be a top-level disjunction in the body (otherwise, don't use
% weights).

% Since backchainE needs to know the predicate on which it is being
% invoked, it seems that we need another backchain expert that can
% examine the atom it is openning.

/* weighty */
ttE     noweight.
eqE     noweight.
andE    noweight noweight noweight.
someE   noweight noweight _.

backchainE Atom noweight (cases Rnd Ws 0) :-
   weights Atom Ws, read_7_bits Rnd.

orE (cases Rnd (W::Ws) Acc) Cert Choice :- Acc' is Acc + W,
  ((Acc'  > Rnd, Choice = left,  Cert = noweight) ;
   (Acc' =< Rnd, Choice = right, Cert = (cases Rnd Ws Acc'))).

weights (nat _)    [32,96].
weights (nlist _)  [32,96].
/* end */

weights (ord_bad _)   [20, 40, 68].
weights (reverse _ _) [20,108].

% DM end of example with weights

% Quick trick: iterate at the starting point of the evaluation
% backchainE (rtries N) random :- iterate N.
% backchainE (rtriesW Id N) (qw Cont) :- qdistr Id Cont, iterate N.
%backchainE (qtriesW Id N) (qw Cont) :- qdistr Id Cont, N > 0.

/* max */
ttE     (max empty).
eqE     (max empty).
orE     (max (choose C M)) (max M) C.
someE   (max (term   T M)) (max M) T.
andE    (max (binary M N)) (max M) (max N).
backchainE _ (max M) (max M).
/* end */
/* pairing */
ttE     (A <c> B) :- ttE A, ttE B.
eqE     (A <c> B) :- eqE A, eqE B.
someE   (A <c> B) (C <c> D) T         :- someE A C T, someE B D T.
orE     (A <c> B) (C <c> D) E         :- orE A C E, orE B D E. 
andE    (A <c> B) (C <c> D) (E <c> F) :- andE A C E, andE B D F. 
backchainE At (A <c> B) (C <c> D)  :-
        backchainE At A C, backchainE At B D.
/* end */

ttE   (l-or-r In In).
eqE   (l-or-r In In).
orE   (l-or-r [C|In] Out)
      (l-or-r In Out) C.
someE (l-or-r In Out) (l-or-r In Out) _.
andE  (l-or-r In Out) (l-or-r In Mid)
                      (l-or-r Mid Out).
backchainE _ (l-or-r In Out) (l-or-r In Out).

/* collect */
ttE   (collect In In).
eqE   (collect In In).
orE   (collect In Out)
      (collect In Out) C.
andE  (collect In Out) (collect In Mid) (collect Mid Out).
backchainE _ (collect In Out) (collect In Out).
someE (collect [(c_nat T) | In] Out)
      (collect In Out) (T : nat).
someE (collect [(c_list_nat T)|In] Out)
      (collect In Out) (T : list nat).

subterm Item Item.
subterm Item (c_nat (succ M)) :- subterm Item (c_nat M).
subterm Item (c_list_nat (Nat::L)) :- subterm Item (c_nat Nat) ;
                                      subterm Item (c_list_nat L).
/* end */
/* huniv */
ttE     (huniv _).
eqE     (huniv _).
orE     (huniv Pred) (huniv Pred) _.
andE    (huniv Pred) (huniv Pred) (huniv Pred).
backchainE _ (huniv Pred) (huniv Pred).
someE (huniv Pred) (huniv Pred) (T:nat) :-
  Pred (c_nat T).
someE (huniv Pred) (huniv Pred) (T:list nat) :-
  Pred (c_list_nat T).
/* end */

/*
/* weights */
ttE     (qw qid).
eqE     (qw qid).
andE    (qw (qand Left Right)) (qw Left) (qw Right).
orE     (qw (qor WL WR L R)) (qw Cont) Side :-
  qchoose WL WR Side,
  ((Side = left, Cont = L);
   (Side = right, Cont = R)).
someE   (qw Cont) (qw Cont) _.
backchainE (qw (qprog Id)) (qw Cont) :- qdistr Id Cont.
/* end */
qchoose WL WR Choice :-
  Rng is WL + WR, term_to_string Rng RngStr,
  Cmd is "shuf -i 1-" ^ RngStr ^ " -n 1 -o rand.txt",
  system Cmd _, system "sed -i 's/$/\\./' rand.txt" _,
  open_in "rand.txt" In, readterm In Rnd, close_in In,
  ((Rnd <= WL, Choice = left); (Rnd > WL, Choice = right)).
%  next_bit I0, next_bit I1, next_bit I2, next_bit I3,
%  next_bit I4, next_bit I5, next_bit I6, next_bit I7,
%  R is I0 + I1 * 2 + I2 * 4 + I3 * 8 + I4 * 16 + I5 * 32 + I6 * 64 + I7 * 128,
%  T is (WL * 256) div (WL + WR),
%  ((R <= T, Choice = left); (R > T, Choice = right)).

*/
\end{lstlisting}
\caption{FPC for pairing}
\label{fig:pairing}
\end{figure}

Various additional examples and experiments using the pairing of FPCs
can be found in~\cite{Pair}. Using similar techniques, it is possible
to define FPCs that target specific types for special treatment: for
example, when generating integers, only (user-defined) small integers
can be inserted into counterexamples.

% AM: done
% DM We should mention somewhere in Section 2 that ``FPCs can be used as
% \emph{proof outlines}~\cite{blanco15wof} by describing some of the
% general shape of a proof: checking such outlines essentially results
% in an attempt to fill in the missing details.''

\section{PBT as proof elaboration in the reasoning logic}
\label{sec:basic}
%% \begin{metanote}
%%   AM: lousy section title
%% \end{metanote}
 
% \subsection{A two-level approach}
%\label{sec:2l}

% We have explored several implementations of FPC, varying in host
% languages and applications~\cite{BlancoPhD}.  

The two-level logic approach resembles the use of meta-interpreters in
logic programming.  Particularly strong versions of such interpreters
have been formalized in~\cite{mcdowell02tocl,gacek12jar}
and exploited in~\cite{baelde14jfr,FeltyM12}.
In our generate-and-test approach
to PBT, the generation phase is controlled by using appropriate FPCs,
and the testing phase is performed by the standard vanilla
meta-interpreter (such as the one in Figure~\ref{fig:interp}).

To illustrate this division between generation and testing, consider
the following two simple examples.  Suppose we want to falsify the
assertion that the reversal of a list is equal to itself. The
generation phase is steered by the predicate \lsti{check}, which uses
a certificate (its first argument) to produce candidate lists
according to a generation strategy. The testing phase performs the
list reversal computation using the meta-interpreter \lsti{interp},
and then negates the conclusion using negation-as-finite-failure,
yielding the clause:
% \begin{metanote}
%   AM: please implement my convention of using \texttt{prop} for tests
% \end{metanote}
\begin{lstlisting}[basicstyle=\small\ttfamily]
prop_rev_id Gen Xs :- 
  check Gen (nlist Xs), 
  interp (rev Xs Ys),
  not (interp (Xs eq Ys)).
\end{lstlisting}
If we set \lsti{Gen} to be  say \lsti{height 3}, the logic
programming engine will return, among others, the answer
\mbox{\lsti{Xs =  s z :: z :: nil}}.
Note that the call to \lsti{not} is safe since, by the totality of
\lsti{rev}, \lsti{Ys} will be ground at calling time.%, which is also
%the reason why we choose not to return it.

As a second simple example, the testing of the symmetry of
\textit{reverse} can be written as:
\begin{lstlisting}[basicstyle=\small\ttfamily]
prop_rev_sym Gen Xs :- 
  check Gen (nlist Xs),
  interp (reverse Xs Ys), 
  not (interp (reverse Ys Xs)).
\end{lstlisting}
Unless one's implementation of \textit{reverse} is grossly mistaken,
the engine should complete its search (according to the prescriptions
of the generator \lsti{Gen}) without finding a counterexample.

We now illustrate how we can capture in our framework various flavors
of PBT\@.

\subsection{Exhaustive generation}
\label{ssec:exh}
While PBT is traditionally associated with random generation, several
tools rely on exhaustive data generation up to a bound~\cite{Sullivan:2004} --- in
fact, such strategy is now the default in Isabelle/HOL's PBT
suite~\cite{BlanchetteBN11}. In particular,
 \begin{enumerate}
 \item (Lazy)SmallCheck~\cite{smallcheck} views the bound as the
   nesting depth of constructors of algebraic data types.
 \item  $\alpha$Check~\cite{Pessina16} employs the  derivation height.
\end{enumerate}
Our \lsti{sze} and \lsti{height} FPCs in Figure~\ref{fig:resources},
respectively, match (1) and (2) and, therefore, can accommodate both.
% But, as mentioned in the previous section, we can go further.
% \begin{metanote}
%   AM: further where?
% \end{metanote}
%

% rewritten by chat
\begin{sloppypar}
One minor limitation of our method is that, although \lsti{check Gen P}
will generate terms up to the specified bound when using either 
\lsti{sze} or \lsti{height} for \lsti{Gen}, the logic programming
engine will enumerate these terms in reverse order, starting from the
bound and going downwards.%, during a PBT query.
%
% A small drawback of our approach is that, while \lsti{check Gen P},
% for \lsti{Gen} using one of \lsti{sze,height}, will generate terms up to
% the given bound, in a PBT query, the logic programming engine will
% enumerate them from that bound downwards.
For example, a query such as
\mbox{\lsti{?- prop_rev_id (height 10) Xs}} will return larger
counterexamples first, starting here with a list of nine $0$ and a
$1$.  This means that if we do not have a good estimate of the
dimension of our counterexample, our query may take an unnecessary
long time or even loop.
\end{sloppypar}

A first fix uses again certificate pairing. The query
\begin{lstlisting}[basicstyle=\small\ttfamily]
?- prop_rev_id ((height 10) <c> (sze 6 _)) Xs
\end{lstlisting}
will converge quickly to the usual minimal answer.
Generally
speaking, constraining the size to $n$ will also effectively
constrain the height to be approximately $O(\log{n})$.
However, we still ought to have some idea about the dimension of the
counterexample beforehand and this is not realistic. Yet, it is
easy, thanks to logic programming, to implement a simple-minded
form of \emph{iterative deepening}, where we backtrack over
an increasing list of bounds:
\begin{lstlisting}[basicstyle=\small\ttfamily]
prop_rev_sym_it Start End Xs :-
  mk_list Start End Range,
  member H Range,
  check (height H) (nlist Xs),
  interp (reverse Xs Ys),
  not (interp (Xs eq Ys)).
\end{lstlisting}
Here, \lsti{mk_list Start End Range} holds  when \lsti{Start,End} are  positive
integers and \lsti{Range} is the list natural numbers  \lsti{[Start,...,End]}.
In addition, we can choose to express size as a function
of height --- of course this can be tuned by the user, depending on
the data they are working with:
\begin{lstlisting}[basicstyle=\small\ttfamily]
prop_rev_sym_it Start End Xs :-
  mk_list Start End Range,
  member H Range, Sh is H * 3,
  check ((height H) <c> (sze Sh _)) (nlist Xs),
  interp (reverse Xs Ys), 
  not (interp (Xs eq Ys)).
\end{lstlisting}

% \begin{metanote}
%   AM: the ../code is in file \verb|fpc-qbound| and it's messy. Remove?
% \end{metanote}
While these test generators are easy to construct, they have the
drawback of recomputing candidates at each level.  A better approach
is to introduce an FPC for a form of iterative deepening for
\emph{exact} bounds, where we output only those candidates requiring
that precise bound. This has some similarity with the approach in
\emph{Feat}~\cite{feat}. We will not pursue this avenue here.
% The interested reader can peruse this FPC in  the ../code accompanying our paper.

\subsection{Random generation}
\label{ssec:rand}

\begin{figure} %,basicstyle=\small\ttfamily
\begin{lstlisting}[linerange={random-end}]
sig fpcs.
accum_sig kernel.
accum_sig random.
accum_sig debug.

/* resources */
type   height   int -> cert.
type   sze     int -> int -> cert.
/* end */

/* random */
type   random   cert.
/* end */
type   rtries    int -> cert.
type   rtriesW   string -> int -> cert.

type   iterate   int -> o.

/* max */
kind max     type.
type max     max    -> cert.
type binary  max    -> max -> max.
type choose  choice -> max -> max.
type term    A      -> max -> max.
type empty   max.
/* end */

/* pairing */
type   <c>     cert ->  cert ->  cert.
infixr <c>     5.
/* end */

type l-or-r    list choice -> 
               list choice -> cert.

kind nat     type.
type zero    nat.
type succ    nat -> nat.

/* collect */
kind item                     type.
type c_nat             nat -> item.
type c_list_nat   list nat -> item.

type subterm      item -> item -> o.
type collect      list item -> list item -> cert.
/* end */
% type c1      list item -> cert.
/* huniv */
type huniv   (item -> o) -> cert.
/* end */
type tries int -> o.

/* weights */
kind qform       type.
%type qtt, qeq    qform.
type qid         qform.
type qand        qform -> qform -> qform.
type qor         int -> int -> qform -> qform -> qform.
%type qsome       qform -> qform.
type qprog       string -> qform.

type qdistr    string -> qform -> o.
type qchoose   int -> int -> choice -> o.

type qw   qform -> cert.
/* end */

% Another approach to weighing clause

/* weight_cert */
type noweight      cert.
type cases         int -> list int -> int -> cert.
/* end */

/* weight_pred */
type weights       sl -> list int -> o.
/* end */
\end{lstlisting}
\begin{lstlisting}[linerange={random-end}]
module fpcs.

%backchainEx A B C :- announce (backchainEx A B C).
%weights A B       :- announce (weights A B).

/* iterate */
iterate N :- N > 0.
iterate N :- N > 0, N' is N - 1, iterate N'.
/* end */

/* resources */
ttE     (height _).
eqE     (height _).
orE     (height H) (height H) _.
someE   (height H) (height H) _.
andE    (height H) (height H) (height H).
backchainE _ (height H) (height H') :- H  > 0, H' is H  - 1.

ttE     (sze In In).
eqE     (sze In In).
orE     (sze In Out) (sze In Out) _.
someE   (sze In Out) (sze In Out) _.
andE    (sze In Out) (sze In Mid) (sze Mid Out).
backchainE _  (sze In Out) (sze In' Out) :-
         In > 0, In' is In - 1.
/* end */
%backchainE (qtriesW Id N) Cont :- N > 0, N' is N - 1, backchainE (qtriesW Id N') Cont.
/* random */
ttE random.
eqE random.
orE random random Choice :- next_bit I,
  ((I = 0, Choice = left); (I = 1, Choice = right)).
someE random random _.
andE  random random random.
backchainE _ random random.
/* end */

% DM Another approach to putting weights on disjunctive cases.
% Assume that when we use case-weight, our clauses are of the form
%      head <>== b1 or (b2 or ... or (bn or tt)).
% Here, the disjunction only appears at the top-level (no bi formula
% contains disjunction) and that the disjunctions are associated to
% the right (which is the default declaration).  Note that the
% disjunction must end with tt.

% Weights for the disjunctions should always add up to 128.  There
% must be a top-level disjunction in the body (otherwise, don't use
% weights).

% Since backchainE needs to know the predicate on which it is being
% invoked, it seems that we need another backchain expert that can
% examine the atom it is openning.

/* weighty */
ttE     noweight.
eqE     noweight.
andE    noweight noweight noweight.
someE   noweight noweight _.

backchainE Atom noweight (cases Rnd Ws 0) :-
   weights Atom Ws, read_7_bits Rnd.

orE (cases Rnd (W::Ws) Acc) Cert Choice :- Acc' is Acc + W,
  ((Acc'  > Rnd, Choice = left,  Cert = noweight) ;
   (Acc' =< Rnd, Choice = right, Cert = (cases Rnd Ws Acc'))).

weights (nat _)    [32,96].
weights (nlist _)  [32,96].
/* end */

weights (ord_bad _)   [20, 40, 68].
weights (reverse _ _) [20,108].

% DM end of example with weights

% Quick trick: iterate at the starting point of the evaluation
% backchainE (rtries N) random :- iterate N.
% backchainE (rtriesW Id N) (qw Cont) :- qdistr Id Cont, iterate N.
%backchainE (qtriesW Id N) (qw Cont) :- qdistr Id Cont, N > 0.

/* max */
ttE     (max empty).
eqE     (max empty).
orE     (max (choose C M)) (max M) C.
someE   (max (term   T M)) (max M) T.
andE    (max (binary M N)) (max M) (max N).
backchainE _ (max M) (max M).
/* end */
/* pairing */
ttE     (A <c> B) :- ttE A, ttE B.
eqE     (A <c> B) :- eqE A, eqE B.
someE   (A <c> B) (C <c> D) T         :- someE A C T, someE B D T.
orE     (A <c> B) (C <c> D) E         :- orE A C E, orE B D E. 
andE    (A <c> B) (C <c> D) (E <c> F) :- andE A C E, andE B D F. 
backchainE At (A <c> B) (C <c> D)  :-
        backchainE At A C, backchainE At B D.
/* end */

ttE   (l-or-r In In).
eqE   (l-or-r In In).
orE   (l-or-r [C|In] Out)
      (l-or-r In Out) C.
someE (l-or-r In Out) (l-or-r In Out) _.
andE  (l-or-r In Out) (l-or-r In Mid)
                      (l-or-r Mid Out).
backchainE _ (l-or-r In Out) (l-or-r In Out).

/* collect */
ttE   (collect In In).
eqE   (collect In In).
orE   (collect In Out)
      (collect In Out) C.
andE  (collect In Out) (collect In Mid) (collect Mid Out).
backchainE _ (collect In Out) (collect In Out).
someE (collect [(c_nat T) | In] Out)
      (collect In Out) (T : nat).
someE (collect [(c_list_nat T)|In] Out)
      (collect In Out) (T : list nat).

subterm Item Item.
subterm Item (c_nat (succ M)) :- subterm Item (c_nat M).
subterm Item (c_list_nat (Nat::L)) :- subterm Item (c_nat Nat) ;
                                      subterm Item (c_list_nat L).
/* end */
/* huniv */
ttE     (huniv _).
eqE     (huniv _).
orE     (huniv Pred) (huniv Pred) _.
andE    (huniv Pred) (huniv Pred) (huniv Pred).
backchainE _ (huniv Pred) (huniv Pred).
someE (huniv Pred) (huniv Pred) (T:nat) :-
  Pred (c_nat T).
someE (huniv Pred) (huniv Pred) (T:list nat) :-
  Pred (c_list_nat T).
/* end */

/*
/* weights */
ttE     (qw qid).
eqE     (qw qid).
andE    (qw (qand Left Right)) (qw Left) (qw Right).
orE     (qw (qor WL WR L R)) (qw Cont) Side :-
  qchoose WL WR Side,
  ((Side = left, Cont = L);
   (Side = right, Cont = R)).
someE   (qw Cont) (qw Cont) _.
backchainE (qw (qprog Id)) (qw Cont) :- qdistr Id Cont.
/* end */
qchoose WL WR Choice :-
  Rng is WL + WR, term_to_string Rng RngStr,
  Cmd is "shuf -i 1-" ^ RngStr ^ " -n 1 -o rand.txt",
  system Cmd _, system "sed -i 's/$/\\./' rand.txt" _,
  open_in "rand.txt" In, readterm In Rnd, close_in In,
  ((Rnd <= WL, Choice = left); (Rnd > WL, Choice = right)).
%  next_bit I0, next_bit I1, next_bit I2, next_bit I3,
%  next_bit I4, next_bit I5, next_bit I6, next_bit I7,
%  R is I0 + I1 * 2 + I2 * 4 + I3 * 8 + I4 * 16 + I5 * 32 + I6 * 64 + I7 * 128,
%  T is (WL * 256) div (WL + WR),
%  ((R <= T, Choice = left); (R > T, Choice = right)).

*/
\end{lstlisting}
\caption{FPC for random generation}
\label{fig:qrandom}
\end{figure}

The FPC setup can be extended to support random generation of
candidates.  The idea is to implement a form of \emph{randomization}
of choice points: when a choice appears, we flip a coin to
decide which case to choose.
There are two major sources of choice in running a logic program:
which disjunct in a disjunction to pick and which clause on which
to backchain.
In this subsection, we will assume that there is only one clause for
backchaining: this can be done by putting clauses into their
Clark-completion format.
Thus, both forms of choice are represented as the selection of a
disjunct within a disjunction. 
% In this way, these two forms of choice are accounted for as a choice
% of a disjunct within a disjunction.
%
For example, we shall write the definitions of the \lsti{isnat} and
\lsti{nlist} predicates from Figure~\ref{fig:examples} as follows.
\begin{lstlisting}[linerange={examples-end},basicstyle=\small\ttfamily]
module interp.

/* interp */
interp tt.
interp (T eq T).
interp (G1 and G2) :- interp G1, interp G2.
interp (G1 or G2)  :- interp G1 ; interp G2.
interp (some G)    :- interp (G T).
interp A           :- (A <>== G), interp G.
/* end */
/* nabla */
interp (nabla G)     :- pi x\ interp (G x).
interp (all G)     :- pi x\ interp (G x).
/* end */
% interp A                  :- progs A Gs, member G GS, interp G.

% (nat z)     <>== tt.
% (nat (s X)) <>== (nat X).
% (nlist nil)     <>== tt.
% (nlist (N::Ns)) <>== ((nat N) and (nlist Ns)).
% (append nil K K)         <>== tt.
% (append (X::L) K (X::M)) <>== (append L K M).
% (reverse L K)    <>== (rev L K nil).
% (rev nil A A)    <>== tt.
% (rev (X::L) K A) <>== (rev L K (X::A)).
/* examples */
(isnat N) <>== (N eq z) or
               (some N'\ N eq (s N') and (isnat N')) or ff.
(nlist L) <>== (L eq nil) or 
               (some N\ some Ns\ L eq (N::Ns) and (isnat N) 
                                   and (nlist Ns)) or ff.
/* end */

% There was a mistake in append (which I think was not intended (X for X').  I fixed it.
% AM: not adding or ff unless it is used within a check in test-lst
(append Xs Ys Zs) <>== ((Xs eq nil) and (Ys eq Zs)) or
                        (some X'\ some Xs'\ some Zs'\ Xs eq (X':: Xs') and (append Xs' Ys Zs') and (Zs eq (X' :: Zs'))).
(revApp L R) <>==
   ((L eq nil) and (R eq nil)) or
    some X\ some Xs\ some Ts\ (L eq (X::Xs)) and ((reverse Xs Ts) and (append Ts (X:: nil) R)).

(reverse L K) <>== (rev_acc L K nil).
(rev_acc L K A) <>== ((L eq nil) and (K eq A)) or 
                   some L'\ ( L eq (X::L')) and (rev_acc L' K (X::A)).

member A (A:: _).
member A (_ :: Rs) :- member A Rs.

(leq N M) <>==
      ( N eq z) or
       some X\ (some Y\ (N eq  (s X)) and  (M eq  (s Y)) and (leq X Y)).

(gt N M) <>==
      (M eq z) and some K\ (N eq (s K)) or
       some X\ some Y\ ((N eq (s X)) and (( M eq (s Y)) and (gt X Y))).

(ord_bad L)  <>==
      (L eq nil) or
      some X\ ((L eq (X :: nil)) and (nat X)) or
      some Y\ some Rs\ ((L eq (X :: ( Y :: Rs))) and (leq X Y) and (ord_bad Rs))  or ff.

(ins X L R) <>==
       ( (L eq nil) and ( (R eq ( X :: nil)) and (nat X))) or
       some Y\ some Ys\ ((L eq ( Y :: Ys)) and (R eq  (X :: ( Y :: Ys)) and (leq X Y))) or
       some Y\ some Ys\  some Rs\ (L eq ( Y ::  Ys)) and
       	    (R eq (X :: Rs)) and (gt X Y)  and (ins X Ys Rs).

\end{lstlisting}
In these two examples, the body of clauses is written as a
\emph{list} of disjunctions: that is, the body of such clauses is written as
\[
  D_1\lsti{ or } D_2\lsti{ or } \cdots D_n\lsti{ or ff},
\]
where $n\ge 1$ and $D_1, \ldots, D_n$ are formulas that are not
disjunctions --- here, false, written as \lsti{ff}, represents the
empty list of disjunctions. This choice of writing the body of
clauses will make it easier to specify a probability distribution to the
disjunctions $D_1, \ldots, D_n$, see the FPC defined in
Figure~\ref{fig:weighted}.

A simple FPC given by the constructor \lsti{random} is described in
Figure~\ref{fig:qrandom}.   Here, we assume that the predicate
\lsti{next_bit} can access a stream of random bits.

A more useful random test generator is based on a certificate 
instructing the kernel to select disjunctions according to certain
probability distributions.  The user can specify such a distribution
in a list of weights assigned to each disjunction.  In the examples we
consider here, these disjuncts appear only at the top level of the
body of the clause defining a given predicate.
When the kernel encounters an atomic formula, the backchain expert
\lsti{backchainE} is responsible for expanding that atomic formula
into its definition, which is why the expert is indexed by an atom. At this stage, it is necessary to consider the
list of weights assigned to individual predicates. 
% such awareness if not possible for the backchain expert shown in
% Figure~\ref{fig:kernel}.
% However, we can
% extend the checker in that figure with the following clause that uses
% another version of that expert, namely \lsti{backchainEx}.
% \lstinputlisting[linerange={checkx-end},basicstyle=\small\ttfamily]{code/sect5/kernel.sig}
% \lstinputlisting[linerange={checkx-end},basicstyle=\small\ttfamily]{code/sect5/kernel.mod}

\begin{figure}[h]
\begin{lstlisting}[linerange={weight_cert-end},basicstyle=\small\ttfamily]
sig fpcs.
accum_sig kernel.
accum_sig random.
accum_sig debug.

/* resources */
type   height   int -> cert.
type   sze     int -> int -> cert.
/* end */

/* random */
type   random   cert.
/* end */
type   rtries    int -> cert.
type   rtriesW   string -> int -> cert.

type   iterate   int -> o.

/* max */
kind max     type.
type max     max    -> cert.
type binary  max    -> max -> max.
type choose  choice -> max -> max.
type term    A      -> max -> max.
type empty   max.
/* end */

/* pairing */
type   <c>     cert ->  cert ->  cert.
infixr <c>     5.
/* end */

type l-or-r    list choice -> 
               list choice -> cert.

kind nat     type.
type zero    nat.
type succ    nat -> nat.

/* collect */
kind item                     type.
type c_nat             nat -> item.
type c_list_nat   list nat -> item.

type subterm      item -> item -> o.
type collect      list item -> list item -> cert.
/* end */
% type c1      list item -> cert.
/* huniv */
type huniv   (item -> o) -> cert.
/* end */
type tries int -> o.

/* weights */
kind qform       type.
%type qtt, qeq    qform.
type qid         qform.
type qand        qform -> qform -> qform.
type qor         int -> int -> qform -> qform -> qform.
%type qsome       qform -> qform.
type qprog       string -> qform.

type qdistr    string -> qform -> o.
type qchoose   int -> int -> choice -> o.

type qw   qform -> cert.
/* end */

% Another approach to weighing clause

/* weight_cert */
type noweight      cert.
type cases         int -> list int -> int -> cert.
/* end */

/* weight_pred */
type weights       sl -> list int -> o.
/* end */
\end{lstlisting}
\begin{lstlisting}[linerange={weighty-end},basicstyle=\small\ttfamily]
module fpcs.

%backchainEx A B C :- announce (backchainEx A B C).
%weights A B       :- announce (weights A B).

/* iterate */
iterate N :- N > 0.
iterate N :- N > 0, N' is N - 1, iterate N'.
/* end */

/* resources */
ttE     (height _).
eqE     (height _).
orE     (height H) (height H) _.
someE   (height H) (height H) _.
andE    (height H) (height H) (height H).
backchainE _ (height H) (height H') :- H  > 0, H' is H  - 1.

ttE     (sze In In).
eqE     (sze In In).
orE     (sze In Out) (sze In Out) _.
someE   (sze In Out) (sze In Out) _.
andE    (sze In Out) (sze In Mid) (sze Mid Out).
backchainE _  (sze In Out) (sze In' Out) :-
         In > 0, In' is In - 1.
/* end */
%backchainE (qtriesW Id N) Cont :- N > 0, N' is N - 1, backchainE (qtriesW Id N') Cont.
/* random */
ttE random.
eqE random.
orE random random Choice :- next_bit I,
  ((I = 0, Choice = left); (I = 1, Choice = right)).
someE random random _.
andE  random random random.
backchainE _ random random.
/* end */

% DM Another approach to putting weights on disjunctive cases.
% Assume that when we use case-weight, our clauses are of the form
%      head <>== b1 or (b2 or ... or (bn or tt)).
% Here, the disjunction only appears at the top-level (no bi formula
% contains disjunction) and that the disjunctions are associated to
% the right (which is the default declaration).  Note that the
% disjunction must end with tt.

% Weights for the disjunctions should always add up to 128.  There
% must be a top-level disjunction in the body (otherwise, don't use
% weights).

% Since backchainE needs to know the predicate on which it is being
% invoked, it seems that we need another backchain expert that can
% examine the atom it is openning.

/* weighty */
ttE     noweight.
eqE     noweight.
andE    noweight noweight noweight.
someE   noweight noweight _.

backchainE Atom noweight (cases Rnd Ws 0) :-
   weights Atom Ws, read_7_bits Rnd.

orE (cases Rnd (W::Ws) Acc) Cert Choice :- Acc' is Acc + W,
  ((Acc'  > Rnd, Choice = left,  Cert = noweight) ;
   (Acc' =< Rnd, Choice = right, Cert = (cases Rnd Ws Acc'))).

weights (nat _)    [32,96].
weights (nlist _)  [32,96].
/* end */

weights (ord_bad _)   [20, 40, 68].
weights (reverse _ _) [20,108].

% DM end of example with weights

% Quick trick: iterate at the starting point of the evaluation
% backchainE (rtries N) random :- iterate N.
% backchainE (rtriesW Id N) (qw Cont) :- qdistr Id Cont, iterate N.
%backchainE (qtriesW Id N) (qw Cont) :- qdistr Id Cont, N > 0.

/* max */
ttE     (max empty).
eqE     (max empty).
orE     (max (choose C M)) (max M) C.
someE   (max (term   T M)) (max M) T.
andE    (max (binary M N)) (max M) (max N).
backchainE _ (max M) (max M).
/* end */
/* pairing */
ttE     (A <c> B) :- ttE A, ttE B.
eqE     (A <c> B) :- eqE A, eqE B.
someE   (A <c> B) (C <c> D) T         :- someE A C T, someE B D T.
orE     (A <c> B) (C <c> D) E         :- orE A C E, orE B D E. 
andE    (A <c> B) (C <c> D) (E <c> F) :- andE A C E, andE B D F. 
backchainE At (A <c> B) (C <c> D)  :-
        backchainE At A C, backchainE At B D.
/* end */

ttE   (l-or-r In In).
eqE   (l-or-r In In).
orE   (l-or-r [C|In] Out)
      (l-or-r In Out) C.
someE (l-or-r In Out) (l-or-r In Out) _.
andE  (l-or-r In Out) (l-or-r In Mid)
                      (l-or-r Mid Out).
backchainE _ (l-or-r In Out) (l-or-r In Out).

/* collect */
ttE   (collect In In).
eqE   (collect In In).
orE   (collect In Out)
      (collect In Out) C.
andE  (collect In Out) (collect In Mid) (collect Mid Out).
backchainE _ (collect In Out) (collect In Out).
someE (collect [(c_nat T) | In] Out)
      (collect In Out) (T : nat).
someE (collect [(c_list_nat T)|In] Out)
      (collect In Out) (T : list nat).

subterm Item Item.
subterm Item (c_nat (succ M)) :- subterm Item (c_nat M).
subterm Item (c_list_nat (Nat::L)) :- subterm Item (c_nat Nat) ;
                                      subterm Item (c_list_nat L).
/* end */
/* huniv */
ttE     (huniv _).
eqE     (huniv _).
orE     (huniv Pred) (huniv Pred) _.
andE    (huniv Pred) (huniv Pred) (huniv Pred).
backchainE _ (huniv Pred) (huniv Pred).
someE (huniv Pred) (huniv Pred) (T:nat) :-
  Pred (c_nat T).
someE (huniv Pred) (huniv Pred) (T:list nat) :-
  Pred (c_list_nat T).
/* end */

/*
/* weights */
ttE     (qw qid).
eqE     (qw qid).
andE    (qw (qand Left Right)) (qw Left) (qw Right).
orE     (qw (qor WL WR L R)) (qw Cont) Side :-
  qchoose WL WR Side,
  ((Side = left, Cont = L);
   (Side = right, Cont = R)).
someE   (qw Cont) (qw Cont) _.
backchainE (qw (qprog Id)) (qw Cont) :- qdistr Id Cont.
/* end */
qchoose WL WR Choice :-
  Rng is WL + WR, term_to_string Rng RngStr,
  Cmd is "shuf -i 1-" ^ RngStr ^ " -n 1 -o rand.txt",
  system Cmd _, system "sed -i 's/$/\\./' rand.txt" _,
  open_in "rand.txt" In, readterm In Rnd, close_in In,
  ((Rnd <= WL, Choice = left); (Rnd > WL, Choice = right)).
%  next_bit I0, next_bit I1, next_bit I2, next_bit I3,
%  next_bit I4, next_bit I5, next_bit I6, next_bit I7,
%  R is I0 + I1 * 2 + I2 * 4 + I3 * 8 + I4 * 16 + I5 * 32 + I6 * 64 + I7 * 128,
%  T is (WL * 256) div (WL + WR),
%  ((R <= T, Choice = left); (R > T, Choice = right)).

*/
\end{lstlisting}
\lstinputlisting[linerange={iterate-end},basicstyle=\small\ttfamily]{code/sect5/fpcs.mod}
\caption{An FPC that selects randomly from a weighted disjunct.}
\label{fig:weighted}
\end{figure}

Consider the FPC specification in Figure~\ref{fig:weighted}.  This
certificate has two constructors: the constant \lsti{noweight}
indicates that no weights are enforced at this part of the
certificate.  The other certificate is of the form 
\lsti{cases Rnd Ws Acc},
where \lsti{Rnd} is a random number (%from
between 0 and 127
inclusively), \lsti{Ws} are the remaining weights for the other
disjunctions, and \lsti{Acc} is the accumulation of the weights that
have been skipped at this point in the proof-checking process.  The
value of this certificate is initialized (by the \lsti{backchainE}
expert) to be \lsti{cases Rnd Ws 0} using the  random 7-bit number
\lsti{Rnd} (which can be computed by calling \lsti{next_bit} seven
times) and a list of weights \lsti{Ws} stored in the \lsti{weights}
predicate associated to the atomic formula that is being unfolded.

The weights (Figure~\ref{fig:weighted}) used here for
\lsti{nat}-atoms select the first disjunction (for zero) one time out
of four and select the successor clause in the remaining cases.  Thus,
this weighting scheme favors selecting small natural numbers.  The
weighting scheme for \lsti{nlist} similarly favors short lists.  For
example, the query\footnote{You can only execute this and the next
  query with an implementation that can access a stream of random
  bits. This is not shown here.}
\begin{lstlisting}[basicstyle=\small\ttfamily]
  ?- iterate 5, check noweight (nlist L).
\end{lstlisting}
would then generate the following stream of five lists of natural
numbers (depending, of course, on the random stream of bits
provided). 
\begin{lstlisting}[basicstyle=\small\ttfamily]
L = nil
L = nil
L = s (s (s (s (s (s z))))) :: s (s (s (s (s (s (s z)))))) :: z :: 
    s (s z) :: s (s (s z)) :: s z :: s z :: 
    s (s (s (s (s (s (s (s (s (s (s (s (s (s z))))))))))))) :: 
    s z :: z :: nil
L = s z :: z :: z :: s z :: s (s (s (s (s (s (s (s z))))))) :: nil
L = s z :: s (s (s (s (s (s z))))) :: nil
\end{lstlisting}
As an example of using such randomize test case generation, the query 
\begin{lstlisting}[basicstyle=\small\ttfamily]
?- iterate 10, check noweight (nlist L), 
               interp (reverse L R), 
               not (interp (reverse R L))).
\end{lstlisting}
will test the property that \lsti{reverse} is a symmetric relation on
10 randomly selected short lists of small numbers.

As we mention in Section~\ref{sec:rel}, this is but one strategy for
random generation and quite possibly not the most efficient one, as
the experiments in~\cite{blanco19ppdp} indicate. In fact,
programming random generators is an
art~\cite{TNIQ,pltredexconstraintlogic,LampropoulosPP18} in every PBT
approach.  We can, of course, use the pairing of FPCs
(Figure~\ref{fig:pairing}) to help filter and fully elaborate structures
generated using the randomization techniques mentioned above.

\subsection{Shrinking}
\label{ssec:shrink}

Randomly generated data that raise counterexamples may be too large to
be the starting point of  the often frustrating process of bug-fixing.  For
a compelling example, look no further than the run of the information-flow abstract
machine described in~\cite{TNIQ}.
For our much simpler example, there is certainly a
smaller counterexample than the above for our running property, say
\lsti{z :: s z :: nil}.

Clearly, it is desirable to find automatically such smaller
counterexamples. This phase is known as \emph{shrinking} and consists
of creating a number of smaller variants of the bug-triggering data. These
variants are then tested to determine if they induce a
failure.  If that is the case, the shrinking process can be repeated
until we get to a local minimum. In the QuickCheck tradition,
shrinkers, as well as custom generators, are the user's
responsibility, in the sense that PBT tools offer little  support
for their formulation. This is particularly painful when we need to
shrink \emph{modulo} some invariant, e.g., well-typed terms or
meaningful sequences of machine instructions.

\begin{figure}
\begin{lstlisting}[linerange={collect-end}]
sig fpcs.
accum_sig kernel.
accum_sig random.
accum_sig debug.

/* resources */
type   height   int -> cert.
type   sze     int -> int -> cert.
/* end */

/* random */
type   random   cert.
/* end */
type   rtries    int -> cert.
type   rtriesW   string -> int -> cert.

type   iterate   int -> o.

/* max */
kind max     type.
type max     max    -> cert.
type binary  max    -> max -> max.
type choose  choice -> max -> max.
type term    A      -> max -> max.
type empty   max.
/* end */

/* pairing */
type   <c>     cert ->  cert ->  cert.
infixr <c>     5.
/* end */

type l-or-r    list choice -> 
               list choice -> cert.

kind nat     type.
type zero    nat.
type succ    nat -> nat.

/* collect */
kind item                     type.
type c_nat             nat -> item.
type c_list_nat   list nat -> item.

type subterm      item -> item -> o.
type collect      list item -> list item -> cert.
/* end */
% type c1      list item -> cert.
/* huniv */
type huniv   (item -> o) -> cert.
/* end */
type tries int -> o.

/* weights */
kind qform       type.
%type qtt, qeq    qform.
type qid         qform.
type qand        qform -> qform -> qform.
type qor         int -> int -> qform -> qform -> qform.
%type qsome       qform -> qform.
type qprog       string -> qform.

type qdistr    string -> qform -> o.
type qchoose   int -> int -> choice -> o.

type qw   qform -> cert.
/* end */

% Another approach to weighing clause

/* weight_cert */
type noweight      cert.
type cases         int -> list int -> int -> cert.
/* end */

/* weight_pred */
type weights       sl -> list int -> o.
/* end */
\end{lstlisting}
\lstinputlisting[linerange={collect-end}]{code/sect5/fpcs.mod}
\caption{An FPC for collecting substitution terms from proof and a
  predicate to compute subterm.}
\label{fig:collect}
\end{figure}

One way to describe shrinking using FPCs is to consider the
following outline.

\paragraph{Step 1: Collect all substitution terms in an existing proof.}
Given a successful proof that a counterexample exists, use  the
\lsti{collect} FPC in Figure~\ref{fig:collect} to extract the list of
terms instantiating the existentials in that proof.
%
% One can also use the pairing FPC to additionally collect a metric (the
% size and/or height) of that proof.
%
Note that this FPC formally collects a list of terms of different
types, in our running example \lsti{nat} and \lsti{list nat}: we
accommodate such a collection by providing constructors (e.g.,
\lsti{c_nat} and \lsti{c_list_nat}) that map each of these types into
the type \lsti{item}.
Since the third argument of the \lsti{someE} expert predicate can be of
any type, we use the \emph{ad hoc polymorphism} available in
\lP~\cite{nadathur92types} to specify different clauses % to use
for
this 
expert depending on the type of the term in that position: this %then
allows us to choose different  coercion constructors to inject all
these terms into the one type \lsti{item}.

% \begin{metanote}
% can we exemplify
% \end{metanote}

% For example the query
% \begin{lstlisting}[basicstyle=\small\ttfamily]
% ?- check (collect Ts _) (prop_rev_idR Xs).
% \end{lstlisting}
% will collect in to a list the answer for \lsti{Xs} we saw before ---
% assuming the same seed.
\smallskip

For the purposes of the next step, it might  be useful to remove
from this list any item that is a subterm of another item in
that list, where
The definition of the subterm relation is given also in
Figure~\ref{fig:collect}.

\begin{figure}
\begin{lstlisting}[linerange={huniv-end}]
sig fpcs.
accum_sig kernel.
accum_sig random.
accum_sig debug.

/* resources */
type   height   int -> cert.
type   sze     int -> int -> cert.
/* end */

/* random */
type   random   cert.
/* end */
type   rtries    int -> cert.
type   rtriesW   string -> int -> cert.

type   iterate   int -> o.

/* max */
kind max     type.
type max     max    -> cert.
type binary  max    -> max -> max.
type choose  choice -> max -> max.
type term    A      -> max -> max.
type empty   max.
/* end */

/* pairing */
type   <c>     cert ->  cert ->  cert.
infixr <c>     5.
/* end */

type l-or-r    list choice -> 
               list choice -> cert.

kind nat     type.
type zero    nat.
type succ    nat -> nat.

/* collect */
kind item                     type.
type c_nat             nat -> item.
type c_list_nat   list nat -> item.

type subterm      item -> item -> o.
type collect      list item -> list item -> cert.
/* end */
% type c1      list item -> cert.
/* huniv */
type huniv   (item -> o) -> cert.
/* end */
type tries int -> o.

/* weights */
kind qform       type.
%type qtt, qeq    qform.
type qid         qform.
type qand        qform -> qform -> qform.
type qor         int -> int -> qform -> qform -> qform.
%type qsome       qform -> qform.
type qprog       string -> qform.

type qdistr    string -> qform -> o.
type qchoose   int -> int -> choice -> o.

type qw   qform -> cert.
/* end */

% Another approach to weighing clause

/* weight_cert */
type noweight      cert.
type cases         int -> list int -> int -> cert.
/* end */

/* weight_pred */
type weights       sl -> list int -> o.
/* end */
\end{lstlisting}
\lstinputlisting[linerange={huniv-end}]{code/sect5/fpcs.mod}
\caption{An FPC for restricting existential choices.}
\label{fig:huniv}
\end{figure}

\paragraph{Step 2: Search again restricting substitution instances.} 
Search again for the proof of a counterexample but this time use the
\lsti{huniv} FPC (Figure~\ref{fig:huniv}) that restricts the
existential quantifiers to use subterms of terms collected in the
first pass.
(The name \lsti{huniv} is  mnemonic for ``Herbrand
universe'': that is, its argument is a predicate that describes the
set of allowed substitution terms within the certificate.)
%
% Pairing with the FPC restricting size and/or height  can
% additionally control the search for a new proof.
%
Replacing the subterm relation with the proper-subterm relation can
further constrain the search for proofs.
For example, consider the following \lP query, where \lsti{G} is a
formula that encodes the generator, \lsti{Is} is the
list of terms (items) collected from the proof of a counterexample,
and \lsti{H} is the height determined for that proof.

\begin{lstlisting}[basicstyle=\small\ttfamily]
?- check ((huniv (T\ sigma I\ member I Is, subterm T I)) <c> 
       (height H) <c> (max Max)) G.
\end{lstlisting}

In this case, the variable \lsti{Max} will record the details of a
proof that satisfies the height bound as well as instantiates the
existential quantifiers with terms that were subterms of the original
proof.
One can also rerun this query with a lower height bound and by
replacing the implemented notion of subterm with ``proper subterm''.
In this way, the search for proofs involving smaller but related
instantiations can be used to shrink a counterexample.

\section{PBT for metaprogramming}
\label{sec:mmm}

We now explore how to extend PBT to the domain of
metaprogramming. This extension will allow us to find counterexamples
to statements about the execution of functional programs or above
desirable relations (such as type preservation) between the static and
the dynamic semantics of a programming language.
% We now describe how we can move PBT into 
% the setting of metaprogramming.  For example, we would like 
% to find counterexamples to claims that might be made about the
% evaluation of a certain functional program or about the type
% preservation of a  programming language.
%
The main difficulty in treating entities such as programs as data
structures within a (meta) programming settings is the treatment of
bound variables.
There have been many approaches to the treatment of term-level
bindings within symbolic systems: they include 
nameless dummies~\cite{debruijn72},
higher-order abstract syntax (HOAS)~\cite{pfenning88pldi},
nominal logic~\cite{pitts03ic}, 
parametric HOAS~\cite{chlipala08icfp}, and 
locally nameless~\cite{chargueraud11jar}.
The approach used in \lP, called \emph{$\lambda$-tree
  syntax}~\cite{miller18jar}, is based on the movement of binders from
term-level abstractions to formula-level abstractions (\ie
quantifiers) to proof-level abstract variables (called
\emph{eigenvariables} in~\cite{gentzen35}).
This approach to bindings is probably the oldest one, since it appears
naturally when organizing Church's Simple Theory of
Types~\cite{church40} within Gentzen's sequent
calculus~\cite{gentzen35}.
As we illustrate in this section, the $\lambda$-tree syntax
approach to bindings allows us to lift PBT to the meta-programming
setting in a simple and modular manner.
In what follows, we assume that the reader has a passing
understanding of how $\lambda$-tree syntax is supported in frameworks
such as \lP or Twelf~\cite{pfenning99cade}. 

The treatment of bindings in \lP is intimately related to
including into $G$-formulas  universal quantification and implications.
While we restricted \spec in the previous two sections to
Horn clauses, we now allow the full set of $D$ and $G$-formulas that
were defined in Section~\ref{sec:spec}.
To that end, we now replace the interpreter code given in
Figure~\ref{fig:interp} with the specification in
Figure~\ref{fig:hh-interp}.
Here, the goal \lsti{interp Ctx G} is intended to capture the fact
that \lsti{G} follows (in \spec) from the union of the atomic formulas
in \lsti{Ctx} and the logic programs defined by the \lsti{<>==}
predicate.

\begin{figure}
\lstinputlisting[linerange={new-end}]{code/sect6.sig}
\lstinputlisting[linerange={interp-end}]{code/sect6.mod}
\caption{A re-implementation of the interpreter  in
  Figure~\ref{fig:interp} that treats implications and universal
  quantifiers in $G$-formulas.}
\label{fig:hh-interp}
% \end{figure}
% \begin{figure}[h]
\lstinputlisting[linerange={check-end}]{code/sect6.sig}
\lstinputlisting[linerange={check-end}]{code/sect6.mod}
\caption{A re-implementation of the FPC checker in
  Figure~\ref{fig:kernel} that treats implications and universal
  quantifiers in $G$-formulas.}
\label{fig:hh-kernel}
% \end{figure}
% \begin{figure}[h]
\begin{lstlisting}[basicstyle=\small\ttfamily]
initE (max empty).
allC  (max C) (x\ max C).
impC  (max C) (max C)
initE (sze In In') :- In > 0, In' is In - 1.
allC  (sze In Out) (x\ sze In Out).
impC  (sze In Out) (sze In Out).
\end{lstlisting}
\caption{Additional clauses for two FPCs.}
\label{fig:additional}
\end{figure}

Similar to the extensions made to \lsti{interp}, we need to extend
the notion of FPC and the \lsti{check} program: this is 
given in Figure~\ref{fig:hh-kernel}.
Three new predicates---\lsti{initE}, \lsti{impC}, and
\lsti{allC}---have been added to FPCs.  Using the terminology of~\cite{chihani17jar}, the last two of these predicates are referred to
as \emph{clerks} instead of \emph{experts}.  This distinction arises
from the fact that no essential information is extracted from a
certificate by these predicates, whereas experts often need to make
such extractions.
In order to use a previously defined FPC in this setting, we simply
need to provide the definition of these three definitions for the
constructors used in that FPC\@.  For example, the \lsti{max} and
\lsti{sze} FPCs (see Section~\ref{ssec:fpc}) are accommodated by the
additional clauses listed in Figure~\ref{fig:additional}.

\newcommand{\squarenip}[7][]{
\draw(#2-0.3,#3+0.75) node{#4};
        \draw[-{Stealth[slant=0]}] (#2,#3+1.5) -- (#2,#3);
        \draw(#2+0.75,#3+1.8) node{#5};
        \draw[-{Stealth[slant=0]}] (#2,#3+1.5) -- (#2+1.5,#3+1.5);
        \draw(#2+0.75, #3-0.3) node{#6};
        \draw[-{Stealth[slant=0]}, style = dashed] (#2,#3) -- (#2+1.5,#3);
        \draw(#2+1.8,#3+0.75) node{#7};
        \draw[-{Stealth[slant=0]}, style = dashed] (#2+1.5,#3+1.5) -- (#2+1.5,#3);
        \draw(#2+0.75, #3-0.75) node{#1};
      }

To showcase the ease with which we handle searching for
counterexamples in binding signatures, we go back in history and
explore a tiny bit of the classical theory of the $\lambda$-calculus,
namely the Church-Rosser theorem and related notions. We recall two
basic definitions, for a binary relation $R$ and its Kleene closure
$R^*$: 
%DM%
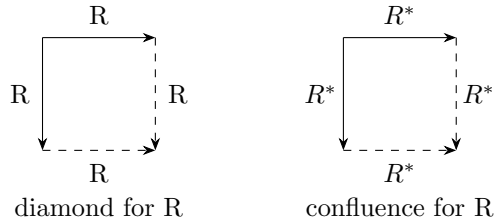
   \begin{center}
     \begin{tikzpicture}
                       \squarenip[diamond for R]{0}{0}{R}{R}{R}{R}
        \squarenip[confluence for R ]{4}{0}{$R^*$}{$R^*$}{$R^*$}{$R^*$}
    \end{tikzpicture}
\end{center}

Given the syntax of the untyped $\lambda$-calculus:
\[
\begin{array}{llcl}
\mbox{Terms} & M & \bnfas &  x \mid \lambda {x}.\ {M} \mid {M_1}\ {M_2}
\end{array}
\]
in Fig.~\ref{fig:beta} we display the standard rules for beta
reduction, consisting of the beta rule itself augmented by
congruences.

Fig.~\ref{fig:lp-beta} displays the $\lambda$-tree encoding of 
the term structure of the untyped $\lambda$-calculus as well as the
specification of the inference rules in Fig.~\ref{fig:beta}.
As it is now a staple of
\lP and similar systems,  we only
note  how in the encoding of the beta rule, substitution is
realized via meta-level application; further, in the $\mathtt{B}-\xi$
rule we descend into an abstraction via \spec-level universal
quantification.  The clause for generating/checking abstractions
features the combination of hypothetical and parametric  judgments.

\begin{figure}[h]
  %\begin{small}
    \[
\begin{array}{c}
  \infer[\mathtt{B-\beta}]{(\lambda x.~M)~N \step [x \mapsto N] M}{}
    \qquad
  \infer[\mathtt{B}-\xi]{\lambda x.M \step \lambda x.M'}{M \step M'}
\medskip\\
  \infer[\mathtt{B-APP1}]{M_1~M_2 \step M_1'~M_2}{M_1 \step M_1'}
  \qquad 
  \infer[\mathtt{B-APP2}]{M_1~M _2\step M_1~M_2'}{M_2 \step M_2'}\\
    \dotfill
  \medskip\\
  \infer[]{\Gamma\vdash x : \mathtt{exp}}{x\in \Gamma}
  \qquad
  \infer[]{\Gamma\vdash M~M' : \mathtt{exp}}{\Gamma\vdash M : \mathtt{exp}\quad
  \Gamma\vdash M' : \mathtt{exp}}
  \qquad
    \infer[]{\Gamma\vdash \lambda x.M : \mathtt{exp}}{\Gamma,x\vdash M : \mathtt{exp}}\\
%\dotfill
\end{array}
\]
%\end{small}
\caption{Specifications of beta reduction and well-formed terms.}
\label{fig:beta}
\lstinputlisting[linerange={lambdas-end},basicstyle=\small\ttfamily]{code/sect6.sig}
\lstinputlisting[linerange={beta-end},basicstyle=\small\ttfamily]{code/sect6.sig}
\lstinputlisting[linerange={beta-end},basicstyle=\small\ttfamily]{code/sect6.mod}
\caption{The \lP specification of the inference rules in Figure~\ref{fig:beta}.}
\label{fig:lp-beta}
\end{figure}

When proving the \emph{confluence} of a (binary) reduction relation, a
key stepping stone is the \emph{diamond property}. % : if
% $M_1 \leftarrow M\rightarrow M_2$, then
% $\exists N, M_1 \rightarrow N\leftarrow M_2$.
In fact, diamond implies
confluence. It is a well-known fact, however, that beta reduction does
\emph{not} satisfy the diamond property, since redexes can be
discarded or duplicated and this is why notions such as \emph{parallel}
reduction have been developed~\cite{Takahashi95}.

To find a counterexample to the claim that beta reduction 
implies the diamond property, we write the following predicates, which abstract
over a binary reduction relation. 

%% DM In fact, we don't need the existential explicitly in joinableS.
%% It's useful to have since its witness will appear in the max
%% certificate.  But from a logical point-of-view, it's not needed. 

%% we need a little care in stating the
%% diamond property, since its consequent is an existential that we
%% cannot readily negate via NAF. We simply hide it within a new
%% predicate (\texttt{joinable}), where we also abstract over the
%% reduction relation and build a harness predicate to try and refute it:

%% \begin{metanote}
%% DM I wonder if we can get rid of the explicit existential for joinable
%% and simplify the discussion about NAF above.  Also, we should
%% \lsti{cex} as a prefix (not suffix) on predicate names used for
%% finding counterexamples.  Mention this convention to the reader.
%% AM: i suggest the prefix prop
%% \end{metanote}

% (joinableS Step M1 M2) <>==   (M1 eq M2) or  some P\ (Step M1 P) and (Step M2 P).

\lstinputlisting[linerange={joinable-end},basicstyle=\small\ttfamily]{code/sect6.sig}
\lstinputlisting[linerange={joinable-end},basicstyle=\small\ttfamily]{code/sect6.mod}

%% \begin{lstlisting}[basicstyle=\small\ttfamily][basicstyle=\small\ttfamily]
%% type joinableS    (exp -> exp -> sl) -> exp -> exp -> sl.
%% type dia_cexS    cert -> (exp -> exp -> sl) -> exp -> o.

%% joinableS Step M M  <>== tt.
%% joinableS Step M1 M2  <>== some P\ (Step M1 P) and (Step M2 P).

%% dia_cexS Cert Step M :-
%%     check Cert nil (is_exp M),
%%     interp nil (Step M M1),
%%     interp nil (Step M M2),
%%     not(interp nil (joinableS Step M1 M2)).
%% \end{lstlisting}

Note that the negation-as-failure call is safe since, when the last
goal is called, 
all variables in it will be bound to closed terms.
A minimal counterexample found by exhaustive generation is:
\begin{lstlisting}[basicstyle=\small\ttfamily][basicstyle=\small\ttfamily]
M = app (lam x\ app x x) (app (lam x\ x) (lam x\ x))
\end{lstlisting}
or, using the identity combinators, the term $(\lambda x.\ x\ x)
(I\ I)$, which beta reduces to $ (I\ I) (I\ I)$ and $ (I\ I)$.

% rewritten by bard
It is worth keeping in mind  that the property holds true had we
defined \texttt{Step} to be the reflexive-transitive closure of beta
reduction, or other relations of interest, such as parallel reduction
or complete developments. The code in the repository provides
implementations of these relations. As expected, such queries do not
report any counterexample, up to a reasonable bound.

% Of
% course, the property would not be falsified had we taken \texttt{Step} to be
% the reflexive-transitive closure of beta reduction, or, for that
% matter, other relations such as parallel reduction and complete
% developments---see the code in the  repository for their
% implementation.

Let us dive further by looking at $\eta$-reduction in a typed setting.
Again, it is well-know (see, e.g.~\cite{Selinger08}) that the diamond
property fails for  $\beta\eta$-reduction for the simply-typed
$\lambda$-calculus, once we add unit and pairs: the main culprit is the
$\eta$ rule for unit, which licenses any term of type unit to
$\eta$-reduce to the empty pair. Verifying the existence of such
counterexamples requires building-in typing obligations in the
reduction semantics, following the style of~\cite{Goguen95}.  In fact,
 it is not enough for the generation phase
to yield only well-typed terms, lest we meet false positives.

Since a counterexample manifests itself considering only $\eta$ and
unit,
% \ednote{In fact, if we put also beta, we keep finding cex to beta,
%   which is a known problem, see Hughes' tool \emph{MoreBugs} that it
%   would nice to address.}
we list in Fig.~\ref{fig:tyred} the $\eta$
rules restricted to arrow and unit; see Fig.~\ref{fig:lp-tyred} for
their implementation.

\begin{figure}[t]
  \centering
\[
\begin{array}{c}
%\multicolumn{1}{l}{\mbox{Type-directed $\eta$-reduction}: \fbox{$\Gamma
%  \vdash M \red N : A$}}\\[0.9em]
\infer[\betar]{\Gamma \vdash \lambda x.(M~x)  \red M : A\arrow B }
    {\Gamma \vdash M : A \arrow B & x \not\in \mbox{FV(M)}}
%\infer[\eta]{\Gamma \vdash M \red \lambda x{:}A.M~x : A \arrow B}{
% M \not= \lambda y{:}A.M'}
\qquad % \\[0.5em]
    \infer[\epr]{\Gamma\vdash M \red \ep:\unit }{\Gamma\vdash M :\unit }
\\[0.7em]
\infer[\kern-2pt\appl]{\Gamma \vdash M\,N \red M'\,N : B}{{\Gamma \vdash M \red M' : A \arrow B} & {\Gamma \vdash N : A}}
\quad
\infer[\kern-2pt\appr]{\Gamma \vdash M\,N \red M\,N' : B}{{\Gamma \vdash M : A \arrow B} & \Gamma \vdash N \red N' : A}
\\[0.5em]
\infer[\rabs]{\Gamma \vdash \lambda x.M \red \lambda x.M' : A \arrow B}{\Gamma, x{:}A \vdash M \red M' : B}
\end{array}
\]
% \end{small}
  \caption{Type-directed $\eta$ reduction}
  \label{fig:tyred}
\lstinputlisting[linerange={wt-end},basicstyle=\small\ttfamily]{code/sect6.sig}
\lstinputlisting[linerange={wt-end},basicstyle=\small\ttfamily]{code/sect6.mod}
\caption{The \lP specification of the inference rules in Figure~\ref{fig:tyred}.}
\label{fig:lp-tyred}
\end{figure}

%% \begin{lstlisting}[basicstyle=\small\ttfamily][basicstyle=\small\ttfamily]
%% type  wt           exp -> ty -> sl.
%% type  teta         exp -> exp -> ty ->  sl.

%% wt ep unitTy          <>== tt.
%% wt (lam M) (arTy A B) <>== all x\ wt x A =o wt (M x) B.
%% wt (app M N) B        <>== wt M (arTy A B) and wt N A.

%% teta (lam x\ (app M x)) M (arTy A B) <>== wt M (arTy A B).
%% teta M ep unitTy                     <>== (wt M unitTy).
%% % cong rules omitted
%% \end{lstlisting}

A first order of business is to ensure that the typing annotations that we
have added  to the reduction semantics are consistent with the
standard typing rules. In other words, we need to verify that eta
reduction preserves typing. The encoding of the property
\[
\Gamma\vdash M \red M' : A \Longrightarrow \Gamma\vdash M  : A \land \Gamma\vdash  M' : A
\]
is the following and does not report any problem.
\begin{lstlisting}[basicstyle=\small\ttfamily]
wt_pres M M' A <>== (wt M A) and (wt M' A).

prop_eta_pres Gen M M' A:-
  check Gen nil (is_exp M),
  interp nil (teta M M' A),
  not (interp nil (wt_pres M M' A)).
\end{lstlisting}
However, suppose we made a small mistake in the rules in Fig.~\ref{fig:tyred}, say forget a typing assumption in a congruence rule:
\[
\infer[\appl-\mathtt{BUG}]{\Gamma \vdash M\,N \red M'\,N : B}{{\Gamma \vdash M \red M' : A \arrow B}}
\]
then type preservation is refuted with the following counterexample.
\begin{lstlisting}[basicstyle=\small\ttfamily][basicstyle=\small\ttfamily]
A = unitTy
N = app (lam (x\ unit)) (lam (x\ x))
M = app (lam (x\ x))  (lam (x\ x))
\end{lstlisting}
A failed attempt of an inductive proof of this property in a proof
assistant would have eventually pointed to the missing assumption, but
testing is certainly a faster way to discover this mistake. 
\smallskip

% CONTINUE HERE
% AM: continued
We can now refute the diamond property for $\eta$.  The harness is the obvious extension of the previous definitions, where to foster  better coverage we only generate well-typed terms:
\lstinputlisting[linerange={dia_teta-end},basicstyle=\small\ttfamily]{code/sect6.mod}

% \begin{lstlisting}[basicstyle=\small\ttfamily][basicstyle=\ttfamily]
% prop_eta_dia Cert  M A :-
%     check Cert nil (wt M A),
%     interp nil (teta M M1 A),
%     interp nil (teta M M2 A),
%     not(interp nil (joinable_teta M1 M2 A)).
% \end{lstlisting}
One counterexample found by exhaustive generation is 
\lsti{lam x\ lam y\ app x y}, which, at type 
\lsti{(unitTy -> unitTy) -> unitTy -> unitTy}, reduces to 
\lsti{lam x\ x} by the $\eta$ rule and
\lsti{lam x\ lam y\ unit} by $\eta-\unit$.

\section{Linear logic as the specification logic}
\label{sec:linear}

One of linear logic's early promises
% One of the earliest promises of linear logic
was that it could help in
specifying computational systems with side-effects, exceptions, and
concurrency~\cite{girard87tcs}.  In support of that
promise, an early application of linear logic was to enhance big-step
operational semantic specifications~\cite{kahn87stacs} for programming
languages that incorporated such features: see, for example,~\cite{andreoli90iclp,Hodas1994,chirimar95phd,miller96tcs,pfenning00ic}.  In this section, we
adapt the work in~\cite{MantovaniM21} to show how PBT can be applied
in the setting where the specification logic is  a fragment of linear logic.

\subsection{\spec as a subset of linear logic}

We extend the definition of \spec given in Section~\ref{sec:spec}
to involve the following %extended
 grammar for $D$ and $G$-formulas.
\[
\begin{array}[t]{rl}
  D~\bnfas& G \limp A \sep \forall x:\tau. D\\
  G~\bnfas& A \sep \true \sep G_1\vee G_2\sep G_1\wedge G_2\sep\exists x:\tau. G
                    \sep\forall x:\tau. G\sep A\imp G\\
          & \kern 8pt\sep A\limp G \sep \bang G
\end{array}
\]
That is, we allow $G$-formulas to be formed using the linear
implications $\limp$ (with atomic antecedents) and the exponential
$\bang$.  As the reader might be aware, linear logic has two
conjunctions ($\with$ and $\otimes$) and two disjunctions ($\parr$ and
$\oplus$).  When we view $G$-formulas in terms of linear logic, we
identify $\vee$ as $\oplus$ and $\wedge$ as $\otimes$ (and $\true$ as
the unit of $\otimes$).  Note that we have also changed the top-level
implication for $D$-formulas into a linear implication: 
this change is actually a refinement in the sense that $\limp$ is
a more precise form of the top-level implications of $D$-formulas.
%
% DM I revised my text here by simplifying it.  I don't think we
% should go into detail.  Hopefully, the reader will accept this
% aspect of the proof theory of linear logic.
%
% distinction between $\limp$ and $\imp$ in this position is not
% apparent\ednote{AM: unclear} until genuine linear logic features have
% been added to $G$-formulas.

\begin{figure}
\[
  \infer{\triple{\Gamma}{\cdot}{\true}}{}
  \qquad
  \infer{\triple{\Gamma}{\Delta_1,\Delta_2}{G_1\wedge G_2}}
        {\triple{\Gamma}{\Delta_1}{G_1} & \triple{\Gamma}{\Delta_2}{G_2}}
  \qquad
  \infer[i\in\{1,2\}]{\triple{\Gamma}{\Delta}{G_1\vee G_2}}{\triple{\Gamma}{\Delta}{G_i}}
\]
\[
  \infer{\triple{\Gamma}{\cdot}{\bang G}}{\triple{\Gamma}{\cdot}{G}}
  \qquad
  \infer[(3)]{\triple{\Gamma}{\Delta}{\forall x:\tau. G}}
             {\triple{\Gamma}{\Delta}{G[y/x]}}
  \qquad
  \infer[(1)]{\triple{\Gamma}{\Delta}{\exists x:\tau. G}}
             {\triple{\Gamma}{\Delta}{G[t/x]}}
\]
\[
  \infer{\triple{\Gamma}{\Delta}{A\imp G}}{\triple{\Gamma,A}{\Delta}{G}}
  \qquad
  \infer{\triple{\Gamma}{\Delta}{A\limp G}}{\triple{\Gamma}{\Delta,A}{G}}
\]
\[
  \infer{\triple{\Gamma}{A}{A}}{}
  \qquad
  \infer{\triple{\Gamma,A}{\cdot}{A}}{}
  \qquad
  \infer[(2)]{\triple{\Gamma}{\Delta}{A}}{\triple{\Gamma}{\Delta}{G}}
\]
\begin{flushleft}
The three provisos are the standard ones.  The first
two are repeated from Figure~\ref{fig:augmented}.
\end{flushleft}
\begin{enumerate}[label=(\arabic*)]
  \item The term $t$ is of type $\tau$.
  \item There is a program clause $\forall \bar x (G' \limp
    A')\in\Pscr$ and a substitution for the variables $\bar x$ such
    that $A$ is $A'\theta$ and $G$ is $G'\theta$.
  \item The eigenvariable $y$ is not free in the formulas in the
    concluding sequent.
\end{enumerate}
\caption{A sequent calculus proof system for our linear \spec.}
\label{fig:linear sl}
\end{figure}

A proof system for this specification logic is given in
Figure~\ref{fig:linear sl}: here, $\Pscr$ is a set of closed
$D$-formulas. The sequent $\triple{\Gamma}{\Delta}{G}$ has a left-hand
side that is divided into two zones% , each of which are multisets of
% atomic formulas
.  The $\Gamma$ zone is the \emph{unbounded} zone,
meaning that the atomic assumptions that it contains can be used \emph{any
number} of times in building this proof; it can be seen as a set.  The $\Delta$ zone is the
\emph{bounded} zone, meaning that its atomic assumptions %that it contains
 must
be used \emph{exactly once} in building this proof; it can be seen as a multiset.
%
% rewritten by LLM
To ensure the accurate accounting of formulas in the bounded zone, the
zone must satisfy three conditions: it must be empty in certain rules
(the empty zone is denoted by $\cdot$), it must contain exactly one
formula (as in the two initial rules displayed in the last row of
inference rules), and it must be split when handling a conjunctive
goal.
% In order to support this strict accounting of formulas in the bounded
% zone, that zone must be empty in certain rules (the $\cdot$ is used to
% denote the empty zone), it must be the multiset containing exactly one
% formula (as in one of the two initial rules displayed in the last row
% of inference rules), and it must be split when dealing with a conjunctive
% goal.
Also note that (when reading inference rules from conclusion to
premises) a goal of the form $A\imp G$ places its assumption $A$ in
the unbounded zone and a goal of the form $A\limp G$ places its
assumption $A$ in the bounded zone.  Finally, the goal $\bang G$ can
only be proved if the bounded zone is empty: this is the
\emph{promotion rule} of linear logic.

\begin{figure}
\[
  \infer{\io{\XXi}{\Delta_I}{\Delta_I}{}{\true}}{}
  \qquad
  \infer{\io{\XXi}{\Delta_I}{\Delta_O}{}{G_1 \wedge G_2}}
        {\io{\XXi_1}{\Delta_I}{\Delta_M}{}{G_1} &
         \io{\XXi_2}{\Delta_M}{\Delta_O}{}{G_2}}
\]\[
  \infer{\io{\XXi}{\Delta_I}{\Delta_O}{}{A \imp G}}
        {\io{\XXi'}{\Delta_I, \bang A}{\Delta_O,\bang A}{}{G}}
  \qquad
  \infer{\io{\XXi}{\Delta_I}{\Delta_O}{}{A \limp G}}
        {\io{\XXi'}{\Delta_I, A}{\Delta_O,\square}{}{G}}
  \qquad
  \infer{\io{\XXi}{\Delta_I}{\Delta_I}{}{\bang G}}
        {\io{\XXi'}{\Delta_I}{\Delta_I}{}G}
\]\[
  \infer[i\in\{1,2\}]{\io{\XXi}{\Delta_I}{\Delta_O}{}{G_1\vee G_2}}
                     {\io{\XXi_1}{\Delta_I}{\Delta_O}{} {G_i}}
  \qquad
  \infer[(1)]{\io{\XXi}{\Delta_I}{\Delta_O}{}{\forall x:\tau.G}}
             {\io{\XXi_1}{\Delta_I}{\Delta_O}{}{G[y/x]}}
  \qquad
  \infer[(2)]{\io{\XXi}{\Delta_I}{\Delta_O}{}{\exists x:\tau.G}}
             {\io{\XXi_1}{\Delta_I}{\Delta_O}{}{G[t/x]}}
\]\[
  \infer{\io{\XXi}{\Delta_I, A,\Delta'_I}{\Delta_I,\square,\Delta'_I}{}A}{}
  \qquad
  \infer{\io{\XXi}{\Delta_I,\bang A,\Delta'_I}
                  {\Delta_I, \bang A,\Delta'_I}{}{A}}{}
  \qquad
  \infer[(3)]{\io{\XXi}{\Delta_I}{\Delta_O}{}A}
             {\io{\XXi'}{\Delta_I}{\Delta_O}{}G}
\]
\begin{flushleft}
The three proviso (1), (2), and (3) are the same as in
Figure~\ref{fig:linear sl}. 
\end{flushleft}

\caption{The I/O proof system% alternative to the proof system inFigure~\ref{fig:linear sl}
  .}
\label{fig:io}
\end{figure}

The inference rule for $\wedge$
% (often written as $\otimes$ in linear logic)
% 
can be expensive to realize in a proof search setting, since,
when we read inference rules from conclusion to premises,
% the rule for $\wedge$
it requires \emph{splitting} the bounded zone into two multisets
before proceeding with the proof.  Unfortunately, at the time that
this split is made, it might not be clear which atoms in the bounded
zone will be needed to prove the left premise and which are needed to
prove the right premise.  If the bounded zone contains $n$ distinct
items, there are $2^n$ possible ways to make such a split: thus,
considering all splittings is far from desirable.  Figure~\ref{fig:io}
presents a different proof system, known as the I/O
system~\cite{Hodas1994}, that is organized around making this split in
a \emph{lazy} fashion.  Here, the sequents are of the form
$\io{}{\Delta_I}{\Delta_O}{}{G}$ where $\Delta_I$ and $\Delta_O$ are
\emph{lists} of items that are of the form $\square$, $A$, and
$\bang A$ (where $A$ is an atomic formula).

The idea behind proving the sequent $\io{}{\Delta_I}{\Delta_O}{}{G}$
is that all the formulas in $\Delta_I$ are \emph{input} to the proof
search process for finding a proof of $G$: in that process, atoms in
$\Delta_I$ that are not marked by a $\bang$ and that are used in
building that proof are then deleted (by replacing them with
$\square$).  That proof search method outputs $\Delta_O$ as a result of such a deletion.
% Such a deletion results in $\Delta_O$ that is output by
% that proof search process.
Thus, the process of proving
$\io{}{\Delta_I}{\Delta_O}{}{G_1\otimes G_2}$ involves sending the
full context $\Delta_I$ into the process of finding a proof of $G_1$,
which  returns the output context $\Delta_M$, which is then made
into the input for the process of finding a proof of $G_2$.  The
correctness and completeness of this alternative proof system follows
directly from results in~\cite{Hodas1994}.  A \lP specification of
this proof system appears in Figure~\ref{fig:llinterp}.  In that
specification, the input and output contexts are represented by a list
of \emph{option-\spec-formulas}, which are of three kinds: \lsti{del}
to denote $\square$, \lsti{(ubnd A)} to denote $\bang A$, and
\lsti{(bnd A)} to denote simply $A$.

Note that if we use this interpreter on the version of \spec described
in Section~\ref{sec:spec} (\ie without occurrences of $\limp$ and
$\bang$ within $G$ formulas), then the first two arguments of
\lsti{llinterp} are  always the same.  If we further restrict
ourselves to having only Horn clauses (\ie they have no occurrences of
implication in $G$ formulas), then those first two arguments of
\lsti{llinterp} are both \lsti{nil}.  Given these observations, the
interpreters in Figures~\ref{fig:interp} and~\ref{fig:hh-interp} can
be derived directly from the one given in Figure~\ref{fig:llinterp}.

\begin{figure}
\lstinputlisting[linerange={llinterp-end},basicstyle=\small\ttfamily]{code/sect7.sig}
\lstinputlisting[linerange={llinterp-end},basicstyle=\small\ttfamily]{code/sect7.mod}
\caption{An interpreter based on the proof system in Figure~\ref{fig:io}.}
\label{fig:llinterp}
\end{figure}

It is a simple matter to modify the interpreter in Figure~\ref{fig:llinterp}
in order to get the corresponding \lsti{llcheck} predicate that works
with proof certificates.  In the process of making that change, we
need to add two new predicates: the clerk predicate \lsti{limpC} to
treat the linear implication and the expert predicate \lsti{bangE} to
treat the $\bang$ operator.  In order to save space, we do not display
the clauses for the \lsti{llcheck} predicate.

%% DM I figure we can just remove the text, as mentioned in the rebuttal.
%% \begin{cancel}
%% A simple linear logic program is the one that turns a switch on and
%% off.
%% \lstinputlisting[linerange={toggle-end},basicstyle=\small\ttfamily]{code/sect7.sig}
%% \lstinputlisting[linerange={toggle-end},basicstyle=\small\ttfamily]{code/sect7.mod}
%%   When attempting to prove \lsti{toggle G} when the bounded zone
%% is $\Delta,\lsti{on}$ should reduce to attempting to prove \lsti{G}
%% where the bounded zone is $(\Delta,\lsti{off})$ (conversely, when the
%% roles of \lsti{on} and \lsti{off} are switched).  This operational
%% reading of these clauses is supported by the following inference rules
%% (following the rules in Figure~\ref{fig:linear sl}).
%% \[
%%   \infer{\triple{\Gamma}{(\Delta,\lsti{on})}{\lsti{toggle G}}}{
%%   \infer{\triple{\Gamma}{(\Delta,\lsti{on})}{\lsti{on}\wedge(\lsti{off}\limp\lsti{G})}}
%%         {\infer{\triple{\Gamma}{\lsti{on}}{\lsti{on}}}{} &
%%          \infer{\triple{\Gamma}{\Delta}{\lsti{off}\limp\lsti{G}}}
%%                {\triple{\Gamma}{(\Delta,\lsti{off})}{\lsti{G}}}}}
%% \]

%% \end{cancel}
To exemplify derivations and refutations with linear logic programming, consider the following
specification of the predicate that relates two lists if and only if
they are permutations of each other.
\lstinputlisting[linerange={perm-end},basicstyle=\small\ttfamily]{code/sect7.sig}
\lstinputlisting[linerange={perm-end},basicstyle=\small\ttfamily]{code/sect7.mod}
As the reader can verify, the goal
\begin{lstlisting}[basicstyle=\small\ttfamily]
?- llinterp nil nil (perm [1,2,3] K).
\end{lstlisting}
will produce six solutions for the list \lsti{K} and they will all be
permutations of \lsti{[1,2,3]}.
%% AM: no further discussion of why we bang
%
% \todo{Modify discussion to account for bang(load)}
% More generally, however, a call to
% the \lsti{perm} predicate will occur in settings where the input and
% output contexts are not necessarily empty, for example, in the query
% \begin{lstlisting}[basicstyle=\small\ttfamily]
% ?- llinterp In Out ((perm [1,2,3] K) and Goal).
% \end{lstlisting}
% where \lsti{Out} might have some entries in \lsti{In} marked as
% deleted and where \lsti{Goal} is some goal.
% In order to ensure that the permutation specification works properly
% in such a situation, we should invoke the following goal instead:
% \begin{lstlisting}[basicstyle=\small\ttfamily]
% ?- llinterp In Out ((bang (perm [1,2,3] K)) and Goal).
% \end{lstlisting}
% This invocation will ensure that all the entries that are in the
% bounded zone are passed to the process of building a proof of \lsti{Goal}.

A property that should hold of any definition of permutation is that the latter preserves list membership:

\begin{lstlisting}[basicstyle=\small\ttfamily]
perm_pres Cert PermDef L K :-
  llcheck Cert nil nil (nlist L),  
  member X L,
  llinterp nil nil (PermDef L K),
  not (member X K).
\end{lstlisting}
%% DM The above might be too cryptic but a simple rewrite should fix
%% it. 
Note the interplay of the various levels in this property:
\begin{enumerate}
\item We generate (ephemeral) lists of natural numbers using the
  interpreter for linear logic (\lsti{llcheck}): however, the
  specification for \lsti{nlist} makes use of only intuitionistic
  connectives.
\item The \lsti{member} predicate is the standard \lP predicate.
\item The test expects a definition of permutation to be interpreted linearly.
\end{enumerate}
Suppose now we made a mistake in the definition of \lsti{load}
confusing linear with intuitionistic implication: 
\begin{lstlisting}[basicstyle=\small\ttfamily]
load'(X::L) K <>== element X =o load' L K. 
                             % bug here, everything else as before
\end{lstlisting}
A call to the checker with goal \lsti{perm_pres (height 2) (perm' L K)} would separate the good from the bad implementation with counterexample:
\begin{lstlisting}[basicstyle=\small\ttfamily]
K = nil
L = z :: nil
\end{lstlisting}

% \begin{metanote}

%% \item shoud we motivate not using additive conjunction? DM I don't
%% think we need to do that.

%% \item should we remove \texttt{tl} from the interpreter (in the ../code)?

%% \item the definition of \verb|is_lexp| should be extended to
%%   \texttt{set, get} etc.

% \item Take some of the motivation from LOPSTR: we do linear PBT
  %    because we have no ``linear'' proof assistants DM I'm not sure
  %    of this motivation.  We have argued that we want to have SL
  %    that are linear but we have no motivation (here) for a linear
  %    RL (which is where a linear proof assistant would be
  %    situationed.

%\end{metanote}
\subsection{The operational semantics of $\lambda$-terms with a counter}
\label{sec:counter}

%% AM: commented out as I merged it with initial paragraph

% It has been known since the 1990s that logic programming based on
% linear logic could be used to capture various notions of
% side-effects~\cite{andreoli90iclp,Hodas1994}.  For example,
% specifications based on big step operational semantic can be extended
% using linear logic to capture notions such as reading and writing to
% memory locations and the creation and communications between
% computation threads~\cite{miller96tcs}.

Figure~\ref{fig:cbnv} contains the \spec specification of
call-by-value and call-by-name big-step operational semantics for a
version of the $\lambda$-calculus to which a single memory location (a
counter) is added%
\footnote{This specification can easily be generalized to finite
registers or to a specification of 
references in functional programming languages~\cite{chirimar95phd,miller96tcs}.}.
% such as in Standard ML~\cite{CervesatoP02}.}. 
% DM Since the reference to Cervesato and Pfenning is to a paper about
%    dependently typed lambda-calculus (and not logic, per sa), I thought
%    it better to reference directly papers on linear logic and
%    references. 
The untyped $\lambda$-calculus of Section~\ref{sec:mmm}, with its two
constructors \lsti{app} and \lsti{lam}, is extended with the
additional four constants.
\lstinputlisting[linerange={newcons-end},basicstyle=\small\ttfamily]{code/sect7.sig}
The specification in Figure~\ref{fig:cbnv} uses \emph{continuations}
to provide for a sequencing of operations.  A continuation is an \spec
goal formula that should be called once the program getting evaluated
is completed.  For example, the attempt to prove the goal
\lsti{cbn M V K} when the bounded zone is the multiset containing only the
formula $\counter{C}$ (for some integer $C$) reduces to an attempt to
prove the goal \lsti{K} with the bounded zone consisting of just
$\counter{D}$ (for some other integer $D$), provided \lsti{V} is the call-by-name value of \lsti{M}.
This situation can be represented as the following (open) derivation
(following the rules in Figure~\ref{fig:linear sl}).
\[
\infer{\triple{\strut\Pscr}{\counter{C}}{\lsti{cbn M V K}}}{
\infer{\strut\vdots}{\triple{\Pscr}{\counter{D}}{\lsti{K}}}}
\]
Such a goal reduction can be captured in \lP using the following
higher-order quantified expression
\begin{lstlisting}[basicstyle=\small\ttfamily]
  ?- (pi k\ (pi I\ pi O\ llinterp I O k) =>
             (llinterp [bnd(counter 0)] _ (cbn M V k)).
\end{lstlisting}
Operationally, \lP introduces a new eigenvariable (essentially a
local constant) \lsti{k} of type \lsti{sl} and assumes that this new
\spec formula can be proved no matter the values of the input and output
contexts.  Once this assumption is made, the linear logic interpreter
is called with the counter given the initial value of 0 and with
the \lsti{cbn} evaluator asked to compute the call-by-name value of
\lsti{M} to be \lsti{V} and with the final continuation being
\lsti{k}.  This hypothetical reasoning can be captured by the
following predicate.
\lstinputlisting[linerange={eval-end},basicstyle=\small\ttfamily]{code/sect7.sig}
\lstinputlisting[linerange={eval-end},basicstyle=\small\ttfamily]{code/sect7.mod}

\begin{figure}
\lstinputlisting[linerange={cbnv-end},basicstyle=\small\ttfamily]{code/sect7.sig}
\lstinputlisting[linerange={cbnv-end},basicstyle=\small\ttfamily]{code/sect7.mod}
\caption{Specifications of call-by-name (\lsti{cbn}) and call-by-value
  (\lsti{cbv}) evaluations.}
\label{fig:cbnv}
\end{figure}

It is well known that if the call-by-name and call-by-value strategies
terminate when evaluating a \emph{pure} untyped $\lambda$-term (those
without side-effects such as our counter), then those two strategies
yield the same value.  One might conjecture that this is also true
once counters are added.  To probe that conjecture, we write the
following logic program:
% \lstinputlisting[linerange={cex_cbnv-end},basicstyle=\small\ttfamily]{code/sect7.sig}
\lstinputlisting[linerange={cex_cbnv-end},basicstyle=\small\ttfamily]{code/sect7.mod}
The query \lsti{?- prop_cbnv (height 3) M V U} returns the smallest
counterexample to the claim that call-by-name and call-by-value
produce the same values in this setting.  In particular, this query
instantiates \lsti{M} with \lsti{app (lam (w\ get)) (set (- 1))}: this
expression has the call-by-name value of 0 while the it has a
call-by-value value of $-1$, given the generator \lsti{is_prog} in
Fig.~\ref{fig:cbnv}.

\subsection{Queries over linear $\lambda$-expressions}
\label{sec:lingen}

A slight variation to  \lsti{is_exp}
(Figure~\ref{fig:lp-beta}) yields the following \spec specification
that succeeds with a $\lambda$-term only when the bindings are used
linearly.

\lstinputlisting[linerange={is_lexp-end},basicstyle=\small\ttfamily]{code/sect7.sig}
\lstinputlisting[linerange={is_lexp-end},basicstyle=\small\ttfamily]{code/sect7.mod}

Using this predicate and others defined in Section~\ref{sec:mmm}, it
is an easy matter to search for untyped $\lambda$-terms with various
properties.  Consider, for example, the following two predicates definitions.

\lstinputlisting[linerange={pres-end},basicstyle=\small\ttfamily]{code/sect7.sig}
\lstinputlisting[linerange={pres-end},basicstyle=\small\ttfamily]{code/sect7.mod}

The \lsti{prop_pres1} predicate is designed to search for linear
$\lambda$-terms that are related by \lsti{Step} to a non-linear
$\lambda$-term.  When \lsti{Step} is \lsti{cbn} or \lsti{cbv}, no such
term is possible.
%
% \begin{metanote}
%   AM: I have a problem with pres2: it does \emph{not} generate linear
%   exp, so it does not fit well with the title of the
%   subsection. Further \lsti{is_exp} is defined on the pure calculus,
%   while \lsti{is_lexp} uses also pairs etc, which do not add much to
%   the example.
% \end{metanote}
The \lsti{prop_pres2} predicate is designed to search for non-linear
$\lambda$-terms that have a simple type (via the \lsti{wt} predicate)
and are related by \lsti{Step} to a linear $\lambda$-term.  When
\lsti{Step} is \lsti{cbn} or \lsti{cbv}, the smallest such terms
(using the certificate \lsti{(height 4)}) are
\begin{lstlisting}[basicstyle=\small\ttfamily]
app (lam x\ lam y\ y) (app (lam x\ x) (lam x\ x))
app (lam x\ lam y\ y) (lam x\ x)
app (lam x\ lam y\ y) (lam x\ lam y\ y)
app (lam x\ lam y\ y) (lam x\ lam y\ x)
\end{lstlisting}
All of these terms evaluates (using either \lsti{cbn} or \lsti{cbv})
to the term \lsti{(lam x\ x)}.

% \subsection{Misc notes}

% AM: merged
% \begin{metanote}
%   TODO
%   \begin{itemize}
%   \item Take some of the motivation from LOPSTR: we do linear PBT
%     because we have no ``linear'' proof assistants
%   \item  Say we can handle
%     more LHHF. Mention Forum and encodings with par? Here? or in conclusion?
%   \end{itemize}
% \end{metanote}

% \begin{metanote}
%   DM I leave AM's comments here in order to remind us about what
%   discussions/examples might be missing, etc.
% \end{metanote}
%\newpage

%%% Local Variables:
%%% mode: latex
%%% TeX-master: "main"
%%% End:

% LocalWords:  rl del ubnd bnd llinterp limpC bangE DM sl eq Xs PermDef
% LocalWords:  nlist llinterp lc iff cst cbn cbv cex VV VN Ou lexp tl
% LocalWords:  lapp llcheck LOPSTR TODO LHHF init basicstyle cbnv
% LocalWords:  newcons isnat StartSize EndSize programmability
% \input linear % linear PBT

\section{Related and future work}
\label{sec:rel}

% DM I add subsections to provide an outline but we might drop them in
% the final draft.

\subsection{Two-level logic approach}

% DM I made the choice to cite only the early work.  Other works can
% be added. Alberto: did I get the right references for the CMU group
% wrt this topic of SL, RL?

First-order Horn clauses have a long tradition, via the Prolog
language, of specifying computation.  Such clauses have also been used
to specify the operational semantics of other programming languages:
see, for example, the early work on \emph{natural
semantics}~\cite{kahn87stacs} and the Centaur system~\cite{borras88}.
The intuitionistic logic setting of \emph{higher-order hereditary Harrop
formulas}~\cite{miller91apal}---a logical framework that significantly
extends the \spec logic in Section~\ref{sec:spec}---has similarly been
used for the past three decades to specify the static and dynamic
semantics of programming language: see, for example,~\cite{hannan92mscs,hannan93jfp}.  
Similar specifications could also be
written in the dependently typed $\lambda$-calculus
LF~\cite{harper93jacm}; see, for example,~\cite{michaylov91}.

Linear logic has been effectively used to enrich the natural semantic
framework.  The Lolli subset of linear logic~\cite{Hodas1994} as well
as the Forum presentation~\cite{miller96tcs} of all of linear logic
have been used to specify the operational semantics of references and
concurrency~\cite{miller96tcs} as well as the behavior of the
sequential and the concurrent (piped-line) operational semantics of
the DLX RISC processor~\cite{chirimar95phd}. Another fruitful example is the specification of \emph{session types}~\cite{CairesPT16}. Ordered logic
programming~\cite{PolakowY00,PfenningS09} has also been investigated.
Similar specifications
are also possible in  linear logic-inspired variants of
LF~\cite{CervesatoP02,Schack-NielsenS08,GeorgesMOP17}.

The use of a separate \emph{reasoning logic} to reason directly on the
kind of logic specifications used in this paper was proposed in~\cite{mcdowell02tocl}.  That logic included certain inductive
principles that could be applied to the definition of sequent calculus
provability.  That framework was challenged, however, by the need to
treat eigenvariables in sequent calculus proof systems.  The
$\nabla$-quantifier, introduced in~\cite{miller05tocl}, provided an
elegant solution to the treatment of eigenvariables~\cite{gacek12jar}.
In this paper, our use of the reasoning logic is mainly to determine
reachability and nonreachability: in those situations, the
$\nabla$-quantifier can be implemented by the universal quantifier in
\lP (see~\cite[Proposition 7.10]{miller05tocl}).  If we were to
investigate PBT examples that go beyond that simple class of queries,
then we would have to abandon \lP for the richer logic that underlies
Abella~\cite{gacek11ic}: see, for example, the specifications of
bisimulation for the $\pi$-calculus in~\cite{tiu10tocl}.
The two-level approach extends to  reasoning over specifications in
dependently typed $\lambda$-calculus, first via
the ${\cal M}_{\omega}$ reasoning logic over LF in~\cite{schurmann00phd} and then more extensively within the
\emph{Beluga} proof environment~\cite{pientka10ijcar,PientkaC15}.

However, formal \emph{verification} by reasoning over a linear logic
frameworks, is still in its
infancy, although the two-level approach is flexible enough to accommodate one reasoning logic over several SL\@.  % Sch\"{u}rmann et
% al.~\cite{McCreightS08} have designed $\cal{L}^+_{\omega}$, a linear
% meta-logics conservatively extending the meta-theory of Twelf and
% Pientka et al.~\cite{GeorgesMOP17} have introduced \emph{LINCX}, a
% linear version of contextual modal type theory to be used within
% Beluga.
The most common  case study is
 type soundness of MiniML with references, first checked in~\cite{mcdowell02tocl} with the $\Pi$ proof checker and then by Martin in his
dissertation~\cite{Martin10}   using
Isabelle/HOL's Hybrid library~\cite{FeltyM12}, in several styles, including linear and
ordered specifications.  More extensive use of Hybrid, this time on
top of Coq, includes the recent verification in a Lolli-like specification
logic of type soundness of the \emph{proto-Quipper}
quantum functional programming language~\cite{MahmoudF19}, as well as the meta-theory of
sequent calculi~\cite{FeltyOX21}.

\subsection{Foundational proof certificates}
\label{ssect:fpc}

The notion of \emph{foundational proof certificates} was introduced in~\cite{chihani17jar} as a means for flexibly describing proofs to a
logic programming base proof checker~\cite{miller17fac}.  In that
setting, proof certificates can include all  or just certain
details, whereby missing details can often be
recreated during checking using unification and backtracking search.
The pairing FPC in Section~\ref{ssec:fpc} can be used to
\emph{elaborate} a proof certificate into one including full details
and to \emph{distill} a detailed proof into a certificate containing
less information~\cite{Pair}.  

Using this style of proof elaboration, it is possible to use the ELPI
plugin to Coq~\cite{tassi:hal-01637063}, which supplies
the Coq computing environment with a \lP implementation, to elaborate
proof certificates from an external theorem prover into fully detailed
certificates that can be checked by the Coq kernel~\cite{Manighetti0M20,manighetti22phd}.  This same
interface between Coq and ELPI allowed Manighetti el al.\ to  illustrate
how PBT could be used to search for counterexamples to conjectures
proposed to Coq.

% \begin{metanote}
%   AM: moving this here from later on. Decide if we want to keep it
% \end{metanote}
Using an FPC as a description of how to traverse a search space bears
some resemblance with principled ways to change the depth-first nature
of search in logic programming implementations. An example is \emph{Tor}~\cite{Tor},
which, however, is unable to account for random search. Similarly to
\emph{Tor}, FPCs would benefit of \emph{partial evaluation} to remove
the meta-interpretive layer.

\subsection{Property based testing for meta-theory model-checking}

% \begin{metanote}
% AM: started some revision, but needs a fresh look. There may be a way
% to do model checking in the K frameworks with binders, but I can't be
% bothered . More emphasis about QuickChick, Coq.
%   \begin{itemize}
% %  \item  give more emphasis to QuickChick-Coq 
%   \item mention K and binders -- does K do testing?
% %  \item on linear spec: stateful PBT (state machines)~\cite{BarrioFHEM21} %done
% %  \item stuff from LOPSTR paper: linear terms generation (Tarau) % done
% \end{itemize}
%\end{metanote}

The literature on PBT is very large and always evolving. We refer
to~\cite{cheney_momigliano_2017} for a partial review with an emphasis to its
interaction with proof assistants and specialized domains such as
programming language meta-theory.

While Isabelle/HOL were at the forefront of combining theorem proving
and refutations~\cite{BlanchetteBN11,Bulwahn12}, nowadays most of the
action is within Coq: \emph{QuickChick}~\cite{QChick} started as a
clone of QuickCheck, but soon evolved in a major research project
involving automatic generation of custom
generators~\cite{LampropoulosPP18}, coverage based improvements of
random generation~\cite{Lampropoulos0P19,GoldsteinHLP21}, as well as going beyond the
executable fragment of Coq~\cite{Paraskevopoulou22}.

Within the confine of model checking PL theory, a major player is
\emph{PLT-Redex}~\cite{PLTbook}, an executable DSL for mechanizing
semantic models built on top of \emph{DrRacket} with support for
random testing \`a la QuickCheck; its usefulness has been demonstrated
in several impressive case studies~\cite{Klein12}. However, Redex has
limited support for relational specifications and none whatsoever for
binding signature. On the other hand,
\emph{$\alpha$Check}~\cite{cheney_momigliano_2017,Pessina16} is built
on top of the nominal logic programming language $\alpha$Prolog and it
checks relational specifications similar to those considered here.
Arguably, $\alpha$Check is less flexible than the FPC-based
architecture, since its generation strategy can be seen as a
fixed choice of experts. The same comment applies to % Indeed, our ``bounding'' FPCs in
% Figure~\ref{fig:resources} have a clear correspondence with the way
% exhaustive term generation is been addressed there, as well as in
(Lazy)SmallCheck~\cite{smallcheck}. In both cases, those strategies
are wired-in and cannot be varied, let alone combined as we can via
pairing. For a small empirical comparison between our approach and
$\alpha$Check on the PLT-Redex benchmark\footnote{\url{http://docs.racket-lang.org/redex/benchmark.html}}, please
see~\cite{blanco19ppdp}.

In the random case, the logic programming paradigm has some advantages
over the labor-intensive QuickCheck approach of writing custom
generators. Moreover, the last few years have witnessed an increasing
interest in the (random) generation of data satisfying some
invariants~\cite{ClaessenDP15,pltredexconstraintlogic,LampropoulosPP18};
in particular well-typed $\lambda$-terms, with an application to testing
optimizing
compilers~\cite{PalkaCRH11,MidtgaardJKNN17,BendkowskiGT18}.  Our
approach can use \emph{judgments} (think typing), as generators,
avoiding the issue of keeping generators and predicates in sync when
checking invariant-preserving properties such as type
preservation~\cite{beginners-luck}. Further, viewing random generation
as expert-driven random back-chaining opens up several
possibilities:
we have chosen just one simple-minded strategy, namely permuting the
predicate definition at each unfolding, but we could easily follow
others, such as the ones described in~\cite{pltredexconstraintlogic}:
permuting the definition just once at the start of the generation
phase, or even changing the weights at the end of the run so as to
steer the derivation towards axioms/facts. Of course, our default
uniform distribution corresponds to QuickCheck's \lsti{oneOf}
combinator, while the weights table to \lsti{frequency}. Moreover,
pairing random and size-based FPC accounts for some of QuickCheck's
configuration options, such as \emph{StartSize} and \emph{EndSize}.

% In
% particular, the most successful generator~\cite{PalkaCRH11} consists
% of over 1500 lines of dense Haskell ../code hard-wired to generate a
% specific object language. Compare this to our $10$ lines of 
% readable clauses. 
% We make no claim about how successfully we could challenge a compiler, but we do
% want to note
%  how flexible our approach 
% is. %, as exemplified earlier.
% Finally, more distant cousins in the logic programming world are
% \emph{declarative debugging}~\cite{naish97declarative} and the
% Logic-Based Model Checking project at Stony Brook
% (\url{http://www.cs.sunysb.edu/~lmc}).

% AM: commented out
% There also seems to be a connection with \emph{probabilistic logic
%   programming}, e.g.,~\cite{FierensBRSGTJR15}, although the inference
% mechanism is very different.

%

% Finally, there is a very long tradition of using logic programming for
% \emph{tests case generation}, starting from~\cite{BougeCFG86}. Of
% some interest is work such as~\cite{FioravantiPS15} that
% advocates the \emph{constrain-and-generate} approach to implement
% invariant-based generators (here over finite domains) and program
% transformations to optimize them via well-know ideas such as
% \emph{filter promotion}.

In mainstream programming, property-based testing of stateful programs
is accomplished via some form of \emph{state
  machine}~\cite{quickcheckfunprofit,BarrioFHEM21}.  The idea of
linear PBT has been proposed in~\cite{MantovaniM21} and applied to
mechanized meta-theory model checking, restricted to first-order
encodings, e.g.\ the compilation of a simple imperative language to a
stack machine.  For efficient generation of typed linear $\lambda$-terms,
albeit limited to the purely implicational propositional fragment,
see~\cite{Tarau20}

\subsection{Future work}

% DM A few sentences about certificates and trust
%  AM: moved to conclusion
While \lP is used here to discover counterexamples, one
does not actually need to trust the logical soundness of
\lP, negation-as-failure making this a complex
issue.
Any identified counterexample can be exported and utilized within a proof assistants such as  Abella.  In fact, it would be a logical future task to
incorporate our perspective on PBT into Abella in order to accommodate
both proofs and disproofs, as many proof helpers frequently
do.
As mentioned in Section~\ref{ssect:fpc}, the  implementation of
the ELPI plugin for the Coq theorem prover~\cite{tassi:hal-01637063}
should allow the PBT techniques described in this paper to be applied
to proposed theorems within Coq: discovering a counterexamples to a
proposed theorem could save a lot of time in attempting the proof of a
non-theorem.

The strategy we have outlined here serves more as a
proof-of-concept or logical reconstruction than as a robust
implementation. %  Any counterexample that is discovered can be output
% and used within, say, Abella.  In fact, integrating our take on PBT in
% Abella, in order to support both proofs and \emph{disproofs}, as many
% proof assistants routinely do, would be a natural future work . The
% approach that we have delineated here is more of a
% proof-of-concept/logical reconstruction than a robust implementation.
%
A natural environment to support PBT for every specification in Abella
is the Bedwyr model-checker~\cite{baelde07cade}, which shares the same
meta-logic, but is more efficient from the point of view of proof
search.

% AM: I'm dropping this as a future work
% Another dimension refers to \emph{coinductive} specifications:
% % , where Abella excels~\cite{tiu10tocl,Howe}:
% consider for example using PBT to
% find programs that refute the equivalence of \emph{ground} and
% \emph{applicative} bisimilarity~\cite{PITobtpe}.

Finally,
we have just hinted at ways for localizing the origin of the bugs
reported by PBT\@. This issue can benefit from research in declarative
debugging as well as in \emph{justification} for logic
programs~\cite{onlinejust}. Coupled with  results in
linking game semantics and focusing~\cite{MillerS06},
this could lead us also to a reappraisal of
techniques for \emph{repairing} (inductive)
proofs~\cite{ireland96jar}.

\section{Conclusions}
\label{sec:concl}

\begin{modified}
We have presented a framework for property-based testing whose design
relies on logic programming and proof theory.  This framework offers a
versatile and consistent foundation for various PBT features based on
relational specifications.  It leverages proof outlines (i.e., incomplete
proof certificates) for flexible counterexample search.  It employs
a single, transparent proof checker for certain subsets of classical,
intuitionistic, and linear logic to elaborate proof outlines
nondeterministically.  Furthermore, this framework easily handles
counterexamples involving bindings, making it possible to apply PBT to
specifications that deal directly with programs as objects.  The
framework's programmability enables specifying diverse search
strategies (exhaustive, bounded, randomized) and counterexample
shrinking.
\end{modified}

%% Given this proof-theoretic pedigree, it was immediate to extend PBT to
%% the metaprogramming setting, inheriting the handling of $\lambda$-tree
%% syntax, which is naturally supported by \lP and notably absent from
%% most other environments for meta-theory model checking.
%% % such as PLT-Redex.  
%% Similarly it was straightforward to apply PBT to specifications in
%% fragments of linear logic.

\smallskip\noindent{\bf Acknowledgments} We would like to thank Rob
Blanco for his major contributions to~\cite{blanco19ppdp}, on which
the present paper is based.  We also thank the reviewers for their
comments on an earlier draft of this paper.

% This paper is a major extension of work-in-progress presented (but not
% published) at \emph{LFMTP'17}. We wish to thank Enrico Tassi for
% extending ELPI with support for random number generation. The third
% author has been partially funded by the INdAM-GNCS project 2018
% ``Metodi di prova orientati al ragionamento automatico per logiche
% non-classiche''.
%
\bibliographystyle{acmtrans}
%\bibliography{../local}

\begin{thebibliography}{}

\bibitem[\protect\citeauthoryear{Andreoli and Pareschi}{Andreoli and
  Pareschi}{1990}]{andreoli90iclp}
{\sc Andreoli, J.-M.} {\sc and} {\sc Pareschi, R.} 1990.
\newblock Linear objects: Logical processes with built-in inheritance.
\newblock In {\em Proceeding of the Seventh International Conference on Logic
  Programming, Jerusalem}. MIT Press.

\bibitem[\protect\citeauthoryear{Apt and Doets}{Apt and
  Doets}{1994}]{apt94jlpb}
{\sc Apt, K.~R.} {\sc and} {\sc Doets, K.} 1994.
\newblock A new definition of {SLDNF}-resolution.
\newblock {\em The Journal of Logic Programming\/}~{\em 18,\/}~2 (Feb.),
  177--190.

\bibitem[\protect\citeauthoryear{Apt and van Emden}{Apt and van
  Emden}{1982}]{apt82jacm}
{\sc Apt, K.~R.} {\sc and} {\sc van Emden, M.~H.} 1982.
\newblock Contributions to the theory of logic programming.
\newblock {\em JACM\/}~{\em 29,\/}~3, 841--862.

\bibitem[\protect\citeauthoryear{Baelde, Chaudhuri, Gacek, Miller, Nadathur,
  Tiu, and Wang}{Baelde et~al\mbox{.}}{2014}]{baelde14jfr}
{\sc Baelde, D.}, {\sc Chaudhuri, K.}, {\sc Gacek, A.}, {\sc Miller, D.}, {\sc
  Nadathur, G.}, {\sc Tiu, A.}, {\sc and} {\sc Wang, Y.} 2014.
\newblock Abella: {A} system for reasoning about relational specifications.
\newblock {\em Journal of Formalized Reasoning\/}~{\em 7,\/}~2, 1--89.

\bibitem[\protect\citeauthoryear{Baelde, Gacek, Miller, Nadathur, and
  Tiu}{Baelde et~al\mbox{.}}{2007}]{baelde07cade}
{\sc Baelde, D.}, {\sc Gacek, A.}, {\sc Miller, D.}, {\sc Nadathur, G.}, {\sc
  and} {\sc Tiu, A.} 2007.
\newblock The {Bedwyr} system for model checking over syntactic expressions.
\newblock In {\em 21th Conf.\ on Automated Deduction (CADE)}, {F.~Pfenning},
  Ed. Number 4603 in LNAI. Springer, New York, 391--397.

\bibitem[\protect\citeauthoryear{Bendkowski, Grygiel, and Tarau}{Bendkowski
  et~al\mbox{.}}{2018}]{BendkowskiGT18}
{\sc Bendkowski, M.}, {\sc Grygiel, K.}, {\sc and} {\sc Tarau, P.} 2018.
\newblock Random generation of closed simply typed {\(\lambda\)}-terms: {A}
  synergy between logic programming and Boltzmann samplers.
\newblock {\em Theory Pract. Log. Program.\/}~{\em 18,\/}~1, 97--119.

\bibitem[\protect\citeauthoryear{Blanchette, Bulwahn, and Nipkow}{Blanchette
  et~al\mbox{.}}{2011}]{BlanchetteBN11}
{\sc Blanchette, J.~C.}, {\sc Bulwahn, L.}, {\sc and} {\sc Nipkow, T.} 2011.
\newblock Automatic proof and disproof in {Isabelle/HOL}.
\newblock In {\em FroCoS}, {C.~Tinelli} {and} {V.~Sofronie-Stokkermans}, Eds.
  Lecture Notes in Computer Science, vol. 6989. Springer, 12--27.

\bibitem[\protect\citeauthoryear{Blanco, Chihani, and Miller}{Blanco
  et~al\mbox{.}}{2017}]{Pair}
{\sc Blanco, R.}, {\sc Chihani, Z.}, {\sc and} {\sc Miller, D.} 2017.
\newblock Translating between implicit and explicit versions of proof.
\newblock In {\em Automated Deduction - {CADE} 26 --- 26th International
  Conference on Automated Deduction}, {L.~de~Moura}, Ed. Lecture Notes in
  Computer Science, vol. 10395. Springer, 255--273.

\bibitem[\protect\citeauthoryear{Blanco and Miller}{Blanco and
  Miller}{2015}]{blanco15wof}
{\sc Blanco, R.} {\sc and} {\sc Miller, D.} 2015.
\newblock Proof outlines as proof certificates: a system description.
\newblock In {\em Proceedings First International Workshop on Focusing},
  {I.~Cervesato} {and} {C.~Sch{\"u}rmann}, Eds. Electronic Proceedings in
  Theoretical Computer Science, vol. 197. Open Publishing Association, 7--14.

\bibitem[\protect\citeauthoryear{Blanco, Miller, and Momigliano}{Blanco
  et~al\mbox{.}}{2019}]{blanco19ppdp}
{\sc Blanco, R.}, {\sc Miller, D.}, {\sc and} {\sc Momigliano, A.} 2019.
\newblock Property-based testing via proof reconstruction.
\newblock In {\em Principles and Practice of Programming Languages 2019 (PPDP
  '19)}, {E.~Komendantskaya}, Ed.

\bibitem[\protect\citeauthoryear{Borras, Cl\'ement, Despeyroux, Incerpi, Kahn,
  Lang, and Pascual}{Borras et~al\mbox{.}}{1988}]{borras88}
{\sc Borras, P.}, {\sc Cl\'ement, D.}, {\sc Despeyroux, T.}, {\sc Incerpi, J.},
  {\sc Kahn, G.}, {\sc Lang, B.}, {\sc and} {\sc Pascual, V.} 1988.
\newblock {Centaur}: the system.
\newblock In {\em Third Annual Symposium on Software Development Environments
  (SDE3)}. ACM, Boston, 14--24.

\bibitem[\protect\citeauthoryear{Bulwahn}{Bulwahn}{2012}]{Bulwahn12}
{\sc Bulwahn, L.} 2012.
\newblock The new Quickcheck for Isabelle --- random, exhaustive and symbolic
  testing under one roof.
\newblock In {\em Certified Programs and Proofs - Second International
  Conference, {CPP} 2012, Kyoto, Japan, December 13---15, 2012. Proceedings},
  {C.~Hawblitzel} {and} {D.~Miller}, Eds. Lecture Notes in Computer Science,
  vol. 7679. Springer, 92--108.

\bibitem[\protect\citeauthoryear{Caires, Pfenning, and Toninho}{Caires
  et~al\mbox{.}}{2016}]{CairesPT16}
{\sc Caires, L.}, {\sc Pfenning, F.}, {\sc and} {\sc Toninho, B.} 2016.
\newblock Linear logic propositions as session types.
\newblock {\em Math. Struct. Comput. Sci.\/}~{\em 26,\/}~3, 367--423.

\bibitem[\protect\citeauthoryear{Cervesato and Pfenning}{Cervesato and
  Pfenning}{2002}]{CervesatoP02}
{\sc Cervesato, I.} {\sc and} {\sc Pfenning, F.} 2002.
\newblock A linear logical framework.
\newblock {\em Inf. Comput.\/}~{\em 179,\/}~1, 19--75.

\bibitem[\protect\citeauthoryear{Chargu{\'e}raud}{Chargu{\'e}raud}{2011}]{chargueraud11jar}
{\sc Chargu{\'e}raud, A.} 2011.
\newblock The locally nameless representation.
\newblock {\em Journal of Automated Reasoning\/}~{\em 49}, 363--408.

\bibitem[\protect\citeauthoryear{Cheney and Momigliano}{Cheney and
  Momigliano}{2017}]{cheney_momigliano_2017}
{\sc Cheney, J.} {\sc and} {\sc Momigliano, A.} 2017.
\newblock $\alpha${C}heck: A mechanized metatheory model checker.
\newblock {\em Theory and Practice of Logic Programming\/}~{\em 17,\/}~3,
  311–352.

\bibitem[\protect\citeauthoryear{Cheney, Momigliano, and Pessina}{Cheney
  et~al\mbox{.}}{2016}]{Pessina16}
{\sc Cheney, J.}, {\sc Momigliano, A.}, {\sc and} {\sc Pessina, M.} 2016.
\newblock Advances in property-based testing for $\alpha${P}rolog.
\newblock In {\em Tests and Proofs - 10th International Conference, {TAP} 2016,
  Vienna, Austria, July 5-7, 2016, Proceedings}, {B.~K. Aichernig} {and} {C.~A.
  Furia}, Eds. Lecture Notes in Computer Science, vol. 9762. Springer, 37--56.

\bibitem[\protect\citeauthoryear{Chihani, Miller, and Renaud}{Chihani
  et~al\mbox{.}}{2017}]{chihani17jar}
{\sc Chihani, Z.}, {\sc Miller, D.}, {\sc and} {\sc Renaud, F.} 2017.
\newblock A semantic framework for proof evidence.
\newblock {\em J. of Automated Reasoning\/}~{\em 59,\/}~3, 287--330.

\bibitem[\protect\citeauthoryear{Chirimar}{Chirimar}{1995}]{chirimar95phd}
{\sc Chirimar, J.} 1995.
\newblock Proof theoretic approach to specification languages.
\newblock Ph.D. thesis, University of Pennsylvania.

\bibitem[\protect\citeauthoryear{Chlipala}{Chlipala}{2008}]{chlipala08icfp}
{\sc Chlipala, A.} 2008.
\newblock Parametric higher-order abstract syntax for mechanized semantics.
\newblock In {\em Proceeding of the 13th {ACM} {SIGPLAN} international
  conference on Functional programming, {ICFP} 2008, Victoria, {BC}, Canada,
  September 20-28, 2008}, {J.~Hook} {and} {P.~Thiemann}, Eds. ACM, 143--156.

\bibitem[\protect\citeauthoryear{Church}{Church}{1940}]{church40}
{\sc Church, A.} 1940.
\newblock A formulation of the {Simple} {Theory} of {Types}.
\newblock {\em J. of Symbolic Logic\/}~{\em 5}, 56--68.

\bibitem[\protect\citeauthoryear{Claessen, Dureg{\aa}rd, and Palka}{Claessen
  et~al\mbox{.}}{2015}]{ClaessenDP15}
{\sc Claessen, K.}, {\sc Dureg{\aa}rd, J.}, {\sc and} {\sc Palka, M.~H.} 2015.
\newblock Generating constrained random data with uniform distribution.
\newblock {\em J. Funct. Program.\/}~{\em 25}, e8:1--e8:31.

\bibitem[\protect\citeauthoryear{Claessen and Hughes}{Claessen and
  Hughes}{2000}]{claessen00icfp}
{\sc Claessen, K.} {\sc and} {\sc Hughes, J.} 2000.
\newblock {QuickCheck}: a lightweight tool for random testing of {Haskell}
  programs.
\newblock In {\em Proceedings of the 2000 {ACM} {SIGPLAN} International
  Conference on Functional Programming (ICFP 2000)}. ACM, 268--279.

\bibitem[\protect\citeauthoryear{Clark}{Clark}{1978}]{clark78}
{\sc Clark, K.~L.} 1978.
\newblock Negation as failure.
\newblock In {\em Logic and Data Bases}, {J.~Gallaire} {and} {J.~Minker}, Eds.
  Plenum Press, New York, 293--322.

\bibitem[\protect\citeauthoryear{de~Barrio, Fredlund, Herranz, Earle, and
  Mari{\~{n}}o}{de~Barrio et~al\mbox{.}}{2021}]{BarrioFHEM21}
{\sc de~Barrio, L. E.~B.}, {\sc Fredlund, L.}, {\sc Herranz, {\'{A}}.}, {\sc
  Earle, C.~B.}, {\sc and} {\sc Mari{\~{n}}o, J.} 2021.
\newblock Makina: a new Quickcheck state machine library.
\newblock In {\em Proceedings of the 20th {ACM} {SIGPLAN} International
  Workshop on Erlang, Erlang@ICFP 2021, Virtual Event, Korea, August 26, 2021},
  {S.~Aronis} {and} {A.~Bieniusa}, Eds. {ACM}, 41--53.

\bibitem[\protect\citeauthoryear{de~Bruijn}{de~Bruijn}{1972}]{debruijn72}
{\sc de~Bruijn, N.~G.} 1972.
\newblock Lambda calculus notation with nameless dummies, a tool for automatic
  formula manipulation, with an application to the {Church-Rosser} theorem.
\newblock {\em Indagationes Mathematicae\/}~{\em 34,\/}~5, 381--392.

\bibitem[\protect\citeauthoryear{Dureg{\aa}rd, Jansson, and Wang}{Dureg{\aa}rd
  et~al\mbox{.}}{2012}]{feat}
{\sc Dureg{\aa}rd, J.}, {\sc Jansson, P.}, {\sc and} {\sc Wang, M.} 2012.
\newblock Feat: functional enumeration of algebraic types.
\newblock In {\em Haskell Workshop}, {J.~Voigtl{\"{a}}nder}, Ed. {ACM}, 61--72.

\bibitem[\protect\citeauthoryear{Felleisen, Findler, and Flatt}{Felleisen
  et~al\mbox{.}}{2009}]{PLTbook}
{\sc Felleisen, M.}, {\sc Findler, R.~B.}, {\sc and} {\sc Flatt, M.} 2009.
\newblock {\em Semantics Engineering with {PLT R}edex}.
\newblock The MIT Press.

\bibitem[\protect\citeauthoryear{Felty and Momigliano}{Felty and
  Momigliano}{2012}]{FeltyM12}
{\sc Felty, A.~P.} {\sc and} {\sc Momigliano, A.} 2012.
\newblock Hybrid - {A} definitional two-level approach to reasoning with
  higher-order abstract syntax.
\newblock {\em J. Autom. Reasoning\/}~{\em 48,\/}~1, 43--105.

\bibitem[\protect\citeauthoryear{Felty, Olarte, and Xavier}{Felty
  et~al\mbox{.}}{2021}]{FeltyOX21}
{\sc Felty, A.~P.}, {\sc Olarte, C.}, {\sc and} {\sc Xavier, B.} 2021.
\newblock A focused linear logical framework and its application to metatheory
  of object logics.
\newblock {\em Math. Struct. Comput. Sci.\/}~{\em 31,\/}~3, 312--340.

\bibitem[\protect\citeauthoryear{Fetscher, Claessen, Palka, Hughes, and
  Findler}{Fetscher et~al\mbox{.}}{2015}]{pltredexconstraintlogic}
{\sc Fetscher, B.}, {\sc Claessen, K.}, {\sc Palka, M.~H.}, {\sc Hughes, J.},
  {\sc and} {\sc Findler, R.~B.} 2015.
\newblock Making random judgments: Automatically generating well-typed terms
  from the definition of a type-system.
\newblock In {\em {ESOP}}. Lecture Notes in Computer Science, vol. 9032.
  Springer, 383--405.

\bibitem[\protect\citeauthoryear{Friedman}{Friedman}{1978}]{friedman78hol}
{\sc Friedman, H.~M.} 1978.
\newblock Classically and intuitionistically provably recursive functions.
\newblock In {\em Higher Order Set Theory}, {G.~H. M{\"u}ller} {and} {D.~S.
  Scott}, Eds. Springer Verlag, Berlin, 21--27.

\bibitem[\protect\citeauthoryear{Gacek, Miller, and Nadathur}{Gacek
  et~al\mbox{.}}{2011}]{gacek11ic}
{\sc Gacek, A.}, {\sc Miller, D.}, {\sc and} {\sc Nadathur, G.} 2011.
\newblock Nominal abstraction.
\newblock {\em Information and Computation\/}~{\em 209,\/}~1, 48--73.

\bibitem[\protect\citeauthoryear{Gacek, Miller, and Nadathur}{Gacek
  et~al\mbox{.}}{2012}]{gacek12jar}
{\sc Gacek, A.}, {\sc Miller, D.}, {\sc and} {\sc Nadathur, G.} 2012.
\newblock A two-level logic approach to reasoning about computations.
\newblock {\em J. of Automated Reasoning\/}~{\em 49,\/}~2, 241--273.

\bibitem[\protect\citeauthoryear{Gentzen}{Gentzen}{1935}]{gentzen35}
{\sc Gentzen, G.} 1935.
\newblock Investigations into logical deduction.
\newblock In {\em {The Collected Papers of Gerhard Gentzen}}, {M.~E. Szabo},
  Ed. North-Holland, Amsterdam, 68--131.

\bibitem[\protect\citeauthoryear{Georges, Murawska, Otis, and Pientka}{Georges
  et~al\mbox{.}}{2017}]{GeorgesMOP17}
{\sc Georges, A.~L.}, {\sc Murawska, A.}, {\sc Otis, S.}, {\sc and} {\sc
  Pientka, B.} 2017.
\newblock {LINCX:} {A} linear logical framework with first-class contexts.
\newblock In {\em Programming Languages and Systems - 26th European Symposium
  on Programming, {ESOP} 2017, Held as Part of the European Joint Conferences
  on Theory and Practice of Software, {ETAPS} 2017, Uppsala, Sweden, April
  22-29, 2017, Proceedings}, {H.~Yang}, Ed. Lecture Notes in Computer Science,
  vol. 10201. Springer, 530--555.

\bibitem[\protect\citeauthoryear{Girard}{Girard}{1987}]{girard87tcs}
{\sc Girard, J.-Y.} 1987.
\newblock Linear logic.
\newblock {\em Theoretical Computer Science\/}~{\em 50,\/}~1, 1--102.

\bibitem[\protect\citeauthoryear{Girard}{Girard}{1992}]{girard92mail}
{\sc Girard, J.-Y.} 1992.
\newblock A fixpoint theorem in linear logic.
\newblock An email posting to the mailing list linear@cs.stanford.edu.

\bibitem[\protect\citeauthoryear{Goguen}{Goguen}{1995}]{Goguen95}
{\sc Goguen, H.} 1995.
\newblock Typed operational semantics.
\newblock In {\em Typed Lambda Calculi and Applications, Second International
  Conference on Typed Lambda Calculi and Applications, {TLCA} '95, Edinburgh,
  UK, April 10-12, 1995, Proceedings}, {M.~Dezani{-}Ciancaglini} {and} {G.~D.
  Plotkin}, Eds. Lecture Notes in Computer Science, vol. 902. Springer,
  186--200.

\bibitem[\protect\citeauthoryear{Goldstein, Hughes, Lampropoulos, and
  Pierce}{Goldstein et~al\mbox{.}}{2021}]{GoldsteinHLP21}
{\sc Goldstein, H.}, {\sc Hughes, J.}, {\sc Lampropoulos, L.}, {\sc and} {\sc
  Pierce, B.~C.} 2021.
\newblock Do judge a test by its cover - combining combinatorial and
  property-based testing.
\newblock In {\em Programming Languages and Systems - 30th European Symposium
  on Programming, {ESOP} 2021, Held as Part of the European Joint Conferences
  on Theory and Practice of Software, {ETAPS} 2021, Luxembourg City,
  Luxembourg, March 27 - April 1, 2021, Proceedings}, {N.~Yoshida}, Ed. Lecture
  Notes in Computer Science, vol. 12648. Springer, 264--291.

\bibitem[\protect\citeauthoryear{Halln{\"{a}}s and
  Schroeder{-}Heister}{Halln{\"{a}}s and
  Schroeder{-}Heister}{1991}]{HallnasS91}
{\sc Halln{\"{a}}s, L.} {\sc and} {\sc Schroeder{-}Heister, P.} 1991.
\newblock A proof-theoretic approach to logic programming. {II.} programs as
  definitions.
\newblock {\em J. Log. Comput.\/}~{\em 1,\/}~5, 635--660.

\bibitem[\protect\citeauthoryear{Hannan}{Hannan}{1993}]{hannan93jfp}
{\sc Hannan, J.} 1993.
\newblock Extended natural semantics.
\newblock {\em J. of Functional Programming\/}~{\em 3,\/}~2 (Apr.), 123--152.

\bibitem[\protect\citeauthoryear{Hannan and Miller}{Hannan and
  Miller}{1992}]{hannan92mscs}
{\sc Hannan, J.} {\sc and} {\sc Miller, D.} 1992.
\newblock From operational semantics to abstract machines.
\newblock {\em Mathematical Structures in Computer Science\/}~{\em 2,\/}~4,
  415--459.

\bibitem[\protect\citeauthoryear{Harper, Honsell, and Plotkin}{Harper
  et~al\mbox{.}}{1993}]{harper93jacm}
{\sc Harper, R.}, {\sc Honsell, F.}, {\sc and} {\sc Plotkin, G.} 1993.
\newblock A framework for defining logics.
\newblock {\em Journal of the ACM\/}~{\em 40,\/}~1, 143--184.

\bibitem[\protect\citeauthoryear{Heath and Miller}{Heath and
  Miller}{2019}]{heath19jar}
{\sc Heath, Q.} {\sc and} {\sc Miller, D.} 2019.
\newblock A proof theory for model checking.
\newblock {\em J. of Automated Reasoning\/}~{\em 63,\/}~4, 857--885.

\bibitem[\protect\citeauthoryear{Hodas and Miller}{Hodas and
  Miller}{1994}]{Hodas1994}
{\sc Hodas, J.} {\sc and} {\sc Miller, D.} 1994.
\newblock Logic programming in a fragment of intuitionistic linear logic.
\newblock {\em Information and Computation\/}~{\em 110,\/}~2, 327--365.

\bibitem[\protect\citeauthoryear{Hritcu, Hughes, Pierce, Spector-Zabusky,
  Vytiniotis, Azevedo~de Amorim, and Lampropoulos}{Hritcu
  et~al\mbox{.}}{2013}]{TNIQ}
{\sc Hritcu, C.}, {\sc Hughes, J.}, {\sc Pierce, B.~C.}, {\sc Spector-Zabusky,
  A.}, {\sc Vytiniotis, D.}, {\sc Azevedo~de Amorim, A.}, {\sc and} {\sc
  Lampropoulos, L.} 2013.
\newblock Testing noninterference, quickly.
\newblock In {\em Proceedings of the 18th ACM SIGPLAN International Conference
  on Functional Programming}. ICFP '13. ACM, New York, NY, USA, 455--468.

\bibitem[\protect\citeauthoryear{Hughes}{Hughes}{2007}]{quickcheckfunprofit}
{\sc Hughes, J.} 2007.
\newblock Quickcheck testing for fun and profit.
\newblock In {\em Practical Aspects of Declarative Languages, 9th International
  Symposium, PADL 2007, Nice, France, January 14-15, 2007}, {M.~Hanus}, Ed.
  Lecture Notes in Computer Science, vol. 4354. Springer, 1--32.

\bibitem[\protect\citeauthoryear{Ireland and Bundy}{Ireland and
  Bundy}{1996}]{ireland96jar}
{\sc Ireland, A.} {\sc and} {\sc Bundy, A.} 1996.
\newblock Productive use of failure in inductive proof.
\newblock {\em Journal of Automated Reasoning\/}~{\em 16}, 79--111.

\bibitem[\protect\citeauthoryear{Kahn}{Kahn}{1987}]{kahn87stacs}
{\sc Kahn, G.} 1987.
\newblock Natural semantics.
\newblock In {\em Proceedings of the Symposium on Theoretical Aspects of
  Computer Science}, {F.-J. Brandenburg}, {G.~Vidal-Naquet}, {and}
  {M.~Wirsing}, Eds. Lecture Notes in Computer Science, vol. 247. Springer,
  22--39.

\bibitem[\protect\citeauthoryear{Klein, Clements, Dimoulas, Eastlund,
  Felleisen, Flatt, McCarthy, Rafkind, Tobin-Hochstadt, and Findler}{Klein
  et~al\mbox{.}}{2012}]{Klein12}
{\sc Klein, C.}, {\sc Clements, J.}, {\sc Dimoulas, C.}, {\sc Eastlund, C.},
  {\sc Felleisen, M.}, {\sc Flatt, M.}, {\sc McCarthy, J.~A.}, {\sc Rafkind,
  J.}, {\sc Tobin-Hochstadt, S.}, {\sc and} {\sc Findler, R.~B.} 2012.
\newblock Run your research: on the effectiveness of lightweight mechanization.
\newblock In {\em Proceedings of the 39th annual ACM SIGPLAN-SIGACT symposium
  on Principles of programming languages}. POPL '12. ACM, New York, NY, USA,
  285--296.

\bibitem[\protect\citeauthoryear{Lampropoulos, Gallois-Wong, Hri\c{t}cu,
  Hughes, Pierce, and Xia}{Lampropoulos et~al\mbox{.}}{2017}]{beginners-luck}
{\sc Lampropoulos, L.}, {\sc Gallois-Wong, D.}, {\sc Hri\c{t}cu, C.}, {\sc
  Hughes, J.}, {\sc Pierce, B.~C.}, {\sc and} {\sc Xia, L.} 2017.
\newblock Beginner's {Luck}: A language for random generators.
\newblock In {\em 44th ACM SIGPLAN Symposium on Principles of Programming
  Languages (POPL)}.

\bibitem[\protect\citeauthoryear{Lampropoulos, Hicks, and Pierce}{Lampropoulos
  et~al\mbox{.}}{2019}]{Lampropoulos0P19}
{\sc Lampropoulos, L.}, {\sc Hicks, M.}, {\sc and} {\sc Pierce, B.~C.} 2019.
\newblock Coverage guided, property based testing.
\newblock {\em Proc. {ACM} Program. Lang.\/}~{\em 3,\/}~{OOPSLA},
  181:1--181:29.

\bibitem[\protect\citeauthoryear{Lampropoulos, Paraskevopoulou, and
  Pierce}{Lampropoulos et~al\mbox{.}}{2018}]{LampropoulosPP18}
{\sc Lampropoulos, L.}, {\sc Paraskevopoulou, Z.}, {\sc and} {\sc Pierce,
  B.~C.} 2018.
\newblock Generating good generators for inductive relations.
\newblock {\em Proc. {ACM} Program. Lang.\/}~{\em 2,\/}~{POPL}, 45:1--45:30.

\bibitem[\protect\citeauthoryear{Lampropoulos and Pierce}{Lampropoulos and
  Pierce}{2023}]{Lampropoulos:SF4}
{\sc Lampropoulos, L.} {\sc and} {\sc Pierce, B.~C.} 2023.
\newblock {\em QuickChick: Property-Based Testing in Coq}. Software
  Foundations, vol.~4.
\newblock Electronic textbook.
\newblock Version 1.3.2, \url{http://softwarefoundations.cis.upenn.edu}.

\bibitem[\protect\citeauthoryear{Mahmoud and Felty}{Mahmoud and
  Felty}{2019}]{MahmoudF19}
{\sc Mahmoud, M.~Y.} {\sc and} {\sc Felty, A.~P.} 2019.
\newblock Formalization of metatheory of the Quipper quantum programming
  language in a linear logic.
\newblock {\em J. Autom. Reason.\/}~{\em 63,\/}~4, 967--1002.

\bibitem[\protect\citeauthoryear{Manighetti}{Manighetti}{2022}]{manighetti22phd}
{\sc Manighetti, M.} 2022.
\newblock Developing proof theory for proof exchange.
\newblock Ph.D. thesis, Institut Polytechnique de Paris.

\bibitem[\protect\citeauthoryear{Manighetti, Miller, and Momigliano}{Manighetti
  et~al\mbox{.}}{2020}]{Manighetti0M20}
{\sc Manighetti, M.}, {\sc Miller, D.}, {\sc and} {\sc Momigliano, A.} 2020.
\newblock Two applications of logic programming to coq.
\newblock In {\em 26th International Conference on Types for Proofs and
  Programs, {TYPES} 2020, March 2-5, 2020, University of Turin, Italy},
  {U.~de'Liguoro}, {S.~Berardi}, {and} {T.~Altenkirch}, Eds. LIPIcs, vol. 188.
  Schloss Dagstuhl - Leibniz-Zentrum f{\"{u}}r Informatik, 10:1--10:19.

\bibitem[\protect\citeauthoryear{Mantovani and Momigliano}{Mantovani and
  Momigliano}{2021}]{MantovaniM21}
{\sc Mantovani, M.} {\sc and} {\sc Momigliano, A.} 2021.
\newblock Towards substructural property-based testing.
\newblock In {\em Logic-Based Program Synthesis and Transformation - 31st
  International Symposium, {LOPSTR} 2021, Tallinn, Estonia, September 7-8,
  2021, Proceedings}, {E.~D. Angelis} {and} {W.~Vanhoof}, Eds. Lecture Notes in
  Computer Science, vol. 13290. Springer, 92--112.

\bibitem[\protect\citeauthoryear{Martin}{Martin}{2010}]{Martin10}
{\sc Martin, A.} 2010.
\newblock Reasoning using higher-order abstract syntax in a higher-order logic
  proof environment: Improvements to {H}ybrid and a case study.
\newblock Ph.D. thesis, University of Ottawa.
\newblock \url{https://ruor.uottawa.ca/handle/10393/19711}.

\bibitem[\protect\citeauthoryear{McDowell and Miller}{McDowell and
  Miller}{2000}]{mcdowell00tcs}
{\sc McDowell, R.} {\sc and} {\sc Miller, D.} 2000.
\newblock Cut-elimination for a logic with definitions and induction.
\newblock {\em Theoretical Computer Science\/}~{\em 232}, 91--119.

\bibitem[\protect\citeauthoryear{McDowell and Miller}{McDowell and
  Miller}{2002}]{mcdowell02tocl}
{\sc McDowell, R.} {\sc and} {\sc Miller, D.} 2002.
\newblock Reasoning with higher-order abstract syntax in a logical framework.
\newblock {\em ACM Trans.\ on Computational Logic\/}~{\em 3,\/}~1, 80--136.

\bibitem[\protect\citeauthoryear{McDowell, Miller, and Palamidessi}{McDowell
  et~al\mbox{.}}{2003}]{mcdowell03tcs}
{\sc McDowell, R.}, {\sc Miller, D.}, {\sc and} {\sc Palamidessi, C.} 2003.
\newblock Encoding transition systems in sequent calculus.
\newblock {\em Theoretical Computer Science\/}~{\em 294,\/}~3, 411--437.

\bibitem[\protect\citeauthoryear{Michaylov and Pfenning}{Michaylov and
  Pfenning}{1992}]{michaylov91}
{\sc Michaylov, S.} {\sc and} {\sc Pfenning, F.} 1992.
\newblock Natural semantics and some of its meta-theory in {Elf}.
\newblock In {\em Extensions of Logic Programming}, {L.-H. Eriksson},
  {L.~Hallnäs}, {and} {P.~Schroeder-Heister}, Eds. Lecture Notes in Computer
  Science. Springer.

\bibitem[\protect\citeauthoryear{Midtgaard, Justesen, Kasting, Nielson, and
  Nielson}{Midtgaard et~al\mbox{.}}{2017}]{MidtgaardJKNN17}
{\sc Midtgaard, J.}, {\sc Justesen, M.~N.}, {\sc Kasting, P.}, {\sc Nielson,
  F.}, {\sc and} {\sc Nielson, H.~R.} 2017.
\newblock Effect-driven quickchecking of compilers.
\newblock {\em {PACMPL}\/}~{\em 1,\/}~{ICFP}, 15:1--15:23.

\bibitem[\protect\citeauthoryear{Miller}{Miller}{1996}]{miller96tcs}
{\sc Miller, D.} 1996.
\newblock Forum: {A} multiple-conclusion specification logic.
\newblock {\em Theoretical Computer Science\/}~{\em 165,\/}~1, 201--232.

\bibitem[\protect\citeauthoryear{Miller}{Miller}{2011}]{miller11cpp}
{\sc Miller, D.} 2011.
\newblock A proposal for broad spectrum proof certificates.
\newblock In {\em CPP: First International Conference on Certified Programs and
  Proofs}, {J.-P. Jouannaud} {and} {Z.~Shao}, Eds. Lecture Notes in Computer
  Science, vol. 7086. 54--69.

\bibitem[\protect\citeauthoryear{Miller}{Miller}{2017}]{miller17fac}
{\sc Miller, D.} 2017.
\newblock Proof checking and logic programming.
\newblock {\em Formal Aspects of Computing\/}~{\em 29,\/}~3, 383--399.

\bibitem[\protect\citeauthoryear{Miller}{Miller}{2019}]{miller18jar}
{\sc Miller, D.} 2019.
\newblock Mechanized metatheory revisited.
\newblock {\em Journal of Automated Reasoning\/}~{\em 63,\/}~3, 625--665.

\bibitem[\protect\citeauthoryear{Miller}{Miller}{2022}]{miller22tplp}
{\sc Miller, D.} 2022.
\newblock A survey of the proof-theoretic foundations of logic programming.
\newblock {\em Theory and Practice of Logic Programming\/}~{\em 22,\/}~6
  (Oct.), 859--904.
\newblock Published online November 2021.

\bibitem[\protect\citeauthoryear{Miller and Nadathur}{Miller and
  Nadathur}{2012}]{miller12proghol}
{\sc Miller, D.} {\sc and} {\sc Nadathur, G.} 2012.
\newblock {\em Programming with Higher-Order Logic}.
\newblock Cambridge University Press.

\bibitem[\protect\citeauthoryear{Miller, Nadathur, Pfenning, and
  Scedrov}{Miller et~al\mbox{.}}{1991}]{miller91apal}
{\sc Miller, D.}, {\sc Nadathur, G.}, {\sc Pfenning, F.}, {\sc and} {\sc
  Scedrov, A.} 1991.
\newblock Uniform proofs as a foundation for logic programming.
\newblock {\em Annals of Pure and Applied Logic\/}~{\em 51}, 125--157.

\bibitem[\protect\citeauthoryear{Miller and Saurin}{Miller and
  Saurin}{2006}]{MillerS06}
{\sc Miller, D.} {\sc and} {\sc Saurin, A.} 2006.
\newblock A game semantics for proof search: Preliminary results.
\newblock {\em Electr. Notes Theor. Comput. Sci.\/}~{\em 155}, 543--563.

\bibitem[\protect\citeauthoryear{Miller and Tiu}{Miller and
  Tiu}{2005}]{miller05tocl}
{\sc Miller, D.} {\sc and} {\sc Tiu, A.} 2005.
\newblock A proof theory for generic judgments.
\newblock {\em ACM Trans.\ on Computational Logic\/}~{\em 6,\/}~4 (Oct.),
  749--783.

\bibitem[\protect\citeauthoryear{Momigliano and Tiu}{Momigliano and
  Tiu}{2012}]{momigliano12jal}
{\sc Momigliano, A.} {\sc and} {\sc Tiu, A.} 2012.
\newblock Induction and co-induction in sequent calculus.
\newblock {\em Journal of Applied Logic\/}~{\em 10}, 330--367.

\bibitem[\protect\citeauthoryear{Nadathur and Pfenning}{Nadathur and
  Pfenning}{1992}]{nadathur92types}
{\sc Nadathur, G.} {\sc and} {\sc Pfenning, F.} 1992.
\newblock The type system of a higher-order logic programming language.
\newblock In {\em Types in Logic Programming}, {F.~Pfenning}, Ed. MIT Press,
  245--283.

\bibitem[\protect\citeauthoryear{Palka, Claessen, Russo, and Hughes}{Palka
  et~al\mbox{.}}{2011}]{PalkaCRH11}
{\sc Palka, M.~H.}, {\sc Claessen, K.}, {\sc Russo, A.}, {\sc and} {\sc Hughes,
  J.} 2011.
\newblock Testing an optimising compiler by generating random lambda terms.
\newblock In {\em Proceedings of the 6th International Workshop on Automation
  of Software Test, {AST} 2011, Waikiki, Honolulu, HI, USA, May 23-24, 2011},
  {A.~Bertolino}, {H.~Foster}, {and} {J.~J. Li}, Eds. {ACM}, 91--97.

\bibitem[\protect\citeauthoryear{Paraskevopoulou, Eline, and
  Lampropoulos}{Paraskevopoulou et~al\mbox{.}}{2022}]{Paraskevopoulou22}
{\sc Paraskevopoulou, Z.}, {\sc Eline, A.}, {\sc and} {\sc Lampropoulos, L.}
  2022.
\newblock Computing correctly with inductive relations.
\newblock In {\em {PLDI} '22: 43rd {ACM} {SIGPLAN} International Conference on
  Programming Language Design and Implementation, San Diego, CA, USA, June 13 -
  17, 2022}, {R.~Jhala} {and} {I.~Dillig}, Eds. {ACM}, 966--980.

\bibitem[\protect\citeauthoryear{Paraskevopoulou, Hritcu, D{\'{e}}n{\`{e}}s,
  Lampropoulos, and Pierce}{Paraskevopoulou et~al\mbox{.}}{2015}]{QChick}
{\sc Paraskevopoulou, Z.}, {\sc Hritcu, C.}, {\sc D{\'{e}}n{\`{e}}s, M.}, {\sc
  Lampropoulos, L.}, {\sc and} {\sc Pierce, B.~C.} 2015.
\newblock Foundational property-based testing.
\newblock In {\em Interactive Theorem Proving - 6th International Conference,
  {ITP} 2015, Proceedings}, {C.~Urban} {and} {X.~Zhang}, Eds. Lecture Notes in
  Computer Science, vol. 9236. Springer, 325--343.

\bibitem[\protect\citeauthoryear{Pemmasani, Guo, Dong, Ramakrishnan, and
  Ramakrishnan}{Pemmasani et~al\mbox{.}}{2004}]{onlinejust}
{\sc Pemmasani, G.}, {\sc Guo, H.-F.}, {\sc Dong, Y.}, {\sc Ramakrishnan,
  C.~R.}, {\sc and} {\sc Ramakrishnan, I.~V.} 2004.
\newblock Online justification for tabled logic programs.
\newblock In {\em Functional and Logic Programming}, {Y.~Kameyama} {and} {P.~J.
  Stuckey}, Eds. Springer Berlin Heidelberg, Berlin, Heidelberg, 24--38.

\bibitem[\protect\citeauthoryear{Pfenning}{Pfenning}{2000}]{pfenning00ic}
{\sc Pfenning, F.} 2000.
\newblock Structural cut elimination {I}. Intuitionistic and classical logic.
\newblock {\em Information and Computation\/}~{\em 157,\/}~1/2 (Mar.), 84--141.

\bibitem[\protect\citeauthoryear{Pfenning and Elliott}{Pfenning and
  Elliott}{1988}]{pfenning88pldi}
{\sc Pfenning, F.} {\sc and} {\sc Elliott, C.} 1988.
\newblock Higher-order abstract syntax.
\newblock In {\em Proceedings of the {ACM}-{SIGPLAN} Conference on Programming
  Language Design and Implementation}. ACM Press, 199--208.

\bibitem[\protect\citeauthoryear{Pfenning and Sch{\"u}rmann}{Pfenning and
  Sch{\"u}rmann}{1999}]{pfenning99cade}
{\sc Pfenning, F.} {\sc and} {\sc Sch{\"u}rmann, C.} 1999.
\newblock System description: {Twelf} --- {A} meta-logical framework for
  deductive systems.
\newblock In {\em 16th Conf.\ on Automated Deduction (CADE)}, {H.~Ganzinger},
  Ed. Number 1632 in LNAI. Springer, Trento, 202--206.

\bibitem[\protect\citeauthoryear{Pfenning and Simmons}{Pfenning and
  Simmons}{2009}]{PfenningS09}
{\sc Pfenning, F.} {\sc and} {\sc Simmons, R.~J.} 2009.
\newblock Substructural operational semantics as ordered logic programming.
\newblock In {\em {LICS}}. {IEEE} Computer Society, 101--110.

\bibitem[\protect\citeauthoryear{Pientka and Cave}{Pientka and
  Cave}{2015}]{PientkaC15}
{\sc Pientka, B.} {\sc and} {\sc Cave, A.} 2015.
\newblock Inductive Beluga: Programming proofs.
\newblock In {\em Automated Deduction - {CADE-25} - 25th International
  Conference on Automated Deduction, Berlin, Germany, August 1-7, 2015,
  Proceedings}, {A.~P. Felty} {and} {A.~Middeldorp}, Eds. Lecture Notes in
  Computer Science, vol. 9195. Springer, 272--281.

\bibitem[\protect\citeauthoryear{Pientka and Dunfield}{Pientka and
  Dunfield}{2010}]{pientka10ijcar}
{\sc Pientka, B.} {\sc and} {\sc Dunfield, J.} 2010.
\newblock Beluga: {A} framework for programming and reasoning with deductive
  systems (system description).
\newblock In {\em Fifth International Joint Conference on Automated Reasoning},
  {J.~Giesl} {and} {R.~H{\"a}hnle}, Eds. Number 6173 in Lecture Notes in
  Computer Science. 15--21.

\bibitem[\protect\citeauthoryear{Pitts}{Pitts}{2003}]{pitts03ic}
{\sc Pitts, A.~M.} 2003.
\newblock Nominal logic, {A} first order theory of names and binding.
\newblock {\em Information and Computation\/}~{\em 186,\/}~2, 165--193.

\bibitem[\protect\citeauthoryear{Polakow and Yi}{Polakow and
  Yi}{2000}]{PolakowY00}
{\sc Polakow, J.} {\sc and} {\sc Yi, K.} 2000.
\newblock Proving syntactic properties of exceptions in an ordered logical
  framework.
\newblock In {\em The First Asian Workshop on Programming Languages and
  Systems, {APLAS} 2000, National University of Singapore, Singapore, December
  18-20, 2000, Proceedings}. 23--32.

\bibitem[\protect\citeauthoryear{Qi, Gacek, Holte, Nadathur, and Snow}{Qi
  et~al\mbox{.}}{2015}]{teyjus.website}
{\sc Qi, X.}, {\sc Gacek, A.}, {\sc Holte, S.}, {\sc Nadathur, G.}, {\sc and}
  {\sc Snow, Z.} 2015.
\newblock The {T}eyjus system -- version 2.
\newblock \url{http://teyjus.cs.umn.edu/}.

\bibitem[\protect\citeauthoryear{Runciman, Naylor, and Lindblad}{Runciman
  et~al\mbox{.}}{2008}]{smallcheck}
{\sc Runciman, C.}, {\sc Naylor, M.}, {\sc and} {\sc Lindblad, F.} 2008.
\newblock Smallcheck and {L}azy {S}mallcheck: automatic exhaustive testing for
  small values.
\newblock In {\em Haskell}. {ACM}, 37--48.

\bibitem[\protect\citeauthoryear{Schack{-}Nielsen and
  Sch{\"{u}}rmann}{Schack{-}Nielsen and
  Sch{\"{u}}rmann}{2008}]{Schack-NielsenS08}
{\sc Schack{-}Nielsen, A.} {\sc and} {\sc Sch{\"{u}}rmann, C.} 2008.
\newblock Celf - {A} logical framework for deductive and concurrent systems
  (system description).
\newblock In {\em {IJCAR}}. Lecture Notes in Computer Science, vol. 5195.
  Springer, 320--326.

\bibitem[\protect\citeauthoryear{Schrijvers, Demoen, Triska, and
  Desouter}{Schrijvers et~al\mbox{.}}{2014}]{Tor}
{\sc Schrijvers, T.}, {\sc Demoen, B.}, {\sc Triska, M.}, {\sc and} {\sc
  Desouter, B.} 2014.
\newblock Tor: Modular search with hookable disjunction.
\newblock {\em Sci. Comput. Program.\/}~{\em 84}, 101--120.

\bibitem[\protect\citeauthoryear{Schroeder-Heister}{Schroeder-Heister}{1993}]{schroeder-heister93lics}
{\sc Schroeder-Heister, P.} 1993.
\newblock Rules of definitional reflection.
\newblock In {\em 8th Symp.\ on Logic in Computer Science}, {M.~Vardi}, Ed.
  IEEE Computer Society Press, IEEE, 222--232.

\bibitem[\protect\citeauthoryear{Sch{\"{u}}rmann}{Sch{\"{u}}rmann}{2000}]{schurmann00phd}
{\sc Sch{\"{u}}rmann, C.} 2000.
\newblock Automating the meta theory of deductive systems.
\newblock Ph.D. thesis, Carnegie Mellon University.
\newblock {CMU-CS-00-146}.

\bibitem[\protect\citeauthoryear{Selinger}{Selinger}{2008}]{Selinger08}
{\sc Selinger, P.} 2008.
\newblock Lecture notes on the lambda calculus.
\newblock \url{https://arxiv.org/abs/0804.3434}.

\bibitem[\protect\citeauthoryear{Sullivan, Yang, Coppit, Khurshid, and
  Jackson}{Sullivan et~al\mbox{.}}{2004}]{Sullivan:2004}
{\sc Sullivan, K.}, {\sc Yang, J.}, {\sc Coppit, D.}, {\sc Khurshid, S.}, {\sc
  and} {\sc Jackson, D.} 2004.
\newblock Software assurance by bounded exhaustive testing.
\newblock In {\em Proceedings of the 2004 ACM SIGSOFT International Symposium
  on Software Testing and Analysis}. ISSTA '04. ACM, New York, NY, USA,
  133--142.

\bibitem[\protect\citeauthoryear{Takahashi}{Takahashi}{1995}]{Takahashi95}
{\sc Takahashi, M.} 1995.
\newblock Parallel reductions in lambda-calculus.
\newblock {\em Inf. Comput.\/}~{\em 118,\/}~1, 120--127.

\bibitem[\protect\citeauthoryear{Tarau and de~Paiva}{Tarau and
  de~Paiva}{2020}]{Tarau20}
{\sc Tarau, P.} {\sc and} {\sc de~Paiva, V.} 2020.
\newblock Deriving theorems in implicational linear logic, declaratively.
\newblock In {\em {ICLP} Technical Communications}. {EPTCS}, vol. 325.
  110--123.

\bibitem[\protect\citeauthoryear{Tassi}{Tassi}{2018}]{tassi:hal-01637063}
{\sc Tassi, E.} 2018.
\newblock {Elpi: an extension language for Coq (Metaprogramming Coq in the Elpi
  $\lambda$Prolog dialect)}.
\newblock Working paper or preprint.

\bibitem[\protect\citeauthoryear{Tiu and Miller}{Tiu and
  Miller}{2010}]{tiu10tocl}
{\sc Tiu, A.} {\sc and} {\sc Miller, D.} 2010.
\newblock Proof search specifications of bisimulation and modal logics for the
  $\pi$-calculus.
\newblock {\em ACM Trans.\ on Computational Logic\/}~{\em 11,\/}~2,
  13:1--13:35.

\end{thebibliography}

\end{document}